\documentclass[12pt]{article}
\usepackage{amsmath}
\usepackage{epic}
\usepackage{graphics, color}
\usepackage{amsthm}
\usepackage{amsmath}
\usepackage{amsfonts}
\usepackage[dvips]{graphicx}
\usepackage{amssymb}

\def\pplogo{\vbox{\kern-\headheight\kern -29pt
\halign{##&##\hfil\cr&{\ppnumber}\cr\rule{0pt}{2.5ex}&\ppdate\cr}}}
\makeatletter
\def\ps@firstpage{\ps@empty \def\@oddhead{\hss\pplogo}
\let\@evenhead\@oddhead
}
\def\maketitle{\par
\begingroup
\def\thefootnote{\fnsymbol{footnote}}
\def\@makefnmark{\hbox{$^{\@thefnmark}$\hss}}
\if@twocolumn \twocolumn[\@maketitle] \else \newpage
\global\@topnum\z@ \@maketitle
\fi\thispagestyle{firstpage}\@thanks
\endgroup
\setcounter{footnote}{0}
\let\maketitle\relax
\let\@maketitle\relax
\gdef\@thanks{}\gdef\@author{}\gdef\@title{}\let\thanks\relax}
\makeatother

\def\beq{\begin{equation}}
\def\eeq{\end{equation}}
\def\bea{\begin{eqnarray}}
\def\eea{\end{eqnarray}}
\def\ba{\begin{array}}
\def\ea{\end{array}}
\def\bec{\begin{center}}
\def\ec{\end{center}}
\def\nl{\nonumber\\}

\numberwithin{equation}{section}

\begin{document}

\setcounter{page}0

\def\ppnumber{\vbox{\baselineskip14pt
}}
\def\ppdate{\footnotesize{
}}
\date{}

\author{\textbf{Saebyok Bae}\footnote{The author wrote
papers such as Phys.~Lett.~{\bf B487}, 299 (2000)
[arXiv:hep-ph/0002224].}
\\
[7mm]
{
\small \it Department of Physics {\rm \&} Science Education Center
for the Gifted,} \\ {\small \it Gyeongsang National University,
Jinju 52828, South Korea}\thanks{{\it Email addresses:}
sbae@gnu.ac.kr, sbae@kaist.edu. One of previous addresses is
Department of Physics, KAIST, Daejeon, Korea.} 
} 


\title{\bf A Particle Model of Our Spacetime\,: \\
Origin of Gravity \vskip 0.4cm}

\maketitle


\begin{abstract}
\normalsize %
We build a model of our spacetime by assuming new particles called
``space quanta." In the ambient or bulk spacetime ${\cal
S}^{D_{\rm amb}}$ ($D_{\rm amb} \ge 4$), a multitude of space
quanta form a nearly three-dimensional object, whose continuum
approximation is called the space 3-brane. The world volume ${\cal
WV}_{\rm sq}$ of this space 3-brane is described by an embedding
$f^A(x^\mu) \in {\cal S}^{D_{\rm amb}}$, which produces the
induced metric $\gamma_{\mu \nu}$ on the world volume ${\cal
WV}_{\rm sq}$. This emergent spacetime (${\cal WV}_{\rm sq},
\gamma_{\mu \nu}$) from the many space quanta is proposed as the
particle model of our spacetime. To study our spacetime (${\cal
WV}_{\rm sq}, \gamma_{\mu \nu}$), we construct what we call the
Aim-At-Target (AAT) method, which introduces an action for a 4D
metric $g_{\mu \nu}$. This metric action from the AAT method can
lead to General Relativity at low enough energies. The spacetime
(${\cal S}_{\rm GR}, {\bf g}_{\mu \nu}$) of General Relativity is,
at least, a good approximation to the exact or true spacetime
(${\cal WV}_{\rm sq}, \gamma_{\mu \nu}$) of our universe.
\end{abstract}

\bigskip

\newpage



\section{\label{sec: intro} Introduction}

The gravitational physics has been successfully understood in
terms of General Relativity \cite{{GR-Schu},{GR-Wein},{GR-Carr}}.
However, since the non-gravitational physics has been accurately
explained by the principles of quantum mechanics, it seems
necessary that General Relativity is merged with quantum mechanics
\cite{QG}. For the quantum theory of gravity \cite{QG}, there have
been attempts such as string theory \cite{string-th}.

In this paper, we present a particle model of our spacetime, and
explain the origin of gravity (i.e., General Relativity), as
follows: in the ambient or bulk spacetime ${\cal S}^{D_{\rm amb}}$
($D_{\rm amb} \ge 4$), there exist new particles called ``space
quanta." A multitude of space quanta form a nearly
three-dimensional object, which is called the ``quasi-3D object."
Within this quasi-3D object, the average distance $d_{\rm \, sq}$
between nearest-neighbor space quanta satisfies $d_{\rm \, sq}
\lesssim O(M_{\rm P}^{-1})$, where $M_{\rm P}$ is the Planck mass
$\approx 10^{19} \, {\rm GeV}$.

At low energies $\lesssim O(0.1) d_{\rm \, sq}^{\, -1}$, we can
use a continuum approximation \cite{conti-mech} that the quasi-3D
object is replaced with a 3D continuum called the ``space
3-brane." Like a bosonic string \cite{{string-th},{Goto}}, this
space 3-brane sweeps out its 4D ``world volume" ${\cal WV}_{\rm
sq}$ in the ambient spacetime ${\cal S}^{D_{\rm amb}}$. This world
volume ${\cal WV}_{\rm sq}$ is described by an embedding
$f^A(x^\mu) \in {\cal S}^{D_{\rm amb}}$, which produces the
induced metric $\gamma_{\mu \nu}$ on ${\cal WV}_{\rm sq}$. This
emergent spacetime (${\cal WV}_{\rm sq}, \gamma_{\mu \nu}$) from
the many space quanta is proposed as the particle model of our
spacetime. The dynamics of the embedding $f^A$ is provided by an
effective theory $S_{\rm univ}^{\rm (3 br)}[f^A, \cdots] = S_{\rm
emb}^{\rm (3 br)}[f^A] + \cdots$, where the latter ellipsis
$\cdots$ denotes the action for the matter sector (e.g., the
Standard Model).

To study our spacetime (${\cal WV}_{\rm sq}, \gamma_{\mu \nu}$),
we construct the ``Aim-At-Target (AAT) method," which introduces
an action for a 4D metric $g_{\mu \nu}$, namely, $S_{\rm
univ}^{\rm (ovlp)}[g_{\mu \nu}, \cdots] = S_{\rm met}^{\rm
(ovlp)}[g_{\mu \nu}] + \cdots$, where the latter ellipsis $\cdots$
denotes the action for the matter sector. This new metric $g_{\mu
\nu}$ is used for finding the embedding $f^A(x^\mu)$ through the
equality $g_{\mu \nu} = \gamma_{\mu \nu}$, as follows:

For an easy understanding of the AAT method, we consider a simple
situation that the universe contains only the space-quantum sector
(i.e., the space 3-brane){---\,}the absence of the matter sector.
In this situation, we only have to study the two simpler actions
$S_{\rm emb}^{\rm (3 br)}[f^A]$ and $S_{\rm met}^{\rm
(ovlp)}[g_{\mu \nu}]$, which are the actions without the matter
sector. For example, when the ambient spacetime ${\cal S}^{D_{\rm
amb}}$ is the Minkowski spacetime $\mathbb{M}^{D_{\rm amb}}$ of
the flat metric $\eta_{AB}^{\rm bulk}$, the induced metric
$\gamma_{\mu \nu}$ is represented as 
\beq \label{intro-ind-met} \gamma_{\mu \nu} \, = \, \partial_\mu
f_{\rm sol}^A \, \partial_\nu f_{\rm sol}^B \, \eta_{AB}^{\rm
bulk} ~, \eeq 
where $f_{\rm sol}^A$ is a solution for the $f^A$ equation of
motion $\delta S_{\rm emb}^{\rm (3 br)} / \delta f^A = 0$.

For the above metric action $S_{\rm met}^{\rm (ovlp)}[g_{\mu
\nu}]$, when a solution $g_{\mu \nu}^{\rm sol}$ of $\delta S_{\rm
met}^{\rm (ovlp)} / \delta g_{\mu \nu} = 0$ satisfies the equality
\beq \label{intro-met=ind} g_{\mu \nu}^{\rm sol} \, = \,
\gamma_{\mu \nu} ~, \eeq 
Eqs.~(\ref{intro-ind-met}) and (\ref{intro-met=ind}) require that
the solution $f_{\rm sol}^A$ of $\delta S_{\rm emb}^{\rm (3 br)} /
\delta f^A = 0$ should also be a solution of the partial
differential equation (PDE) for $f^A$ 
\beq \label{intro-rifle-PDE} \partial_\mu f^A \, \partial_\nu f^B
\, \eta_{AB}^{\rm bulk} \, = \, g_{\mu \nu}^{\rm sol} ~. \eeq 
In other words, when the new metric $g_{\mu \nu}^{\rm sol}$
satisfies $g_{\mu \nu}^{\rm sol} = \gamma_{\mu \nu}$, the solution
$f_{\rm sol}^A$ of the equation of motion $\delta S_{\rm emb}^{\rm
(3 br)} / \delta f^A = 0$ can be found by solving the PDE
$\partial_\mu f^A \partial_\nu f^B \eta_{AB}^{\rm bulk} = g_{\mu
\nu}^{\rm sol}$. Note that the embedding $f^A$ is similar to the
locally inertial coordinates $\xi^{\hat{\alpha}}$ of General
Relativity, because the PDE $\partial_\mu f^A \partial_\nu f^B
\eta_{AB}^{\rm bulk} = g_{\mu \nu}^{\rm sol}$ is similar in form
to $\partial_\mu \xi^{\hat{\alpha}} \partial_\nu \xi^{\hat{\beta}}
\eta_{\hat{\alpha} \hat{\beta}} = {\bf g}_{\mu \nu}$, where
$\partial_\mu \xi^{\hat{\alpha}}$ is the vierbein \cite{GR-Wein}.

To sum up, the AAT method using the metric action $S_{\rm met}
[g_{\mu \nu}]$ consists of two main steps: (i) finding a solution
$g_{\mu \nu}^{\rm sol}$ of $(\delta S_{\rm met} / \delta g_{\mu
\nu})[g_{\mu \nu}] = 0$, and next (ii) finding a solution $f_{\rm
PDE \cdot sol}^A$ of $\partial_\mu f^A \partial_\nu f^B
\eta_{AB}^{\rm bulk} = g_{\mu \nu}^{\rm sol}$. In case of $g_{\mu
\nu}^{\rm sol} = \gamma_{\mu \nu}$ ($= \partial_\mu f_{\rm sol}^A
\partial_\nu f_{\rm sol}^B \eta_{AB}^{\rm bulk}$), we can find a
solution $f_{\rm PDE \cdot sol}^A$ satisfying $f_{\rm PDE \cdot
sol}^A = f_{\rm sol}^A$, where $f_{\rm sol}^A$ is what we really
want to know.

Then, as far as the equality $g_{\mu \nu}^{\rm sol} = \gamma_{\mu
\nu}$ remains true, the ``combination" of 
\beq \label{combi} ~~~~ ({\rm a}) ~~ {\rm the ~ metric ~ action} ~
S_{\rm met} [g_{\mu \nu}] ~~~\, {\rm and} ~~~ ({\rm b}) ~\, {\rm
the ~ PDE} ~ \partial_\mu f^A \partial_\nu f^B \eta_{AB}^{\rm
bulk} = g_{\mu \nu}^{\rm sol} \eeq 
can be used instead of the 3-brane action $S_{\rm emb}^{\rm (3
br)}[f^A]$. This is the essential feature of the AAT method.

At low enough energies, the metric action $S_{\rm met} [g_{\mu
\nu}]$ in Eq.~(\ref{combi}) can be well approximated by the
Einstein-Hilbert action $S_{\rm EH}$ of General Relativity. Then,
this Einstein-Hilbert action $S_{\rm EH}$ can be a good low-energy
description for the 3-brane action $S_{\rm emb}^{\rm (3 br)}[f^A]$
in the absence of the matter sector. When the matter sector is
present, the whole action of General Relativity can be a good
low-energy description for the above ``universe action" $S_{\rm
univ}^{\rm (3 br)}[f^A, \cdots]$. In this manner, the AAT method
can produce General Relativity at low enough energies{---\,}this
explains the origin of gravity (i.e., General Relativity).

Since, in the AAT method, General Relativity can be subsidiary to
the universe action $S_{\rm univ}^{\rm (3 br)}[f^A, \cdots]$ (see
around Eq.~(\ref{combi})), we must not forget that, at the most
fundamental level, our universe should be studied in terms of {\it
physical laws within the ambient spacetime} ${\cal S}^{D_{\rm
amb}}$ which govern both the space-quantum and matter sectors of
our universe. These physical laws within the ambient spacetime
${\cal S}^{D_{\rm amb}}$ can be represented as quantum field
theories defined on ${\cal S}^{D_{\rm amb}}${---\,}this may shed
some light on the quantum theory of gravity \cite{QG}.

Meanwhile, like usual many-particle systems (e.g.,
superconductors), our universe as a system in the ambient
spacetime ${\cal S}^{D_{\rm amb}}$ consists of an enormous number
of particles such as space quanta. Thus, useful ideas for the
study of our universe in ${\cal S}^{D_{\rm amb}}$ can be found by
surveying {\it physics in our spacetime} ${\cal WV}_{\rm sq}$, for
example, condensed matter physics \cite{cond-matt}.

The rest of this paper is organized, as follows: in Sec.~\ref{sec:
sp-qu-hypothesis}, the wave-particle duality of quantum mechanics
is applied to the gravitational field. Since the particle nature
of the gravitational field implies the existence of the new
particle (i.e., space quantum), the spacetime manifold ${\cal
S}_{\rm GR}$ of General Relativity is assumed to consist of (very
many) space quanta{---\,}this is called the space-quantum
hypothesis.

In Sec.~\ref{sec: cont-approx}, to maintain the wave nature of a
single space quantum (even without any other space quanta), we
assume that there exists the ambient spacetime ${\cal S}^{D_{\rm
amb}}$, which surrounds the spacetime ${\cal S}_{\rm GR}$ of
General Relativity. To explain the 3D space part of the GR
spacetime ${\cal S}_{\rm GR}$, we assume that space quanta in
${\cal S}^{D_{\rm amb}}$ form the quasi-3D object, whose continuum
limit is the space 3-brane.

In Sec.~\ref{sec: 3br-eff-action}, we deal with the kinematics of
the space 3-brane, whose world volume ${\cal WV}_{\rm sq}$ is
described by an embedding $f^A(x^\mu)$. The induced metric
$\gamma_{\mu \nu}$ on the world volume ${\cal WV}_{\rm sq}$ can be
approximated by the GR metric ${\bf g}_{\mu \nu}$. For simplicity,
we consider the effective theory $S_{\rm emb}^{\rm (3 br)}[f^A]$
{\it only} for the space 3-brane (the action for matter will be
studied in Sec.~\ref{sec: EFT-univ}).

In Sec.~\ref{sec: AAT-method}, we present the Aim-At-Target (AAT)
method for studying the effective theory $S_{\rm emb}^{\rm (3
br)}[f^A]$ of the space 3-brane. This AAT method using a metric
action $S_{\rm met} [g_{\mu \nu}]$ contains the coupled equations
$\delta S_{\rm met} / \delta g_{\mu \nu} = 0$ and $\partial_\mu
f^A \partial_\nu f^B \eta_{AB}^{\rm bulk} = g_{\mu \nu}$. As far
as $g_{\mu \nu} = \gamma_{\mu \nu}$ holds good, the metric action
$S_{\rm met} [g_{\mu \nu}]$ can replace the 3-brane action $S_{\rm
emb}^{\rm (3 br)}[f^A]$.

In Sec.~\ref{sec: form-met}, in terms of symmetries, we study the
forms of the metric action $S_{\rm met} [g_{\mu \nu}]$ used in the
AAT method. The ${\rm Diff}(4)$-invariant action $S_{\rm
met}[g_{\mu \nu}]$ can explain the Einstein-Hilbert action $S_{\rm
EH}[{\bf g}_{\mu \nu}]$, which is an essential part of General
Relativity.

In Sec.~\ref{sec: EFT-univ}, since the universe contains the
matter sector, we consider the more general action $S_{\rm
univ}^{\rm (3 br)}[f^A, \cdots] = S_{\rm emb}^{\rm (3 br)}[f^A] +
\cdots$ for the inclusion of matter. By applying the AAT method
similarly, we can obtain General Relativity at low enough
energies.

\section{\label{sec: sp-qu-hypothesis} Applying Quantum
Mechanics to Gravity: Space as a Discrete System of Particles}

Quantum mechanics explains many phenomena of nature very well.
Thus, we can try to combine gravity with it (i.e., a quantum
theory of gravitation). Because quantum mechanics has the
wave-particle duality as its signature property, we further think
about the basic concept {\bf particle}: since the wave-particle
duality has been successfully applied to ordinary sensible objects
like light and matter, considering these ordinary objects (rather
than graviton) is helpful in understanding the concept of
particle.

An ordinary material object (e.g., a bearing ball) has a
``substance" characterized by (i) stuff material (e.g., metal) and
(ii) shape in space (e.g., ball or sphere), which may correspond
to ``matter" ({\it hyle} in Greek) and ``form" ({\it eidos} or
{\it morphe}) of the ancient Greek philosophy, respectively.
Therefore, an object is called a ``particle" if the shape of its
substance is point-like in space while the object exists. The
particle nature of material objects like electron is evident.

Because the important quantum phenomena like the photoelectric
effect and the Davisson-Germer experiment have been observed by
laboratory frames (e.g., $O_{\rm rest}$ of Fig.~1(a)) under the
influence of gravity, the wave-particle duality of quantum
mechanics must be observed by the rest frame $O_{\rm rest}$. In
addition, through the general covariance, the wave-particle
duality is also observed by the freely-falling frame $O_{\rm FF}$
of Fig.~1(a).

In the weak-field situation of General Relativity (GR)
\cite{{GR-Schu},{GR-Wein},{GR-Carr}}, there exists a ``nearly
Lorentz (NL) coordinate system" $x_{\rm NL}^{\, \mu}$ relative to
which the metric ${\bf g}_{\mu \nu}$ of a slightly curved GR
spacetime ${\cal S}_{\rm GR}^{\rm weak}$ has the components at
every point $p$ of the spacetime ${\cal S}_{\rm GR}^{\rm weak}$ 
\beq \label{weakfield} ~~~~~~~~~~~~~ {\bf g}_{\mu \nu} \, = \,
\eta_{\mu \nu} \, + \, {\bf h}_{\mu \nu} ~~~ {\rm with} ~~ |\,
{\bf h}_{\mu \nu} | \, \ll \, 1 ~~~~ {\rm at} ~ {\rm every} ~ p
\in {\cal S}_{\rm GR}^{\rm weak} ~, \eeq 
where $\eta_{\mu \nu} = {\rm diag}(-1, +1, +1, +1)$ in the {\it
mostly plus} convention is called the flat ``background metric,"
and ${\bf h}_{\mu \nu}$ a small ``perturbation" \cite{GR-Schu}.

\begin{figure}
\centering
\begin{minipage}{0.9\textwidth}
\includegraphics[width=9cm]{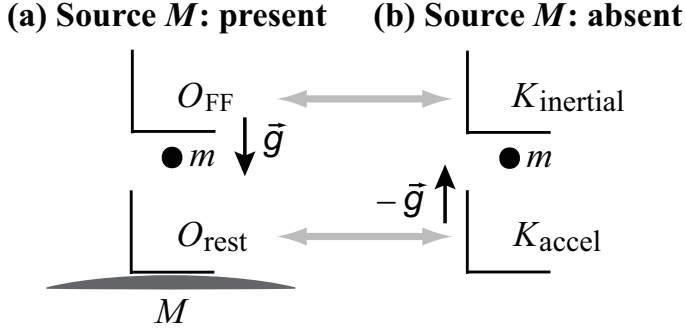}
\caption{\label{equiv} The correspondence between two different
situations distinguished by the existence of a gravitational
source $M$ (e.g., the earth), namely, (a) the {\it
source-present}, and (b) the {\it source-absent} situations. There
are two kinds of ``two-frame equivalences," (i) the ``non-inertial
equivalence" between $O_{\rm rest}$ and $K_{\rm accel}$, and (ii)
the ``inertial equivalence" between $O_{\rm FF}$ and $K_{\rm
inertial}$.}
\end{minipage} \end{figure}

For the above NL coordinates $x_{\rm NL}^{\, \mu}$, the Einstein
tensor $\textit{\textbf{G}}_{\mu \nu} ({\bf g}_{\rho \sigma}) =
\textit{\textbf{R}}_{\mu \nu} - \textit{\textbf{R}} \, {\bf
g}_{\mu \nu} / 2$ has the series expansion in powers of the
perturbation ${\bf h}_{\mu \nu}$
\cite{{GR-Schu},{GR-Wein},{GR-Carr}} 
\beq \label{series-G} ~~~~~~~~~~~~ \textit{\textbf{G}}_{\mu \nu}
(\eta_{\rho \sigma} + {\bf h}_{\rho \sigma}) ~ = ~
\textit{\textbf{G}}_{\mu \nu}^{\, (1)} \, + \, O({\bf h}^2) ~~~~
{\rm with} ~~ \textit{\textbf{G}}_{\mu \nu}^{\, (1)} \,
\stackrel{\rm def}{=} \, (\partial_\rho \partial_\nu {\bf
h}_\mu^\rho + \cdots) / 2 ~. \eeq 
Then, the {\it vacuum} Einstein's equation
$\textit{\textbf{G}}_{\mu \nu} = 0$ has its first-order
approximation 
\beq \label{lin-E-eq} \textit{\textbf{G}}_{\mu \nu}^{\, (1)} ~ = ~
0 ~. \eeq 

Under a ``background Lorentz transformation" with $\frac{\,
\partial x_{\rm NL}^{\prime \mu}}{\, \partial x_{\rm NL}^\rho} \in
SO(1, 3)$, the perturbation ${\bf h}_{\mu \nu}$ in
Eq.~(\ref{weakfield}) transforms like ${\bf h}_{\mu \nu}^\prime =
\frac{\, \partial x_{\rm NL}^\rho}{\, \partial x_{\rm NL}^{\prime
\mu}} \, \frac{\, \partial x_{\rm NL}^\sigma}{\, \partial x_{\rm
NL}^{\prime \nu}} \, {\bf h}_{\rho \sigma}$ as if it were a
Lorentz tensor defined on the flat Minkowski spacetime
$\mathbb{M}^4$. This leads to the ``flat-spacetime fiction" that
the tensor ${\bf h}_{\mu \nu}$ belongs to a {\it theory in the
flat spacetime} $\mathbb{M}^4$ \cite{GR-Schu}. This fiction is
supported by the Fierz-Pauli (F-P) theory, where gravity is
described by a symmetric tensor on the flat spacetime
$\mathbb{M}^4$ \cite{FP-action}.

Because the F-P theory shares $\textit{\textbf{G}}_{\mu \nu}^{\,
(1)} = 0$ with General Relativity, the curved spacetime ${\cal
S}_{\rm GR}^{\rm weak} = (\, \mathbb{R}^4 , \, \eta_{\mu \nu} +
{\bf h}_{\mu \nu})$ of General Relativity can be interpreted as
the combination of (i) the flat spacetime $\mathbb{M}^4 = (\,
\mathbb{R}^4 , \, \eta_{\mu \nu})$, and (ii) the field ${\bf
h}_{\mu \nu}$ propagating in this $\mathbb{M}^4$. This
interpretation about ${\cal S}_{\rm GR}^{\rm weak}$ is expressed
as 
\beq \label{cl-equiv} {\cal S}_{\rm GR}^{\rm weak} ~ \, \equiv ~
\, \mathbb{M}^4 ~ \oplus ~ {\bf h}_{\mu \nu} ~. \eeq 

Since the linearized vacuum Einstein's equation
$\textit{\textbf{G}}_{\mu \nu}^{\, (1)} = 0$ in
Eq.~(\ref{lin-E-eq}) has plane-wave solutions, its solution ${\bf
h}_{\mu \nu}$ in Eq.~(\ref{cl-equiv}) can be the superposition of
plane-wave solutions 
\beq \label{pl-wave} ~~ {\bf h}_{\mu \nu}(x_{\rm NL}) ~ = ~
\sum_{\sigma} \int d^{\, 3} k \left[ \, {\bf a}(\vec{k}, \sigma)
\, e_{\mu \nu}(\vec{k}, \sigma) \, e^{+ \, i \, k_\rho x_{\rm
NL}^\rho} \, + \, {\bf a}^*(\vec{k}, \sigma) \, e_{\mu
\nu}^*(\vec{k}, \sigma) \, e^{- \, i \, k_\rho x_{\rm NL}^\rho} \,
\right] \, , \eeq 
where $e_{\mu \nu}(\vec{k}, \sigma)$ is a polarization tensor for
wave vector $\vec{k}$ and helicity $\sigma$ \cite{GR-Wein}.

As in the field quantization of the F-P theory, the amplitudes
${\bf a}(\vec{k}, \sigma)$ and ${\bf a}^*(\vec{k}, \sigma)$ in
Eq.~(\ref{pl-wave}) are replaced with the annihilation and
creation operators $\widehat{{\bf a}}(\vec{k}, \sigma)$ and
$\widehat{{\bf a}}^{\, \dagger}(\vec{k}, \sigma)$ for a
``particle" called the graviton{---\,}the wave-particle duality is
applied to the ``wave" ${\bf h}_{\mu \nu}$. The graviton for the
field operator $\widehat{{\bf h}}_{\mu \nu}(x_{\rm NL})$ is a
massless spin-2 particle moving in the flat spacetime
$\mathbb{M}^4$.

After the field quantization, the ``classical field" ${\bf h}_{\mu
\nu}$ in the expression ${\cal S}_{\rm GR}^{\rm weak} \equiv
\mathbb{M}^4 \, \oplus \, {\bf h}_{\mu \nu}$ corresponds to
``gravitons," whose number is denoted by $N_{\rm gr}$ ($\ge 1$).
Then, the classical relation ${\cal S}_{\rm GR}^{\rm weak} \equiv
\mathbb{M}^4 \, \oplus \, {\bf h}_{\mu \nu}$ in
Eq.~(\ref{cl-equiv}) is replaced with its semi-classical
counterpart 
\beq \label{semi-equiv} {\cal S}_{\rm GR}^{\rm weak} ~ \, \equiv ~
\, \mathbb{M}^4 ~ \oplus ~ {\rm gravitons} ~, \eeq 
which means that the curved spacetime ${\cal S}_{\rm GR}^{\rm
weak}$ of General Relativity is the combination of (i) the flat
spacetime $\mathbb{M}^4$ and (ii) the $N_{\rm gr}$ gravitons
moving in this $\mathbb{M}^4$.

In other words, the curved GR spacetime ${\cal S}_{\rm GR}^{\rm
weak}$ is formed by adding the gravitons (i.e., {\it particles})
to the flat spacetime $\mathbb{M}^4$. The ``gravitons" in
Eq.~(\ref{semi-equiv}) can be regarded as the building blocks of
the difference between ${\cal S}_{\rm GR}^{\rm weak}$ and
$\mathbb{M}^4$. For example, the  GR spacetime ${\cal S}_{\rm
GR}^{\rm weak}$ depends on the number $N_{\rm gr}$ and the
locations of the gravitons.

Of course, the flat spacetime $\mathbb{M}^4$ in
Eq.~(\ref{semi-equiv}) may be such a bizarre entity that it does
not contain any particles unlike the curved spacetime ${\cal
S}_{\rm GR}^{\rm weak}$. However, this $\mathbb{M}^4$ shares the
same name ``spacetime manifold" with the ${\cal S}_{\rm GR}^{\rm
weak}$, which surely contains particles (i.e., the $N_{\rm gr}$
gravitons in Eq.~(\ref{semi-equiv})). Thus, the analogical
reasoning based on its sharing the same name favors the opposite
opinion that the $\mathbb{M}^4$ contains particles like the ${\cal
S}_{\rm GR}^{\rm weak}$. Moreover, since the quantum theory of
gravitons is possible for non-flat ``background spacetimes" (e.g.,
an expanding universe) \cite{graviton}, the flat spacetime
$\mathbb{M}^4$ cannot be the ``only" background spacetime for the
definition of gravitons.

Therefore, we assume that the background spacetime $\mathbb{M}^4$
is composed of {\it particles}, whose number is denoted by $N_{\rm
BS}$ ($\ge 1$). This implies, through ``${\cal S}_{\rm GR}^{\rm
weak} \equiv \mathbb{M}^4 \, \oplus \, {\rm gravitons}$" in
Eq.~(\ref{semi-equiv}), that the ``full GR spacetime" ${\cal
S}_{\rm GR}^{\rm weak}$ is also composed of particles (e.g., the
$N_{\rm BS}$ particles$\,+\,$the $N_{\rm gr}$ gravitons).

This conclusion that ${\cal S}_{\rm GR}^{\rm weak}$ consists of
particles is based on the particular form of the metric ${\bf
g}_{\mu \nu}$ in Eq.~(\ref{weakfield}), which is unchanged only
for special types of coordinate transformations among all the
transformations of General Relativity \cite{GR-Schu}. Despite
this, the conclusion about ${\cal S}_{\rm GR}^{\rm weak}$ can be
true for all the other coordinate transformations, because our
theory can produce General Relativity as a prediction (see
Sec.~\ref{sec: EFT-univ}).

To explain that the flat and curved spacetimes $\mathbb{M}^4$ and
${\cal S}_{\rm GR}^{\rm weak}$ of General Relativity are composed
of particles, we make a hypothesis that 
\beq \label{grav-hyp} {\rm every ~ spacetime ~ manifold ~ {\cal
S}_{\rm GR} ~ of ~ General ~ Relativity ~ is ~ composed ~ of ~
{\it particles}} , \eeq 
which has the {\it meaning} that 
\bea \label{grav-hyp2} && {\rm every ~ point} ~ p ~ {\rm of ~ the
~ GR ~ spacetime ~ {\cal S}_{\rm GR} ~ has ~ a ~
three\textrm{-}dimensional ~ (3D) ~ spacelike} ~~~~~~ \nl && {\rm
neighborhood} ~ {\cal N}_{\rm space}^{\rm \, 3D}(p) ~ {\rm which ~
is ~ a ~ ``continuum ~ approximation" ~ to ~ a ~ discrete} \nl &&
{\rm system ~ composed ~ of ~ particles} . \eea 

Since the concepts like substance and shape are basically defined
at a constant time, the ``3D spacelike neighborhood ${\cal N}_{\rm
space}^{\rm \, 3D}(p)$" in Eq.~(\ref{grav-hyp2}) can represent
(partially) the {\it substance of the spacetime} ${\cal S}_{\rm
GR}$. For example, the substance of a Robertson-Walker spacetime
${\cal S}_{\rm GR}^{\rm (RW)}$ is wholly represented by the 3D
spacelike hypersurface of a constant cosmic time $t$, which
approximately describes a discrete system composed of particles
according to Eq.~(\ref{grav-hyp2}).

Next, we consider a question: ``Is graviton a {\it fundamental}
building block for the substance of the GR spacetime ${\cal
S}_{\rm GR}$?" According to Eq.~(\ref{grav-hyp2}), the substance
of the spacetime ${\cal S}_{\rm GR}$ is a discrete system like
solid materials (cf.~Sec.~\ref{sec: cont-approx}). Thus, for
analysis, we can use an analogy that the substance of the
spacetime ${\cal S}_{\rm GR}$ corresponds to a crystal composed of
many lattice atoms. This atomic crystal can experience a
large-scale deformation of its lattice. In the quantum-mechanical
framework, the lattice deformation of longer wavelengths than the
lattice spacing(s) can be analyzed by introducing a quantized
normal mode called the ``phonon" \cite{cond-matt}. This bosonic
quasi-particle, phonon, differs much in moving range from the
lattice atom, which is confined to a small region around its
equilibrium position.

If the graviton corresponds to the lattice atom in the above
analogy, then (i) the graviton (e.g., a plane-wave solution moving
at the speed of light) should be confined to a small region like
the lattice atom, and (ii) there exist collective vibrational
motions of many ``lattice gravitons," i.e., the lower-energy
excitations corresponding to the phonon in the analogy. Since
these two conclusions do not seem plausible, the graviton does not
correspond to the lattice atom but to the phonon in the analogy.

Therefore, we formulate the ``space-quantum hypothesis" that 
\bea \label{sq-hyp} && {\rm every ~ point} ~ p ~ {\rm of ~ the ~
GR ~ spacetime ~ {\cal S}_{\rm GR} ~ has ~ a ~ 3D ~ spacelike ~
neighborhood} ~ {\cal N}_{\rm space}^{\rm \, 3D}(p) \nl && {\rm
which ~ is ~ a ~ continuum ~ approximation ~ to ~ a ~ discrete ~
system ~ \textbf{\textit{Syst}}_{\rm sq} ~ composed ~ of} ~ \nl &&
{\rm particles ~ called ~ {\bf space ~ quanta}}, \eea 
which is the final meaning of the hypothesis in
Eq.~(\ref{grav-hyp}). Like the phonon, the graviton is an {\it
emergent phenomenon} arising through interactions among space
quanta (see Secs.~\ref{sec: form-met} and \ref{sec: EFT-univ}),
implying each of these space quanta is different and more
fundamental than the graviton.

If the substance of every space quantum has a point-like shape,
the space quantum is a particle. However, the point-like shape of
the space quantum (and the other kinds of quanta) may be only an
approximation based on the smallness of its substance compared
with the observational precision. Then, the space quantum may be a
spatially $p_{\, \rm br}$-dimensional object such as a string
($p_{\, \rm br}=1$), or a composite system made up of two or more
objects which interact weakly and/or strongly. However, in this
paper, the space quantum is regarded as a particle of point-like
shape (i.e., $p_{\, \rm br}=0$), if the assumption of $p_{\, \rm
br}=0$ produces General Relativity as a low-energy effective
theory (see Secs.~\ref{sec: form-met} and \ref{sec: EFT-univ}).

Since the space-quantum hypothesis implies the space part of the
GR spacetime ${\cal S}_{\rm GR}$ is essentially a discrete object,
the hypothesis is different from the basic axiom of General
Relativity that spacetime is a differentiable manifold (i.e., a
continuous object). This difference may not be sufficiently
studied when the particle nature of gravity receives much less
attention than its wave nature.

However, the discrete system of many space quanta (e.g.,
$\textbf{\textit{Syst}}_{\rm sq}$) can be approximated by a 3D
continuous object, when the precision of length measurement is
sufficiently larger than the average distance $d_{\rm \, sq}$
between nearest-neighbor space quanta (see Sec.~\ref{sec:
cont-approx}). This philosophy has been successfully used in the
continuum mechanics \cite{conti-mech}.

\section{\label{sec: cont-approx} The Continuum Approximation
of a Space-Quantum System in the Ambient Spacetime}

When space quanta forming the discrete system
$\textbf{\textit{Syst}}_{\rm sq}$ in Eq.~(\ref{sq-hyp}) change
their positions, the system $\textbf{\textit{Syst}}_{\rm sq}$
undergoes a deformation. This implies, due to the space-quantum
hypothesis, that the spacelike subset ${\cal N}_{\rm space}^{\rm
\, 3D}(p)$ also deforms. This deformation of the subset ${\cal
N}_{\rm space}^{\rm \, 3D}(p)$ is similarly found in General
Relativity (e.g., the Schwarzschild metric)
\cite{{GR-Schu},{GR-Wein},{GR-Carr}}. In addition, the deformation
of the system $\textbf{\textit{Syst}}_{\rm sq}$ can affect the
motions of other objects (e.g., lights and matters) {\it within}
the system $\textbf{\textit{Syst}}_{\rm sq}$. This is similar to
the deflection of light in General Relativity. These two
similarities to General Relativity suggest the relationship
between the space-quantum hypothesis and General Relativity (see
Secs.~\ref{sec: form-met} and \ref{sec: EFT-univ}).

When $N_{\rm sq}$ space quanta form a GR spacetime ${\cal
S}_{N_{\rm sq}}$, the motion of a {\it single} space quantum
${\cal P}$ within ${\cal S}_{N_{\rm sq}}$ can be described by its
background spacetime ${\cal S}_{\rm bkgd}$ ($={\cal S}_{N_{\rm
sq}-1}$), which is formed by the {\it other} $N_{\rm sq}-1$ space
quanta. However, if we consider the limiting case that there are
no space quanta except the single quantum ${\cal P}$ (i.e.,
$N_{\rm sq} = 1$), then the wave kinematics using the background
spacetime ${\cal S}_{\rm bkgd}$ ($={\cal S}_{0}$) is not possible
any longer, implying the space quantum ${\cal P}$ loses its wave
nature. In other words, the wave-particle duality (and thus
quantum mechanics) cannot be applied to the single particle ${\cal
P}$ in this limiting case.

If we want to maintain the quantum mechanics (e.g., the wave
nature) of the particle ${\cal P}$, a simple solution to the
wave-nature problem for ${\cal P}$ is to introduce another
spacetime ${\cal S}^{D_{\rm amb}}$ of dimension $D_{\rm amb}$
($\ge 4$) within which the space quantum ${\cal P}$ moves like a
particle moving within a GR spacetime ${\cal S}_{\rm GR}$. In
other words, the motion of the single space quantum ${\cal P}$ is
defined by the {\bf ambient (i.e., surrounding) spacetime} ${\cal
S}^{D_{\rm amb}}$, without considering any other space quanta. In
general, any number of space quanta can occupy the ambient
spacetime ${\cal S}^{D_{\rm amb}}$.

Since the spacetime ${\cal S}_{\rm GR}$ of General Relativity has
the metric ${\bf g}_{\mu \nu}$ of the Lorentzian signature
($-,+,+,+$), the ambient spacetime ${\cal S}^{D_{\rm amb}}$ can
have its own $D_{\rm amb}$-dimensional Lorentzian metric
$g_{AB}^{\rm bulk}$ ($A, B = 0, \ldots, D_{\rm amb} - 1$). For
simplicity, the ambient spacetime ${\cal S}^{D_{\rm amb}}$ with
the bulk metric $g_{AB}^{\rm bulk}$ is assumed to be the $D_{\rm
amb}$-dimensional Minkowski spacetime $\mathbb{M}^{D_{\rm amb}} =
(\mathbb{R}^{D_{\rm amb}}, \eta_{AB}^{\rm bulk})$, where the flat
bulk metric $\eta_{AB}^{\rm bulk}$ is the diagonal matrix in the
{\it mostly plus} convention 
\beq \label{bulk-metric} \eta_{AB}^{\rm bulk} \, = \, {\rm
diag}(-1, +1, \ldots, +1 \,) \eeq 
everywhere for the inertial ``bulk-coordinates" $Y^A$ ($\in
\mathbb{R}^{D_{\rm amb}}$). These bulk-coordinates $Y^A$ are used
by an inertial ``bulk observer" $O_{\rm bulk}$ who lives in the
ambient spacetime $\mathbb{M}^{D_{\rm amb}}$.

Because it is natural that any particle performs a time evolution
in every Minkowski spacetime, all space quanta occupying the
ambient spacetime $\mathbb{M}^{D_{\rm amb}}$ must execute time
evolutions, producing their own world lines ${\cal WL}_{\rm sq}$
in the spacetime $\mathbb{M}^{D_{\rm amb}}$. Here, the physics of
space quantum is studied in the ambient spacetime
$\mathbb{M}^{D_{\rm amb}}$.

To explain the observation that the space part of our universe is
three-dimensional, we assume that space quanta in the spacetime
$\mathbb{M}^{D_{\rm amb}}$ form a nearly 3D object, which is
called the {\bf quasi-3D object} of the many space quanta. If the
average distance $d_{\rm \, sq}$ between nearest-neighbor space
quanta is sufficiently smaller than the precision $\Delta L_{\rm
obs}$ of the length measurement, we can apply the continuum
approximation to the quasi-3D object in the spacetime
$\mathbb{M}^{D_{\rm amb}}$, as in the continuum mechanics
\cite{conti-mech}.

The ``validity condition" of the continuum approximation
\cite{conti-mech} is 
\beq \label{valid} d_{\rm \, sq}^{\, 3} ~ \ll ~ \delta V_{\rm sq}
~ \ll ~ (\Delta L_{\rm obs})^3 ~, \eeq 
where $\delta V_{\rm sq}$ is the volume of a 3D spacelike region
$\delta {\cal R}_{\rm sq}$ ($\subset \mathbb{M}^{D_{\rm amb}}$)
which contains space quanta. Since there are $\delta N_{\rm sq} =
O(\delta V_{\rm sq} / d_{\rm \, sq}^{\, 3})$ space quanta inside
the region $\delta {\cal R}_{\rm sq}$, the validity condition in
Eq.~(\ref{valid}) implies the region $\delta {\cal R}_{\rm sq}$
contains many space quanta (i.e., $\delta N_{\rm sq} \gg 1$).

We assume that the quasi-3D object satisfying the validity
condition behaves like a 3D continuously-distributed system, which
is called the {\bf space 3-brane} (i.e., another name of space).
In other words, the space 3-brane in the spacetime
$\mathbb{M}^{D_{\rm amb}}$ is the continuum approximation of the
quasi-3D object having many space quanta, as in the continuum
mechanics for solids and fluids. Mathematically, the space 3-brane
composed of many space quanta is represented by a 3D spacelike
submanifold of the ambient spacetime $\mathbb{M}^{D_{\rm amb}}$.

Since the continuum approximation is applied to both of solids and
fluids, we need to discuss whether the quasi-3D object (or its
space 3-brane) is like a solid or a fluid: because space quanta in
the fluid phase move faster, the Brownian motion can be a crucial
criterion distinguishing between the two phases of the quasi-3D
object. In the Brownian motion, the root-mean-square displacement
$\Delta r_{\rm rms}$ of a ``big" particle (e.g., proton) colliding
with quick space quanta can be proportional to the square root of
the elapsed time $\tau_{\rm E}$, namely, 
\beq \Delta r_{\rm rms} \, \propto \, \tau_{\rm E}^{1/2} ~ . \eeq
This long-term behavior implies that the quasi-3D object (i.e.,
space) behaves like a medium which exerts random forces on the
above big particle.

However, the Brownian motion caused by the fluid phase of space
quanta is rejected by (i) Newton's first law imposing $\Delta
r_{\rm rms} = ({\rm initial ~ speed}) \times \tau_{\rm E}$ on
every free particle, and (ii) the rectilinear propagation of light
in vacuum. For example, if lights from (more) distant stars were
(more) deflected by the above Brownian motion, we would observe
the (larger) disk-like images of the stars. In fact, the images of
stars are point-like.

Therefore, the quasi-3D object of many space quanta is like a
solid in our observation region. This {\it solid-like} quasi-3D
object (a) can have a crystal lattice (e.g., simple cubic) or a
non-crystalline structure like an amorphous glass, and (b) can
deform elastically or plastically in response to stimuli like
ordinary solid materials \cite{cond-matt}. The deformations or
strains of the quasi-3D object can be (approximately) determined
by Einstein's equation $\textit{\textbf{G}}_{\mu \nu} = 8 \pi
G_{\rm N} \textit{\textbf{T}}_{\mu \nu}$ (see Secs.~\ref{sec:
AAT-method}, \ref{sec: form-met} and \ref{sec: EFT-univ}).

Since a space quantum within the solid-like quasi-3D object is not
an isolated particle in the ambient spacetime $\mathbb{M}^{D_{\rm
amb}}$, the mass $m_{\rm sq}$ of the space quantum may differ
considerably from its effective mass $m_{\rm sq}^{\rm (eff)}$
which is affected by interactions like (i) the effective mass of
electron in a solid and (ii) the constituent quark mass in a
hadron.

Because space quanta in the quasi-3D object have inter-particle
spacings of $O(d_{\rm \, sq})$, the physical quantities of the
space quanta (e.g., energy density) can vary significantly over
spatial distances $\lesssim  O(d_{\rm \, sq})$. Thus, since the
quasi-3D object resembles a discontinuously-distributed system at
a high observational precision $\Delta L_{\rm obs} \lesssim
O(d_{\rm \, sq})$, the above continuum approximation breaks down
at the high precision $\Delta L_{\rm obs} \lesssim O(d_{\rm \,
sq})$. This requires the lower bound $\Delta L_{\rm obs}^{({\rm
c})}$ ($\le \Delta L_{\rm obs}$) in order for the continuum
approximation to be acceptable.

Thus, the continuum approximation considers only the larger-scale
(i.e., lower-energy) behaviors of the space-quantum system,
ignoring its smaller-scale (i.e., higher-energy) physics. Then,
the critical precision $\Delta L_{\rm obs}^{({\rm c})}$ for the
continuum approximation plays a similar role to the ``UV cutoff\,"
$\Lambda_{\rm UV}$ of an effective field theory \cite{EFT}, whose
example is the Wilsonian effective action obtained by integrating
out higher-energy modes than a UV cutoff.

Therefore, the continuum approximation of the quasi-3D object can
be regarded as a low-energy effective theory of the space-quantum
system, which has its own UV cutoff $\Lambda_{\rm cont}$ ($\sim
1/\Delta L_{\rm obs}^{({\rm c})}$) satisfying 
\beq \label{cut-cont} \Lambda_{\rm cont} \, = \, \epsilon_{\rm
cont} \times (1/d_{\rm \, sq}) ~~~{\rm with} ~~ \epsilon_{\rm
cont} \lesssim O(10^{-1}) ~ . \eeq 

\section{\label{sec: 3br-eff-action} The Effective
Theory for the Space 3-Brane: the Bottom-Up Approach}

The space 3-brane corresponds to the ``continuum limit" $d_{\rm \,
sq} \rightarrow 0$ of the quasi-3D object which consists of many
space quanta. Then, like a bosonic string
\cite{{string-th},{Goto}}, the space 3-brane sweeps out a 4D
manifold ${\cal WV}_{\rm sq}$ in the ambient spacetime
$\mathbb{M}^{D_{\rm amb}}$. The {\bf world volume} ${\cal WV}_{\rm
sq}$ of the space 3-brane is a continuum approximation to the
discrete set of the world lines ${\cal WL}_{\rm sq}$ of all space
quanta forming the space 3-brane.

As in the case of the 2D world sheet ${\cal WS}$ of a bosonic
string \cite{{string-th},{Goto}}, we can assume that the world
volume ${\cal WV}_{\rm sq}$ of the space 3-brane is a {\it 4D
submanifold} of the ambient spacetime $\mathbb{M}^{D_{\rm amb}}$
(see Refs.~\cite{{GR-Carr},{Math}} for mathematical treatments):
since this submanifold ${\cal WV}_{\rm sq}$ is a subset of
$\mathbb{M}^{D_{\rm amb}}$ (i.e., ${\cal WV}_{\rm sq} \subset
\mathbb{M}^{D_{\rm amb}}$), there exists the inclusion map
$\textbf{\textit{i}}$ of the world volume ${\cal WV}_{\rm sq}$,
which is defined as a function 
\bea \label{inc-map0} && \textbf{\textit{i}}: ~ {\cal WV}_{\rm sq}
~ (\subset \mathbb{M}^{D_{\rm amb}}) ~ \rightarrow ~
\mathbb{M}^{D_{\rm amb}} \, , ~ {\rm satisfying} \\
\label{inc-map} && \textbf{\textit{i}}(p) \, = \, p ~ \in \,
\mathbb{M}^{D_{\rm amb}} ~~~ {\rm for ~ every ~ point} ~ p \in
{\cal WV}_{\rm sq} ~. \eea 

Since ${\cal WV}_{\rm sq}$ is a submanifold of $\mathbb{M}^{D_{\rm
amb}}$, the inclusion map $\textbf{\textit{i}}$ is an immersion,
i.e., 
\beq \label{immers-inc} ~~~~~ {\rm its ~ derivative ~ at} ~ p , \,
\textbf{\textit{i}}_p^{\, \prime} : \, T_p {\cal WV}_{\rm sq}
\rightarrow T_{p \,} \mathbb{M}^{D_{\rm amb}} , ~ {\rm is ~
injective ~ for ~ every} ~ p \in {\cal WV}_{\rm sq} ~, \eeq 
where $T_{p \,} {\cal M}$ (${\cal M} = {\cal WV}_{\rm sq}, \,
\mathbb{M}^{D_{\rm amb}}$) denotes the tangent vector space of
${\cal M}$ at $p$. In addition, the inclusion map
$\textbf{\textit{i}}$ is of constant rank 4 everywhere on ${\cal
WV}_{\rm sq}$, i.e., 
\beq \label{rank-inc} ~~~~~~~~~ {\rm rank}( \textbf{\textit{i}}(p)
) ~ \stackrel{\rm def}{=} ~ {\rm rank}( \textbf{\textit{i}}_p^{\,
\prime} ) ~ = ~ 4 ~~~~ {\rm for ~ every} ~ p \in {\cal WV}_{\rm
sq} ~, \eeq 
where ${\rm rank}(\textbf{\textit{i}}_p^{\, \prime}) \stackrel{\rm
def}{=} {\rm dim}( {\rm Im}(\textbf{\textit{i}}_p^{\, \prime}) )$.
Then, ${\rm rank}( \textbf{\textit{i}}_p^{\, \prime} ) = 4$ in
Eq.~(\ref{rank-inc}) guarantees that the image
$\textbf{\textit{i}}_p^{\, \prime}( T_p {\cal WV}_{\rm sq})$ of
the ``brane tangent space" $T_p {\cal WV}_{\rm sq}$ under the map
$\textbf{\textit{i}}_p^{\, \prime}$ in Eq.~(\ref{immers-inc}) 
\beq \label{pf-tang-sp} ~~~~~~~~~~ \textbf{\textit{i}}_p^{\,
\prime}( T_p {\cal WV}_{\rm sq}) ~ \stackrel{\rm def}{=} ~ \{ \,
\textbf{\textit{i}}_p^{\, \prime}(v) ~~ {\rm for} ~ \forall \, v
\in T_p {\cal WV}_{\rm sq} \, \} ~~~ \subset ~ T_{p \,}
\mathbb{M}^{D_{\rm amb}} \eeq 
is a 4D subspace of the ``bulk tangent space" $T_{p \,}
\mathbb{M}^{D_{\rm amb}}$.

We are studying the submanifold ${\cal WV}_{\rm sq}$ within its
ambient manifold $\mathbb{M}^{D_{\rm amb}}$, which has a
coordinate chart $Y^A$ at every point $P \in \mathbb{M}^{D_{\rm
amb}}$: since the submanifold ${\cal WV}_{\rm sq}$ is also a
manifold, the set ${\cal WV}_{\rm sq}$ as a 4D manifold has its
own coordinate chart $x^\mu$ ($\mu = 0, \ldots, 3$) at every point
$p \in {\cal WV}_{\rm sq}$. Therefore, we need to consider {\it
two} kinds of charts at every point $p$ of ${\cal WV}_{\rm sq}$:
(i) a ``brane-chart" $x^\mu$ of ${\cal WV}_{\rm sq}$, and (ii) a
``bulk-chart" $Y^A$ of $\mathbb{M}^{D_{\rm amb}}$. Then, a
coordinate transformation $x^\mu \rightarrow x^{\prime \mu}$
between two brane-charts $x^\mu$ and $x^{\prime \mu}$ of ${\cal
WV}_{\rm sq}$ is called a ``brane-to-brane (b$\Rightarrow$${\rm
b}^\prime$) transformation." Moreover, a coordinate transformation
$Y^A \rightarrow Y^{\prime A}$ between two bulk-charts $Y^A$ and
$Y^{\prime A}$ of $\mathbb{M}^{D_{\rm amb}}$ is called a
``bulk-to-bulk (B$\Rightarrow$${\rm B}^\prime$) transformation."
None of these coordinate transformations $x^\mu \rightarrow
x^{\prime \mu}$ and $Y^A \rightarrow Y^{\prime A}$ change the
point $p$ of ${\cal WV}_{\rm sq}$ at all{---\,}this
passive-viewpoint property is shared by any coordinate
transformation between two charts in differential geometry.

Relative to the brane-chart $x^\mu$ of ${\cal WV}_{\rm sq}$, and
the bulk-chart $Y^A$ of $\mathbb{M}^{D_{\rm amb}}$, the equality
$p = \textbf{\textit{i}}(p)$ in Eq.~(\ref{inc-map}) has its
coordinate representation 
\beq \label{point-inc-map} Y^A(p) ~ = ~ \left( Y^A \circ
\textbf{\textit{i}} \circ (x^\mu)^{-1} \right) (x^\mu(p)) ~, \eeq
where $x^\mu(p) \in \mathbb{R}^4$ and $Y^A(p) \in
\mathbb{R}^{D_{\rm amb}}$.

Then, for the $x^\mu$-and-$Y^A$ coordinate representation of
$\textbf{\textit{i}}$ (see Eq.~(\ref{point-inc-map})) 
\beq \label{rep-inc-map} f^A ~ \stackrel{\rm def}{=} ~ Y^A \circ
\textbf{\textit{i}} \circ (x^\mu)^{-1} ~, \eeq 
Eq.~(\ref{point-inc-map}) defines a new kind of transformation
$x^\mu \rightarrow Y^A$, called the ``brane-to-bulk
(b$\Rightarrow$B) transformation," 
\beq \label{b-B-tr} ~~~~~~~~~~~~~~~~ Y^A ~ = ~ f^A(x^\mu) ~
\stackrel{\rm def}{=} ~ f^A \circ x^\mu ~~~~~ {\rm at ~ the ~
point} ~ p ~ {\rm of} ~ {\cal WV}_{\rm sq} ~. \eeq 
Through the representation $f^A = Y^A \circ \textbf{\textit{i}}
\circ (x^\mu)^{-1}$ in Eq.~(\ref{rep-inc-map}), ${\rm rank}(
\textbf{\textit{i}}_p^{\, \prime} )$ in Eq.~(\ref{rank-inc}) has
its $x^\mu$-and-$Y^A$ coordinate representation 
\beq \label{rep-rank} {\rm rank}( \textbf{\textit{i}}_p^{\,
\prime} ) ~ = ~ {\rm rank}(\partial_\mu f^A) |_{\, {\rm at} \,
x^\nu(p) \,} ~, \eeq 
where the $D_{\rm amb} \times 4$ matrix $\partial_\mu f^A$ is the
Jacobian matrix of the above b$\Rightarrow$B transformation $Y^A =
f^A(x^\mu)$.

By using the metric bulk-tensor $\eta^{\rm bulk}$ of the ambient
manifold $\mathbb{M}^{D_{\rm amb}}$, the {\it pullback map}
$\textbf{\textit{i}}^*$ of the inclusion map $\textbf{\textit{i}}$
in Eqs.~(\ref{inc-map0}) and (\ref{inc-map}) induces a symmetric
tensor $\gamma_{\mu \nu}$ on the submanifold ${\cal WV}_{\rm sq}$
in the ``brane-coordinates" $x^\mu$ 
\bea \label{ind-metric} && \gamma_{\mu \nu} ~ \stackrel{\rm
def}{=} ~ ( \textbf{\textit{i}}^* \eta^{\rm bulk} )_{\mu \nu} ~ =
~ ( f^* \eta_{AB}^{\rm bulk} )_{\mu \nu} ~ = ~ \partial_\mu f^A \,
\partial_\nu f^B \, \eta_{AB}^{\rm bulk} ~~~~ {\rm satisfying} ~ \\
\label{ind-metric2} && \gamma_{\mu \nu} \, v^\mu w^\nu ~ = ~
\eta^{\rm bulk} (\textbf{\textit{i}}_* v, \textbf{\textit{i}}_* w)
~ = ~ \eta_{AB}^{\rm bulk} \, (f_* v)^A (f_* w)^B ~~~~ {\rm for}
~\, \forall \, v, \, w \in T_p {\cal WV}_{\rm sq} ~, ~~~~~~ \eea
where the two maps $f^*$ and $f_*$ from $f^A = Y^A \circ
\textbf{\textit{i}} \circ (x^\mu)^{-1}$ are the coordinate
representations of the pullback and the pushforward maps
$\textbf{\textit{i}}^*$ and $\textbf{\textit{i}}_*$ (e.g.,
$\textbf{\textit{i}}_p^{\, \prime}$ in Eq.~(\ref{pf-tang-sp})) in
the brane- and bulk-charts $x^\mu$ and $Y^A$
(cf.~Refs.~\cite{{GR-Carr},{Math}}). Then, $\gamma_{\mu
\nu}(x^\rho(p))$ is a tensor defined on the tangent space $T_p
{\cal WV}_{\rm sq}$ at a point $p \in {\cal WV}_{\rm sq}$, whereas
$\eta_{AB}^{\rm bulk}(Y^C(p))$ is a tensor defined on $T_{p \,}
\mathbb{M}^{D_{\rm amb}}$ at the same point $p =
\textbf{\textit{i}}(p)$ as an element of $\mathbb{M}^{D_{\rm
amb}}$.

Then, besides the constraint in Eq.~(\ref{rank-inc})
(equivalently, ${\rm rank}(\partial_\mu f^A) = 4$), we assume
another constraint on $f^A(x^\mu)$ that the symmetric tensor
$\gamma_{\mu \nu} = \partial_\mu f^A \, \partial_\nu f^B \,
\eta_{AB}^{\rm bulk}$ in Eq.~(\ref{ind-metric}) is non-degenerate
everywhere on the world volume ${\cal WV}_{\rm sq}$, i.e., 
\beq \label{nondeg} {\rm det}(\gamma_{\mu \nu}) \, \ne \, 0 ~,
\eeq 
which means that the pullback $\gamma_{\mu \nu}$ becomes a
``metric tensor" on ${\cal WV}_{\rm sq}$ (called the ``induced
metric"). Note that ${\rm det}(\gamma_{\mu \nu}) \ne 0$ is a
sufficient condition for ${\rm rank}(\textbf{\textit{i}}) = {\rm
rank}(\partial_\mu f^A) = 4$ in Eq.~(\ref{rank-inc}).

Relative to the bulk metric $\eta^{\rm bulk}$, the 4D subspace
$\textbf{\textit{i}}_p^{\, \prime}( T_p {\cal WV}_{\rm sq})$ of
the bulk tangent space $T_{p \,} \mathbb{M}^{D_{\rm amb}}$ in
Eq.~(\ref{pf-tang-sp}) contains both
\begin{itemize}
\item {\it timelike} bulk-vectors $V_{\rm t} =
\textbf{\textit{i}}_* v_{\rm t}$ (i.e., $\eta^{\rm bulk}
(\textbf{\textit{i}}_* v_{\rm t}, \textbf{\textit{i}}_* v_{\rm t})
< 0$) due to the time evolution in the ambient spacetime
$\mathbb{M}^{D_{\rm amb}}$, and
\item {\it spacelike} bulk-vectors $V_{\rm s} =
\textbf{\textit{i}}_* v_{\rm s}$ (i.e., $\eta^{\rm bulk}
(\textbf{\textit{i}}_* v_{\rm s}, \textbf{\textit{i}}_* v_{\rm s})
> 0$) due to the three-brane nature of the space 3-brane.
\end{itemize}
Thus, the restriction $\left. \eta^{\rm bulk}
\right|_{\textbf{\textit{i}}_p^{\, \prime}( T_p {\cal WV}_{\rm sq}
)}$ of the bulk-tensor $\eta^{\rm bulk}$ to the subspace
$\textbf{\textit{i}}_p^{\, \prime}( T_p {\cal WV}_{\rm sq} )$ has
the (3+1)-dimensional Lorentzian signature $(-,+,+,+)$. This
signature $(-,+,+,+)$ is shared by the induced metric $\gamma_{\mu
\nu}$, because $\gamma_{\mu \nu}$ as the pullback of
$\eta_{AB}^{\rm bulk}$ satisfies, for example, $\gamma_{\mu \nu}
\, v_{\rm k}^\mu \, v_{\rm k}^\nu = \eta^{\rm bulk}
(\textbf{\textit{i}}_* v_{\rm k}, \textbf{\textit{i}}_* v_{\rm
k})$ with ${\rm k} = {\rm t}, {\rm s}$ (see
Eq.~(\ref{ind-metric2})).

Therefore, through Eq.~(\ref{ind-metric2}), the induced metric
$\gamma_{\mu \nu}$ (i.e., the pullback $f^*(\eta_{AB}^{\rm bulk})$
of $\eta_{AB}^{\rm bulk}$ by $f^A$) becomes a Lorentzian metric
having the 4D Lorentzian signature $(-,+,+,+)$. Then, the 4D
Lorentzian manifold (${\cal WV}_{\rm sq}, \gamma_{\mu \nu}$) is
interpreted as a (3+1)-dimensional spacetime. This spacetime
manifold (${\cal WV}_{\rm sq}, \gamma_{\mu \nu}$) is an ``emergent
entity," because (${\cal WV}_{\rm sq}, \gamma_{\mu \nu}$) arises
through interactions among many space quanta which occupy the
ambient spacetime $\mathbb{M}^{D_{\rm amb}}$.

To sum up, the 4D manifold (${\cal WV}_{\rm sq}, \gamma_{\mu
\nu}$) is the (3+1)-dimensional emergent spacetime which occupies
the ambient spacetime $\mathbb{M}^{D_{\rm amb}}$. Since the
spacetime of our universe is (3+1)-dimensional like (${\cal
WV}_{\rm sq}, \gamma_{\mu \nu}$), we assume that the emergent
spacetime (${\cal WV}_{\rm sq}, \gamma_{\mu \nu}$) occupying
$\mathbb{M}^{D_{\rm amb}}$ forms the spacetime of our
universe{\,---\,}(${\cal WV}_{\rm sq}, \gamma_{\mu \nu}$) is the
model of our spacetime.

Note that the emergent spacetime (${\cal WV}_{\rm sq}, \gamma_{\mu
\nu}$) is the {\it exact or true} spacetime of our universe. Thus,
when we say that a spacetime and a metric of the universe are
observed (or measured), the observed spacetime and the observed
metric should be identical to ${\cal WV}_{\rm sq}$ and
$\gamma_{\mu \nu}$ within the measurement precisions.

Then, since General Relativity has accurately explained the
spacetime of our universe, we can think that the spacetime ${\cal
S}_{\rm GR}$ and the metric ${\bf g}_{\mu \nu}$ of General
Relativity are at least the good approximations of the world
volume ${\cal WV}_{\rm sq}$ and the induced metric $\gamma_{\mu
\nu}$ (see around Eqs.~(\ref{GR-met=g-2}) and (\ref{GR-ST=ST-2})),
i.e., 
\bea \label{S-GR=WV} {\cal S}_{\rm GR} ~ &\approx& ~ {\cal
WV}_{\rm sq} ~, \\ \label{GR-met=ind} ~~~~~~~~~~ {\bf g}_{\mu \nu}
~ &\approx& ~ \gamma_{\mu \nu} ~~ (= \partial_\mu f^A
\partial_\nu f^B \eta_{AB}^{\rm bulk}) ~. \eea 
Because Einstein developed General Relativity without considering
the space quanta and the ambient spacetime, General Relativity is
a {\it phenomenological} theory of spacetime like the meson theory
which Yukawa developed without considering quarks and gluons.

Until now, we have established the kinematics for the space
3-brane: the world volume ${\cal WV}_{\rm sq}$ of the space
3-brane in the ambient spacetime $\mathbb{M}^{D_{\rm amb}}$ is a
4D submanifold of $\mathbb{M}^{D_{\rm amb}}$, which is described
by the brane-to-bulk transformation $Y^A = f^A(x^\mu)$ satisfying
\bea \label{Lemb-emb} & ({\rm i}) & \, {\rm an ~ embedding ~
(i.e., an ~ immersion ~ and ~ an ~ injection)} \, , \\
\label{Lemb-sign} & ({\rm ii}) & \, {\rm the ~ 4D ~ Lorentzian ~
signature ~ of ~ the ~ induced ~ metric} ~ \gamma_{\mu \nu} =
\partial_\mu f^A \partial_\nu f^B \eta_{AB}^{\rm bulk} \, .
~~~~~~~ \eea 
In Eq.~(\ref{Lemb-emb}), the immersion condition is replaced with
${\rm det}(\gamma_{\mu \nu}) \ne 0$ contained in
Eq.~(\ref{Lemb-sign}) (see below Eq.~(\ref{nondeg})), and the
injection condition may be omitted in the case of eccentric
behaviors of the space 3-brane (e.g., self-intersections). Note
that the function $f^A(x^\mu)$ describing the world volume ${\cal
WV}_{\rm sq}$ is neither an arbitrary function of $x^\mu$ nor a
general embedding, but it is a special kind of embedding called
the ``4D-Lorentzian (4DL) embedding," which is defined as a
function satisfying the conditions in Eqs.~(\ref{Lemb-emb}) and
(\ref{Lemb-sign}).

Based on the above kinematics for the space 3-brane, we have to
consider its dynamics: an effective theory for the space 3-brane
can be defined by the ``3-brane action" 
\beq \label{3br-eff-action} S_{\rm emb}^{\rm (3 br)}[f^A] ~ = \,
\int_{{\cal WV}_{\rm sq}} d^{\, 4} x ~ \widehat{{\cal L}}_{\rm
emb} (f^A, \partial_\mu f^A, \ldots \,) ~, \eeq 
where the Lagrangian density $\widehat{{\cal L}}_{\rm emb}$ can
contain the UV cutoff $\Lambda_{\rm cont}$ in
Eq.~(\ref{cut-cont}). In Eq.~(\ref{3br-eff-action}), the symbol
$\widehat{~}$ in $\widehat{{\cal L}}_{\rm emb}$ does not denote
the operator nature of $\widehat{{\cal L}}_{\rm emb}$ unlike the
same symbol used in, e.g., $\widehat{a}(\vec{k}, \sigma)$ of
Sec.~\ref{sec: sp-qu-hypothesis}. Note ${\cal L}_{\rm emb}$
without the symbol $\widehat{~}$ is called the ``Lagrangian" (see
below Eq.~(\ref{Jac-det})).

Because the full theory of the effective theory $S_{\rm emb}^{\rm
(3 br)}[f^A]$ is not known, we use the bottom-up approach to
building an effective theory, i.e., writing out the most general
set of Lagrangians consistent with the {\it symmetries} of the
theory \cite{EFT}. Then, the crucial step is to find the
symmetries satisfied by the effective action $S_{\rm emb}^{\rm (3
br)}[f^A]$ for the embedding $f^A(x)$.

To find the symmetries of the 3-brane action $S_{\rm emb}^{\rm (3
br)}[f^A]$, we will use a generalization based on the special case
of a 0-brane (i.e., point particle) in the 4D Minkowski spacetime
$\mathbb{M}^4$, as follows: similarly to the space 3-brane in the
ambient spacetime $\mathbb{M}^{D_{\rm amb}}$, the 0-brane in
$\mathbb{M}^4$ produces a 1D world line ${\cal WL}$ in
$\mathbb{M}^4$, which is described by a 1D-Lorentzian embedding
$h^\mu(\tau)$ of the world-line parameter $\tau$ ($\in
\mathbb{R}$).

It is well known that the effective action $S_{\rm emb}^{({\rm 0
br})}[h^\mu]$ for the 0-brane has two kinds of symmetries under
(i) the 4D Poincar$\acute{\textrm{e}}$ group $ISO(1, 3)$ with
$h^{\prime \mu}(\tau) = \Lambda^\mu_\nu h^\nu(\tau) + c^{\, \mu}$,
and (ii) the 1D diffeomorphism group ${\rm Diff}(1)$ with
$h^{\prime \mu}(\tau^{\, \prime}) = h^\mu(\tau)$, where $\tau^{\,
\prime} = \Phi_{\rm 1D}(\tau)$ for $\Phi_{\rm 1D} \in {\rm
Diff}(1)$. Note that the $ISO(1, 3)$ symmetry is required by the
Special Principle of Relativity (i.e., special covariance) in
$\mathbb{M}^4$.

Therefore, by using the generalization from the ``\,0-brane in
$\mathbb{M}^{\, 4}$\," to a ``\,$p_{\, \rm br \,}$-brane in
$\mathbb{M}^D$," the effective action $S_{\rm emb}^{\rm (3
br)}[f^A]$ for the space 3-brane in $\mathbb{M}^{D_{\rm amb}}$
(i.e., $p_{\, \rm br}=3$ and $D=D_{\rm amb}$) has two
corresponding symmetries: the first symmetry is the invariance of
the 3-brane action $S_{\rm emb}^{\rm (3 br)}[f^A]$ under the bulk
Poincar$\acute{\textrm{e}}$ group $ISO(1, D_{\rm amb}-1)$ with 
\bea \label{bulk-Poin} ~~~~~ f^{\prime A}(x^\mu(p)) ~ = ~
\Lambda_B^A ~ f^B(x^\mu(p)) ~ + ~ c^{\, A} ~~~ {\rm at ~ a ~
point} ~ p ~{\rm of} ~ {\cal WV}_{\rm sq} && \\ \label{bulk-Poin2}
{\rm for} ~~ f^{\prime A} \, = \, Y^{\prime A} \circ
\textbf{\textit{i}} \circ (x^\mu)^{-1} ~~ {\rm and} ~~ f^A \, = \,
Y^A \circ \textbf{\textit{i}} \circ (x^\mu)^{-1} ~, && \eea 
where $\Lambda_B^A$ and $c^{\, A}$ denote each element of the bulk
Lorentz group $SO(1, D_{\rm amb}-1)$, and each translation in
$\mathbb{M}^{D_{\rm amb}}$, respectively.

Due to Eq.~(\ref{bulk-Poin2}), the transformation $f^A \rightarrow
f^{\prime A}$ in Eq.~(\ref{bulk-Poin}) uses only the
B$\Rightarrow$${\rm B}^\prime$ transformation $Y^A \rightarrow
Y^{\prime A}$ while keeping the brane-chart $x^\mu$ fixed. In
other words, the above $ISO(1, D_{\rm amb}-1)$ is exactly the same
as the set of all coordinate transformations $Y^A \rightarrow
Y^{\prime A}$ of the ambient manifold $\mathbb{M}^{D_{\rm amb}}$.

The second one is the invariance of the 3-brane action $S_{\rm
emb}^{\rm (3 br)}[f^A]$ under the 4D {\it
local}\,-reparametrization group ${\rm Diff}(4)$ (i.e., the
symmetry group of General Relativity) with 
\bea \label{diff4} x^{\prime \mu}(p) ~ = ~ \Phi_{\rm
4D}^\mu(x^\nu(p)) ~~ {\rm and} ~~ f^{\prime A}(x^{\prime \mu}(p))
~ = ~ f^A(x^\mu(p)) ~~~ {\rm at ~ the ~ same ~ point} ~ p && ~~~ \\
\label{diff4-2} {\rm for} ~~ f^{\prime A} \, = \, Y^A \circ
\textbf{\textit{i}} \circ (x^{\prime \mu})^{-1} ~~ {\rm and} ~~
f^A \, = \, Y^A \circ \textbf{\textit{i}} \circ (x^\mu)^{-1} ~, &&
~~~ \eea 
where $\Phi_{\rm 4D} \in {\rm Diff}(4)$ corresponds to every
general coordinate transformation of General Relativity.

Due to Eq.~(\ref{diff4-2}), the transformation $f^{\prime
A}(x^\prime) = f^A(x)$ in Eq.~(\ref{diff4}) uses only the
b$\Rightarrow$${\rm b}^\prime$ transformation $x^\mu \rightarrow
x^{\prime \mu}$ while keeping the bulk-chart $Y^A$ fixed. In other
words, the above ${\rm Diff}(4)$ is exactly the same as the set of
all coordinate transformations $x^\mu \rightarrow x^{\prime \mu}$
of the submanifold ${\cal WV}_{\rm sq}$. The insertion of
Eq.~(\ref{diff4-2}) into Eq.~(\ref{diff4}) produces $Y^A(p) =
f^{\prime A}(x^\prime(p)) = f^A(x(p))$, which means the invariance
of $Y^A(p)$ under $x(p) \rightarrow x^\prime(p)$. This
transformation law $f^{\prime A}(x^\prime) = f^A(x)$ under $x
\rightarrow x^\prime$ implies that each of $f^{A = \, 0, \ldots ,
D_{\rm amb}-1}$ is a scalar field under ${\rm Diff}(4)$. Note that
the Nambu-Goto action for a bosonic string is the $p_{\, \rm
br}=1$ case in the above generalization, having the similar kinds
of symmetries \cite{{string-th},{Goto}}.

First, we deal with the ${\rm Diff}(4)$ invariance of the 3-brane
action $S_{\rm emb}^{\rm (3 br)}[f^A]$ more closely: since the
world volume ${\cal WV}_{\rm sq}$ exists in the ambient spacetime
$\mathbb{M}^{D_{\rm amb}}$ irrespective of the b$\Rightarrow$${\rm
b}^\prime$ transformation $x \rightarrow x^\prime = \Phi_{\rm
4D}(x)$ in Eq.~(\ref{diff4}), the pair $f^{\prime A}(x^\prime)$
and $f^A(x)$ should be simultaneously the solutions for the
equation of motion. Thus, if the unprimed map $f^A(x)$ is an
extremum point of the unprimed action $S_{\rm emb}^{\rm (3
br)}[f^A]$ (i.e., $f^A(x)$ obeys Hamilton's principle $\delta
S_{\rm emb}^{\rm (3 br)}[f^A] = 0$), then the primed map
$f^{\prime A}(x^\prime)$ is an extremum point of the primed action
$S_{\rm emb}^{({\rm 3 br}) \, \prime}[f^{\prime A}]$ in the primed
system ($x^{\prime \rho}, f^{\prime A}, \widehat{{\cal L}}_{\rm
emb}^{\, \prime}$) 
\beq \label{primed-3br-eff-action} S_{\rm emb}^{({\rm 3 br}) \,
\prime}[f^{\prime A}] ~ = \, \int_{{\cal WV}_{\rm sq}} d^{\, 4}
x^\prime ~ \widehat{{\cal L}}_{\rm emb}^{\, \prime}(f^{\prime A},
\partial_\rho^{\, \prime} f^{\prime A}, \ldots \,) ~, \eeq 
and vice versa.

The situation that both of $f^{\prime A}(x^\prime)$ and $f^A(x)$
are the solutions can be realized by the sameness in the values of
the two actions (called the ``value invariance of the action") 
\beq \label{value-inv} S_{\rm emb}^{({\rm 3 br}) \,
\prime}[f^{\prime A}] ~ = ~ S_{\rm emb}^{\rm (3 br)}[f^A] ~, \eeq
which leads to 
\beq \label{transf-Lag} \widehat{{\cal L}}_{\rm emb}^{\, \prime}
(f^{\prime A}, \partial_\rho^{\, \prime} f^{\prime A}, \ldots \,)
~ = ~ {\rm det}(\partial x^\mu / \partial x^{\prime \rho}) \times
\widehat{{\cal L}}_{\rm emb}(f^A, \partial_\mu f^A, \ldots \,) ~.
\eeq 

Due to $f^{\prime A}(x^\prime) = f^A(x)$ in Eq.~(\ref{diff4}), the
primed metric $\gamma_{\rho \sigma}^{\, \prime} \stackrel{\rm
def}{=} \partial_\rho^{\, \prime} f^{\prime A} \partial_\sigma^{\,
\prime} f^{\prime B} \eta_{AB}^{\rm bulk}$ follows the usual
transformation law $\gamma_{\rho \sigma}^{\, \prime} = \frac{\,
\partial x^\mu}{\, \partial x^{\prime \rho}} \frac{\, \partial
x^\nu}{\, \partial x^{\prime \sigma}} \, \gamma_{\mu \nu}$ for a
($0,2$)-type tensor, which together with ${\rm det}(\gamma_{\mu
\nu}) \ne 0$ in Eq.~(\ref{nondeg}) implies 
\beq \label{Jac-det} {\rm det}(\partial x^\mu / \partial x^{\prime
\rho}) ~ = ~ \sqrt{|\, {\rm det}(\gamma^{\, \prime}_{\rho
\sigma})|} ~ / \sqrt{|\, {\rm det}(\gamma_{\mu \nu})|} ~. \eeq 
Then, the Lagrangian ${\cal L}_{\rm emb}$ defined as ${\cal
L}_{\rm emb} \stackrel{\rm def}{=} \widehat{{\cal L}}_{\rm emb} \,
/ \sqrt{|\, {\rm det}(\gamma_{\mu \nu})|}$ is a scalar under ${\rm
Diff}(4)$ due to ${\cal L}_{\rm emb}^{\prime} (f^{\prime A},
\partial_\rho^{\, \prime} f^{\prime A}, \ldots \,) = {\cal L}_{\rm
emb}(f^A, \partial_\mu f^A, \ldots \,)$ unlike the scalar density
$\widehat{{\cal L}}_{\rm emb}$ of weight $-1$. Note that
$\sqrt{|\, {\rm det}(\gamma_{\mu \nu})|}$ is a function of
$\partial_\mu f^A$ due to the definition $\gamma_{\mu \nu} =
\partial_\mu f^A \partial_\nu f^B \eta_{AB}^{\rm bulk}$ in
Eq.~(\ref{ind-metric}).

In addition, when the ``form invariance of the Lagrangian density"
\beq \label{form-inv} \widehat{{\cal L}}_{\rm emb}^{\,
\prime}(f^{\prime A}, \partial_\rho^{\, \prime} f^{\prime A},
\ldots \,) ~ = ~ \widehat{{\cal L}}_{\rm emb} (f^{\prime A},
\partial_\rho^{\, \prime} f^{\prime A}, \ldots \,) \eeq 
(thus ${\cal L}_{\rm emb}^{\prime}(f^{\prime A}, \partial_\rho^{\,
\prime} f^{\prime A}, \ldots \,) = {\cal L}_{\rm emb} (f^{\prime
A}, \partial_\rho^{\, \prime} f^{\prime A}, \ldots \,)$) is
fulfilled, the Euler-Lagrange equation for the primed map
$f^{\prime A}(x^\prime)$ has the same form as that for the
unprimed map $f^A(x)$.

Similarly, the $ISO(1, D_{\rm amb}-1)$ invariance of $S_{\rm
emb}^{\rm (3 br)}[f^A]$ consists of two parts, (a) the value
invariance of the action, and (b) the form invariance of the
Lagrangian density. Due to these value and form invariances, the
invariance under a translation $f^A \rightarrow f^A + c^{\, A}$
from $ISO(1, D_{\rm amb}-1)$ means that the Lagrangian density
$\widehat{{\cal L}}_{\rm emb}$ (thus ${\cal L}_{\rm emb}$) does
not contain any derivative-free terms of $f^A$ (e.g., $f^A f^B
\eta_{AB}^{\rm bulk}$), implying $\widehat{{\cal L}}_{\rm emb} =
\widehat{{\cal L}}_{\rm emb}(\partial_\mu f^A, \ldots \,)$.

From now on, the effective action $S_{\rm emb}^{\rm (3 br)}[f^A]$
in Eq.~(\ref{3br-eff-action}) is expressed as the form using the
${\rm Diff}(4)$-invariant volume element $d^{\, 4} x \sqrt{|\,
{\rm det}(\gamma_{\mu \nu})|}$ 
\beq \label{final-eff-S} S_{\rm emb}^{\rm (3 br)}[f^A] ~ = \,
\int_{{\cal WV}_{\rm sq}} d^{\, 4} x \sqrt{|\, {\rm
det}(\gamma_{\mu \nu})|} ~ {\cal L}_{\rm emb} (\partial_\mu f^A,
\ldots \,) ~, \eeq 
where the Lagrangian ${\cal L}_{\rm emb}$ can contain the UV
cutoff $\Lambda_{\rm cont}$ in Eq.~(\ref{cut-cont}). The
Lagrangian ${\cal L}_{\rm emb}$ of the 3-brane action $S_{\rm
emb}^{\rm (3 br)}[f^A]$ can have the form of 
\beq \label{f-Lag} {\cal L}_{\rm emb} (\partial_\mu f^A, \ldots
\,) ~ = \, - \, {\cal T}_{\rm 3 br} \, + \, {\cal L}_{\rm
emb}^{\rm (der)}(\partial_\mu f^A, \ldots \,) ~, \eeq 
where ${\cal T}_{\rm 3 br}$ is the ``energy density" or
``three-brane tension" of the space 3-brane, which corresponds to
the Nambu-Goto action for a three-brane \cite{{string-th},{Goto}}.

The ``derivative Lagrangian" ${\cal L}_{\rm emb}^{\rm
(der)}(\partial_\mu f^A, \ldots \,)$ in Eq.~(\ref{f-Lag}) does not
have any constant term. This derivative Lagrangian ${\cal L}_{\rm
emb}^{\rm (der)}$ can be originated from (i) internal interactions
between space quanta (e.g., elastic forces), and/or (ii) external
interactions of space quanta with other kind(s) of particles. This
Lagrangian ${\cal L}_{\rm emb}^{\rm (der)}$ can contain the
Einstein-Hilbert term $d_{\, 2} \, \Lambda_{\rm cont}^2 R$, where
$R$ is the Ricci scalar built from the induced metric $\gamma_{\mu
\nu}$ ($d_{\, 2}$: constant).

The ``embedding scalars" $f^A$ of the 3-brane action $S_{\rm
emb}^{\rm (3 br)}[f^A]$ are {\it different} in two ways from
ordinary scalars (e.g., pions $\pi^{\pm, \, 0}$) in General
Relativity, as follows:

First, unlike the ordinary scalars, the embedding scalars $f^A(x)$
appear in the metric tensor $\gamma_{\mu \nu}$ of the world volume
${\cal WV}_{\rm sq}$ through the definition $\gamma_{\mu \nu} =
\partial_\mu f^A \partial_\nu f^B \eta_{AB}^{\rm bulk}$. As a result,
the embedding scalars $f^A$ also appear in the quantities
depending on $\gamma_{\mu \nu}$, for example, (i) $\sqrt{|\, {\rm
det}(\gamma_{\mu \nu})|}$ in the action $S_{\rm emb}^{\rm (3
br)}[f^A]$, (ii) the Christoffel symbols $\Gamma_{\mu \nu}^\rho$
for the covariant derivative $\nabla_\mu$, and (iii) the inverse
metric $\gamma^{\rho \sigma}$ for contraction.

Second, unlike the ordinary scalars, the solution $f_{\rm
sol}^A(x^\mu)$ of the equation $\delta S_{\rm emb}^{\rm (3
br)}[f^A] = 0$ is not an arbitrary function, but a 4DL embedding.
This 4DL embedding $f_{\rm sol}^A(x^\mu)$ makes the induced metric
$\gamma_{\mu \nu}$ a 4D metric of the signature $(-, +, +, +)$.
Then, the non-zero value of the composite field $\gamma_{\mu \nu}
=
\partial_\mu f_{\rm sol}^A \partial_\nu f_{\rm sol}^B
\eta_{AB}^{\rm bulk}$ may be interpreted as the ``condensation"
for the covariant four-vectors $\partial_\mu f^A$.

Now, we want to find the 4D-Lorentzian embedding $f^A(x)$ which
makes the world volume ${\cal WV}_{\rm sq}$ {\it globally flat},
that is, the induced metric 
\beq \label{emb-flat-sol} ~~~~~~~~~~~~~~~~ \gamma_{\mu
\nu}(x^\rho(p)) \, = \, \eta_{\mu \nu} ~~~ {\rm at ~ every ~
point} ~ p ~ {\rm of} ~ {\cal WV}_{\rm sq} \, . \eeq 
Due to the definition of $\gamma_{\mu \nu}$, the equality
$\gamma_{\mu \nu} = \eta_{\mu \nu}$ in Eq.~(\ref{emb-flat-sol})
can be represented as the partial differential equation (PDE) for
the 4DL embedding $f^A$ 
\beq \label{f-PDE} \partial_\mu f^A \, \partial_\nu f^B \,
\eta_{AB}^{\rm bulk} ~ = ~ \eta_{\mu \nu} ~. \eeq 

This PDE for the embedding $f^A$ can be solved, when its
derivatives $\partial_\mu f^A$ satisfy 
\beq \label{df-conden} ~~~~~~~~~~ \partial_\mu f^A \, = \,
\Lambda_{B^\mu}^A ~~~ {\rm everywhere ~ on} ~ {\cal WV}_{\rm sq}
\, , \eeq 
where $\Lambda_B^A \in SO(1, D_{\rm amb}-1)$, $B^{\, 0}=0$ (i.e.,
the bulk time), and three different $B^{\, i=1, \, 2, \, 3} \in
\{1, \ldots, D_{\rm amb} -1 \}$. A simple example is
$\Lambda_{B^\mu}^A = \delta^A_\mu$, where $\delta^A_\mu$ comes
from the bulk Kronecker delta. For $\partial_\mu f^A =
\Lambda_{B^\mu}^A$ in Eq.~(\ref{df-conden}), the induced metric is
expressed as 
\beq \label{gl-flat-ind-met} \gamma_{\mu \nu} \, = \, \eta_{B^\mu
B^\nu}^{\rm bulk} ~, \eeq 
which corresponds to the 4D Minkowski spacetime $\mathbb{M}^4$.

Because $\partial_\nu \Lambda_{B^\mu}^A = 0$ everywhere on ${\cal
WV}_{\rm sq}$, the integration of Eq.~(\ref{df-conden}) leads to a
linear function of $x^\mu$ 
\beq \label{lin-emb} f_{\rm lin}^A(x) ~ = ~ \Lambda_{B^\mu}^A \,
x^\mu \, + \, D^A ~, \eeq 
where $D^A$ are independent of $x^\mu$.

Then, our remaining task is to check whether this linear embedding
$f_{\rm lin}^A(x)$ is a solution of the Euler-Lagrange (E-L)
equation, as follows: Hamilton's principle using the Lagrangian
${\cal L}_{\rm emb} = - \, {\cal T}_{\rm 3 br} + {\cal L}_{\rm
emb}^{\rm (der)}$ in Eq.~(\ref{f-Lag})
\beq \label{Hamil-emb} \frac{\delta S_{\rm emb}^{\rm (3
br)}}{\delta f^A} \, [f^A] ~ = ~ 0 \eeq 
produces the equation of motion for the space 3-brane (i.e., the
E-L equation) 
\beq \label{f-eq} \left\{ \, \partial_\mu \left[ \, {\cal T}_{\rm
3 br} \sqrt{|\, {\rm det}(\gamma_{\rho \sigma})|} ~ \gamma^{\mu
\nu} \partial_\nu f^B \eta_{AB}^{\rm bulk} \, \right] \right\} \,
+ \, \cdots ~ = ~ 0 ~~~ {\rm with} ~ \gamma_{\rho \sigma} =
\partial_\rho f^A \partial_\sigma f^B \eta_{AB}^{\rm bulk}
~, \eeq 
where the ellipsis $\cdots$ denotes the contribution of the
derivative Lagrangian ${\cal L}_{\rm emb}^{\rm (der)}$.

If the energy density ${\cal T}_{\rm 3 br}$ of the space 3-brane
is independent of $x^\mu$, then each term of Eq.~(\ref{f-eq}) can
contain only the second or higher derivatives of $f^A$ (e.g.,
$\partial_\mu \partial_\nu f^A$). As a result, each term of
Eq.~(\ref{f-eq}) vanishes for any linear function of $x^\mu$, for
example, the linear embedding $f_{\rm lin}^A(x) =
\Lambda_{B^\mu}^A x^\mu + D^A$ in Eq.~(\ref{lin-emb}).

Therefore, for the $x^\mu$-independent energy density ${\cal
T}_{\rm 3 br}$, the linear embedding $f_{\rm lin}^A(x)$ is the
solution of the E-L equation in Eq.~(\ref{f-eq}), implying the
world volume ${\cal WV}_{\rm sq}$ is the 4D Minkowski spacetime
$\mathbb{M}^4$ due to the flat metric $\gamma_{\mu \nu} = \,
\eta_{B^\mu B^\nu}^{\rm bulk}$ induced by $f_{\rm lin}^A(x)$.

Since the linear embedding $f_{\rm lin}^A(x)$ satisfies the E-L
equation {\it irrespective of the value} of the
$x^\mu$-independent ${\cal T}_{\rm 3 br}$, the flat metric
$\gamma_{\mu \nu} = \eta_{\mu \nu}$ of the world volume ${\cal
WV}_{\rm sq}$ exists for any value of the uniform energy density
${\cal T}_{\rm 3 br}$. This interesting feature distinguishes the
3-brane action $S_{\rm emb}^{\rm (3 br)}[f^A]$ from General
Relativity.

\section{\label{sec: AAT-method} The Aim-At-Target (AAT) Method
for Studying the World Volume of the Space 3-Brane}

In Sec.~\ref{sec: 3br-eff-action}, since the space 3-brane
occupies the ambient spacetime $\mathbb{M}^{D_{\rm amb}}$, the
effective theory $S_{\rm emb}^{\rm (3 br)}[f^A]$ of the space
3-brane was built for the ambient spacetime $\mathbb{M}^{D_{\rm
amb}}$ through the field $f^A(x^\mu) \in \mathbb{M}^{D_{\rm
amb}}$. The world volume (${\cal WV}_{\rm sq}, \gamma_{\mu \nu}$)
of the space 3-brane is described by the solution $f_{\rm sol}^A$
for the equation of motion $\delta S_{\rm emb}^{\rm (3 br)} /
\delta f^A = 0$. Since this world volume (${\cal WV}_{\rm sq},
\gamma_{\mu \nu}$) is the exact or true spacetime of our universe,
it is important to know the solution $f_{\rm sol}^A$ describing
our spacetime (${\cal WV}_{\rm sq}, \gamma_{\mu \nu}$).

In this section, we want to show a methodology for studying the
world volume (${\cal WV}_{\rm sq}, \gamma_{\mu \nu}$) by using a
{\it metric action} $S_{\rm met}[g_{\mu \nu}]$ (see Table
\ref{out-AAT}). The key point of this methodology is that the
solution $f_{\rm sol}^A$ of $\delta S_{\rm emb}^{\rm (3 br)} /
\delta f^A = 0$  can be found by solving the different equation
$\partial_\mu f^A \partial_\nu f^B \eta_{AB}^{\rm bulk} = g_{\mu
\nu}$ when the new metric $g_{\mu \nu}$ satisfies $g_{\mu \nu} =
\gamma_{\mu \nu}$ ($= \partial_\mu f_{\rm sol}^A \partial_\nu
f_{\rm sol}^B \eta_{AB}^{\rm bulk}$). An example of the
methodology is the flat-metric case $\gamma_{\mu \nu} = \eta_{\mu
\nu}$, whose treatment is shown below Eq.~(\ref{emb-flat-sol}).
The details of our methodology are shown, as follows:

To study the world volume (${\cal WV}_{\rm sq}, \gamma_{\mu \nu}$)
of the space 3-brane, we consider the solution set $\Sigma_{\rm
target}$ for the equation of motion $\frac{\delta S_{\rm emb}^{\rm
(3 br)}}{\delta f^A}[f^A] = 0$ in Eq.~(\ref{Hamil-emb}) 
\beq \label{target-set} \Sigma_{\rm target} \, \stackrel{\rm
def}{=} \, \{\, f_{\rm sol}^A : {\rm 4DL ~ embedding} \,| ~
(\delta S_{\rm emb}^{\rm (3 br)} / \delta f^A)[f_{\rm sol}^A] = 0
\, \} ~~~ \subset ~ {\cal F}_{\rm space} ~, \eeq 
where ${\cal F}_{\rm space} = \{ \phi^A \}$ is the function space,
and the element $f_{\rm sol}^A$ is called the ``4D-Lorentzian
(4DL) solution." For the given {\bf original action} $S_{\rm
emb}^{\rm (3 br)}[f^A]$, the solution set $\Sigma_{\rm target} =
\{ f_{\rm sol}^A \}$ is a {\it fixed} set in the function space
${\cal F}_{\rm space}$. Note that $\Sigma_{\rm target} \owns
f_{\rm lin}^A$ for a uniform energy density ${\cal T}_{\rm 3 br}$
(see around Eq.~(\ref{Hamil-emb})).

By the way, since our methodology to study the solution set
$\Sigma_{\rm target} = \{ f_{\rm sol}^A \}$ is similar to ``a
bullet fired at a fixed target in space" (see the discussions
around Eq.~(\ref{methodology})), the solution set $\Sigma_{\rm
target} = \{ f_{\rm sol}^A \}$ is called the {\bf target} (or
target set) in the {\bf space} ${\cal F}_{\rm space} = \{ \phi^A
\}$. For an easy understanding, we show the tenors and the
vehicles in the bullet-target (B-T) metaphor 
\beq \label{metaphor} \langle ~\, \Sigma_{\rm bullet} \, , ~
\Sigma_{\rm target} \, , ~ {\cal F}_{\rm space} ~\, \rangle ~~
\Longleftrightarrow ~~ \langle \langle ~\, {\rm bullet} \, , ~
{\rm target} \, , ~ {\rm space} ~\, \rangle \rangle ~, \eeq 
where the ``bullet" $\Sigma_{\rm bullet}$ will be defined in
Eq.~(\ref{bullet-set}). Moreover, there are discussions about the
``rifle" (above Eq.~(\ref{rifle-PDE})) and the ``aim-of-the-rifle"
(above Eq.~(\ref{methodology})).

Through the definition $\gamma_{\mu \nu} = \partial_\mu f^A
\partial_\nu f^B \eta_{AB}^{\rm bulk}$ in Eq.~(\ref{ind-metric}),
the target $\Sigma_{\rm target} = \{ f_{\rm sol}^A \}$ in
Eq.~(\ref{target-set}) produces the set $\Sigma_{\rm ind}$ of the
induced metrics $\gamma_{\mu \nu}$ with the signature $(-, +, +,
+)$ 
\beq \label{set-ind} \Sigma_{\rm ind} \, \stackrel{\rm def}{=} \,
\{\, \gamma_{\mu \nu} \,| ~ \gamma_{\mu \nu} = \partial_\mu f_{\rm
sol}^A \, \partial_\nu f_{\rm sol}^B \, \eta_{AB}^{\rm bulk} ~~
{\rm for ~ every} ~ f_{\rm sol}^A \in \Sigma_{\rm target} \, \} ~.
\eeq 
Note that $\Sigma_{\rm ind} \owns \eta_{\mu \nu}$ for a uniform
energy density ${\cal T}_{\rm 3 br}$ (see around
Eq.~(\ref{gl-flat-ind-met})).

Conversely, this ``induced-metric set" $\Sigma_{\rm ind} = \{
\gamma_{\mu \nu} \}$ can produce the target set $\Sigma_{\rm
target} = \{ f_{\rm sol}^A \}$, because (i) the former set
$\Sigma_{\rm ind} = \{ \gamma_{\mu \nu} \}$ produces the solution
set $\Sigma_{\rm PDE}^{\rm (sol)}$ of the partial differential
equation (PDE) for the embedding $f^A$ 
\beq \label{sol-PDE} \Sigma_{\rm PDE}^{\rm (sol)} \, \stackrel{\rm
def}{=} \, \{\, f^A \,| ~ \partial_\mu f^A
\partial_\nu f^B \eta_{AB}^{\rm bulk} \, = \, \gamma_{\mu \nu} ~~
{\rm for ~ every} ~ \gamma_{\mu \nu} \in \Sigma_{\rm ind} \, \} ~,
\eeq 
and (ii) this ``PDE solution set" $\Sigma_{\rm PDE}^{\rm (sol)}$
contains the target set $\Sigma_{\rm target} = \{ f_{\rm sol}^A
\}$, i.e., 
\beq \label{ind-cont-emb} \Sigma_{\rm PDE}^{\rm (sol)} ~ \supset ~
\Sigma_{\rm target} ~. \eeq 

The result $\Sigma_{\rm PDE}^{\rm (sol)} \supset \Sigma_{\rm
target}$ in Eq.~(\ref{ind-cont-emb}) suggests a hint about how to
know the target set $\Sigma_{\rm target}$ ($\subset {\cal F}_{\rm
space}$), implying the importance of studying the induced-metric
set $\Sigma_{\rm ind} = \{ \gamma_{\mu \nu} \}$. To study this set
$\Sigma_{\rm ind} = \{ \gamma_{\mu \nu} \}$, since its element
$\gamma_{\mu \nu}$ is a 4D Lorentzian metric, we consider the
theory $S_{\rm met}[g_{\mu \nu}]$ of a 4D Lorentzian metric
$g_{\mu \nu}$ (rather than $\gamma_{\mu \nu}$) 
\beq \label{met-action} S_{\rm met} [\, g_{\mu \nu}] ~ = \,
\int_{{\cal S}_{\rm met}^{\rm 4D}} d^{\, 4} x \sqrt{|\, {\rm
det}(g_{\mu \nu})|} ~ {\cal L}_{\rm met} (\Lambda_{\rm met};
g_{\mu \nu}) ~, \eeq 
where (${\cal S}_{\rm met}^{\rm \, 4D}, g_{\mu \nu}$) is a 4D
spacetime manifold, ${\cal L}_{\rm met} (\Lambda_{\rm met}; g_{\mu
\nu})$ is the Lagrangian containing the derivatives of $g_{\mu
\nu}$, and $\Lambda_{\rm met}$ is the UV cutoff of the metric
action $S_{\rm met}[g_{\mu \nu}]$ (see Table \ref{out-AAT} for the
role of $S_{\rm met}[g_{\mu \nu}]$).

Since the metric action $S_{\rm met}[g_{\mu \nu}]$ in
Eq.~(\ref{met-action}) neglects the {\it microscopic} behaviors of
individual space quanta (i.e., the underlying discreteness of the
space 3-brane) like the original action $S_{\rm emb}^{\rm (3
br)}[f^A]$, we can assume the UV cutoff $\Lambda_{\rm met}$ of the
metric action $S_{\rm met}[g_{\mu \nu}]$ satisfies 
\beq \label{cut-met<cont} \Lambda_{\rm met} \, \lesssim \,
O(\Lambda_{\rm cont}) ~, \eeq 
where $\Lambda_{\rm cont}$ is the UV cutoff of the original action
$S_{\rm emb}^{\rm (3 br)}[f^A]$ (see Eqs.~(\ref{cut-cont}) and
(\ref{3br-eff-action})). The spacetime manifold (${\cal S}_{\rm
met}^{\rm \, 4D}, g_{\mu \nu}$) for the metric action $S_{\rm met}
[g_{\mu \nu}]$ can be a good approximation of the exact or true
spacetime manifold (${\cal WV}_{\rm sq}, \gamma_{\mu \nu}$), which
is an emergent object arising from many space quanta in
$\mathbb{M}^{D_{\rm amb}}$ (see Eqs.~(\ref{met-sol=ind}),
(\ref{ST-met=WV}) and (\ref{equal-ST})).

Then, we can approach the induced-metric set $\Sigma_{\rm ind} =
\{ \gamma_{\mu \nu} \}$ by using the solution set $\Sigma_{\rm
met}^{\rm (sol)}$ of the equation $\frac{\delta S_{\rm
met}}{\delta g_{\mu \nu}}[g_{\mu \nu}] = 0$ 
\beq \label{sol-met-set} \Sigma_{\rm met}^{\rm (sol)} \,
\stackrel{\rm def}{=} \, \{\, g_{\mu \nu}^{\rm sol} \,| ~ (\delta
S_{\rm met} / \delta g_{\mu \nu}) [g_{\mu \nu}^{\rm sol}] = 0 \,
\} ~, \eeq 
which is called the {\bf cartridge} (see above
Eq.~(\ref{rifle-PDE})). The solution $g_{\mu \nu}^{\rm sol}$ in
Eq.~(\ref{sol-met-set}) is called the ``solution metric."

In order to study the target $\Sigma_{\rm target} = \{ f_{\rm
sol}^A \}$ in the space ${\cal F}_{\rm space}$, we define the {\bf
bullet} $\Sigma_{\rm bullet}$ (corresponding to the PDE solution
set $\Sigma_{\rm PDE}^{\rm (sol)}$ in Eq.~(\ref{sol-PDE})) 
\beq \label{bullet-set} \Sigma_{\rm bullet} \, \stackrel{\rm
def}{=} \, \{\, f^A \,| ~ \partial_\mu f^A \partial_\nu f^B
\eta_{AB}^{\rm bulk} = g_{\mu \nu}^{\rm sol} ~~ {\rm for ~ every}
~ g_{\mu \nu}^{\rm sol} \in \Sigma_{\rm met}^{\rm (sol)} \, \} ~~~
\subset ~ {\cal F}_{\rm space} \eeq 
by using the cartridge $\Sigma_{\rm met}^{\rm (sol)} = \{ g_{\mu
\nu}^{\rm sol} \}$. Since the bullet set $\Sigma_{\rm bullet} = \{
f_{\rm bul}^A \}$ can overlap the target set $\Sigma_{\rm target}
= \{ f_{\rm sol}^A \}$ in the function space ${\cal F}_{\rm space}
= \{ \phi^A \}$ like the ``bullet" of the B-T metaphor ``a bullet
fired at a fixed target in space" (see Eq.~(\ref{metaphor})), the
set $\Sigma_{\rm bullet} = \{ f_{\rm bul}^A \}$ is called the
bullet.

In Eq.~(\ref{bullet-set}), the PDE $\partial_\mu f^A \partial_\nu
f^B \eta_{AB}^{\rm bulk} = g_{\mu \nu}^{\rm sol}$ defines a
transformation $\Sigma_{\rm met}^{\rm (sol)} \rightarrow
\Sigma_{\rm bullet} = \Psi_{\rm PDE}(\Sigma_{\rm met}^{\rm
(sol)})$ like the rifle of the above B-T metaphor, which
transforms its cartridge into the metallic bullet. Thus, the
``solution-metric set" $\Sigma_{\rm met}^{\rm (sol)} = \{ g_{\mu
\nu}^{\rm sol} \}$ and the PDE $\partial_\mu f^A \partial_\nu f^B
\eta_{AB}^{\rm bulk} = g_{\mu \nu}^{\rm sol}$ are called the
cartridge (see below Eq.~(\ref{sol-met-set})), and the {\bf rifle}
firing the bullet $\Sigma_{\rm bullet} = \{ f_{\rm bul}^A \}$,
respectively. This ``rifle PDE," and the PDE for $\Sigma_{\rm
PDE}^{\rm (sol)}$ in Eq.~(\ref{sol-PDE}) have the same form 
\beq \label{rifle-PDE} ~~~~~~~~~~ \partial_\mu f^A \, \partial_\nu
f^B \, \eta_{AB}^{\rm bulk} \, = \, q_{\mu \nu} ~~~ (q_{\mu \nu} =
\, g_{\mu \nu}^{\rm sol}, ~ \gamma_{\mu \nu}) ~, \eeq 
which is invariant under the bulk Poincar$\acute{\textrm{e}}$
group $ISO(1, D_{\rm amb}-1)$ with $f^{\prime A} = \Lambda_B^A f^B
+ c^{\, A}$ (see Sec.~\ref{sec: 3br-eff-action}).

Thus, if the ``metric intersection (MI)" $\Sigma_{\rm MI}
\stackrel{\rm def}{=} \Sigma_{\rm met}^{\rm (sol)} \cap
\Sigma_{\rm ind} = \{ {\rm g}_{\mu \nu}^{\rm MI} \}$ contains an
element 
\beq \label{g=gamma} ~~~~~ {\rm g}_{\mu \nu}^{\rm MI \otimes} \, =
\, g_{\mu \nu}^{\rm sol \otimes} \, = \, \gamma_{\mu \nu}^\otimes
~~ \left( = \partial_\mu f_{\rm sol \otimes}^A \, \partial_\nu
f_{\rm sol \otimes}^B \, \eta_{AB}^{\rm bulk} ~~~ {\rm with} ~
f_{\rm sol \otimes}^A \in \Sigma_{\rm target} = \{ f_{\rm sol}^A
\} \right) \, , \eeq 
then the bullet $\Sigma_{\rm bullet} = \{ f_{\rm bul}^A \}$ shares
the 4DL solution $f_{\rm sol \otimes}^A$ with the target
$\Sigma_{\rm target} = \{ f_{\rm sol}^A \}$. Since the converse of
this proposition is true, we have the equivalence that 
\beq \label{MI-BT} \Sigma_{\rm MI} \, = \, \Sigma_{\rm met}^{\rm
(sol)} \, \cap \, \Sigma_{\rm ind} ~ \ne ~ \varnothing ~~~~ {\it
if ~ and ~ only ~ if} ~~~~ \Sigma_{{\rm B} \cap {\rm T}} \,
\stackrel{\rm def}{=} \, \Sigma_{\rm bullet} \, \cap \,
\Sigma_{\rm target} ~ \ne ~ \varnothing ~, \eeq 
where $\Sigma_{{\rm B} \cap {\rm T}}$ is called the
``bullet-target (B-T) overlap."

In Eq.~(\ref{MI-BT}), the B-T overlap $\Sigma_{{\rm B} \cap {\rm
T}} = \Sigma_{\rm bullet} \cap \Sigma_{\rm target}$ describes the
manner in which the target $\Sigma_{\rm target} = \{ f_{\rm sol}^A
\}$ of the original action $S_{\rm emb}^{\rm (3 br)}[f^A]$ is
overlapped by the bullet $\Sigma_{\rm bullet} = \{ f_{\rm bul}^A
\}$ of the metric action $S_{\rm met}[g_{\mu \nu}]$. Since the
original action $S_{\rm emb}^{\rm (3 br)}[f^A]$ is given to us
(i.e., not changed arbitrarily by us), the target $\Sigma_{\rm
target}$ in the B-T overlap $\Sigma_{{\rm B} \cap {\rm T}} =
\Sigma_{\rm bullet} \cap \Sigma_{\rm target}$ is treated as a {\it
fixed set} in the function space ${\cal F}_{\rm space} = \{ \phi^A
\}$. Then, the B-T overlap $\Sigma_{{\rm B} \cap {\rm T}}$
represents the {\it maximum knowledge about the fixed target
$\Sigma_{\rm target}$} which we can obtain by using the bullet
$\Sigma_{\rm bullet}$ of the chosen action $S_{\rm met}[g_{\mu
\nu}]$.

For example, the maximum B-T overlap $\Sigma_{{\rm B} \cap {\rm
T}}^{\rm (max)} = \Sigma_{\rm target}$ (i.e., $\Sigma_{\rm
bullet}^{\rm (max)} \supset \Sigma_{\rm target}$) means that we
can know the whole of the target $\Sigma_{\rm target}$ by using
the maximum bullet $\Sigma_{\rm bullet}^{\rm (max)}$. However, the
minimum overlap $\Sigma_{{\rm B} \cap {\rm T}}^{\rm (min)} =
\varnothing$ (i.e., $\Sigma_{\rm bullet}^{\rm (min)} \cap
\Sigma_{\rm target} = \varnothing$) means that we cannot know the
target $\Sigma_{\rm target}$ through the minimum bullet
$\Sigma_{\rm bullet}^{\rm (min)}$.

Fortunately, unlike the target $\Sigma_{\rm target}$, the bullet
$\Sigma_{\rm bullet}$ changes {\it depending on} which metric
action $S_{\rm met}[g_{\mu \nu}]$ we choose (see
Eqs.~(\ref{sol-met-set}) and (\ref{bullet-set})). Thus, this
metric action $S_{\rm met}[g_{\mu \nu}]$ corresponds to ``the aim
of the rifle at the target" of the above B-T metaphor ``a bullet
fired at a fixed target in space." Then, the metric action $S_{\rm
met}[g_{\mu \nu}]$ is called the {\bf aim-of-the-rifle}.

Through the dependence of $\Sigma_{\rm bullet}$ on $S_{\rm
met}[g_{\mu \nu}]$, the B-T overlap $\Sigma_{{\rm B} \cap {\rm T}}
= \Sigma_{\rm bullet} \cap \Sigma_{\rm target}$ in
Eq.~(\ref{MI-BT}) depends on the metric action $S_{\rm met}[g_{\mu
\nu}]$. In other words, the metric action $S_{\rm met}[g_{\mu
\nu}]$ produces the bullet $\Sigma_{\rm bullet}$, and this bullet
$\Sigma_{\rm bullet}$ produces the B-T overlap $\Sigma_{{\rm B}
\cap {\rm T}}$, i.e., 
\beq \label{methodology} ~~~~~~~~~~~ S_{\rm met}[g_{\mu \nu}] ~~~~
\longrightarrow ~~~~ \Sigma_{\rm bullet} ~~~~ \longrightarrow ~~~~
\Sigma_{{\rm B} \cap {\rm T}} ~ \, (= \, \Sigma_{\rm bullet} \,
\cap \, \Sigma_{\rm target}) ~. \eeq 
Since this sequence in Eq.~(\ref{methodology}) implies that the
metric action $S_{\rm met}[g_{\mu \nu}]$ determines the B-T
overlap $\Sigma_{{\rm B} \cap {\rm T}}$, various metric actions
$S_{\rm met}[g_{\mu \nu}]$ are classified by their B-T overlap
$\Sigma_{{\rm B} \cap {\rm T}}$ into two types, (i) the
``overlapping" metric action $S_{\rm met}^{\rm (ovlp)}[g_{\mu
\nu}]$ satisfying $\Sigma_{{\rm B} \cap {\rm T}} \ne \varnothing$,
and (ii) the ``non-overlapping" metric action $S_{\rm
met}^{(\not{\rm ovlp})}[g_{\mu \nu}]$ satisfying $\Sigma_{{\rm B}
\cap {\rm T}} = \varnothing$.

Because a non-overlapping action $S_{\rm met}^{(\not{\rm
ovlp})}[g_{\mu \nu}]$ has its empty B-T overlap $\Sigma_{{\rm B}
\cap {\rm T}}^{(\not{\rm ovlp})} = \varnothing$ ($= \Sigma_{{\rm
B} \cap {\rm T}}^{\rm (min)}$), the solution set $\Sigma_{\rm
target} = \{ f_{\rm sol}^A \}$ of the original action $S_{\rm
emb}^{\rm (3 br)}[f^A]$ cannot be known by using the
non-overlapping bullet $\Sigma_{\rm bullet}^{(\not{\rm ovlp})}$
(see above). Thus, this {\it undesirable} action $S_{\rm
met}^{(\not{\rm ovlp})}[g_{\mu \nu}]$ should be modified.

Then, through the dependence of $\Sigma_{\rm bullet}$ on $S_{\rm
met}[g_{\mu \nu}]$, we can find a suitable overlapping action
$S_{\rm met}^{\rm (ovlp)}[g_{\mu \nu}]$ by making $\Sigma_{{\rm B}
\cap {\rm T}} \ne \varnothing$, namely, by changing (i) the form
of the metric action $S_{\rm met}[g_{\mu \nu}]$ and (ii) the
values of its parameters. As a result, we can (partially) know the
target $\Sigma_{\rm target} = \{ f_{\rm sol}^A \}$ by using the
bullet $\Sigma_{\rm bullet}^{\rm (ovlp)}$ of the overlapping
action $S_{\rm met}^{\rm (ovlp)}[g_{\mu \nu}]$. This methodology
for knowing the target $\Sigma_{\rm target}$ by trying the
aim-of-the-rifle $S_{\rm met}[g_{\mu \nu}]$ is called the
``Aim-At-Target (AAT) method."

\begin{table} \centering \caption{The Outline of the
Aim-At-Target (AAT) Method for $\Sigma_{{\rm B} \cap {\rm T}} \ne
\varnothing$} \vspace*{0.3cm} \begin{tabular}{cll} \hline
\vspace*{-0.35cm} \\ {\sc Action} & {\sc Name} & {\sc Role}
\vspace*{0.05cm} \\ \hline \vspace*{-0.3cm} \\ $S_{\rm emb}^{\rm
(3 br)}[f^A]$ & Original action & The ``true theory" of the space
3-brane within $\mathbb{M}^{D_{\rm amb}}$ \vspace*{0.1cm} \\
\vspace*{-0.42cm} \\ $S_{\rm met}^{\rm (ovlp)}[g_{\mu \nu}]$ &
Overlapping action & $\left\{ \begin{array}{l} {\rm (a) \, ME : a
~ ``mere ~ tool" ~ for ~ knowing ~ \Sigma_{\rm target} ~ of} ~
S_{\rm emb}^{\rm (3 br)} \vspace*{0.2cm} \\ {\rm (b) \, PE :
producing ~ a ~ ``constitutive ~ equation"} \\ \end{array}
\right.$ \vspace*{0.2cm} \\ \hline \vspace*{-0.12cm}
\end{tabular} \label{out-AAT} {\small $\left[ \begin{array}{l} ~~
{\it Caution \, \textit{:}} ~~ {\rm the ~ motion ~ of ~ the ~
space ~ 3\textrm{-}brane ~ in} ~ \mathbb{M}^{D_{\rm amb}} ~ {\rm
is ~ described ~ by ~ either} ~~ \vspace*{0.05cm} \\ ~~~~~~~~ {\rm
(a^\prime) ~ the ~ target} ~ \Sigma_{\rm target} ~ {\rm or ~
(b^\prime) ~ the ~ B\textrm{-}T ~ overlap} ~ \Sigma_{{\rm B} \cap
{\rm T}}^{\rm (ovlp) \cdot PE} , ~ {\rm depending ~ on} ~
\vspace*{0.05cm} \\ ~~~~~~~~ {\rm the ~ role ~ of ~ the ~
overlapping ~ action} ~ S_{\rm met}^{\rm (ovlp)}[g_{\mu \nu}] ~
({\rm see ~ the ~ text}) . \\ \end{array} \right]$}
\end{table}

For a better understanding of this AAT method, we summarize it for
a non-empty B-T overlap $\Sigma_{{\rm B} \cap {\rm T}} \ne
\varnothing$ (see Table \ref{out-AAT}), as follows: first, the AAT
method uses the two actions $S_{\rm emb}^{\rm (3 br)}[f^A]$ and
$S_{\rm met}[g_{\mu \nu}]$. Second, since the original action
$S_{\rm emb}^{\rm (3 br)}[f^A]$ shows the truth (e.g., the
equation of motion in $\mathbb{M}^{D_{\rm amb}}$) about the space
3-brane occupying $\mathbb{M}^{D_{\rm amb}}$, the 3-brane action
$S_{\rm emb}^{\rm (3 br)}[f^A]$ is the ``true theory" of the space
3-brane in $\mathbb{M}^{D_{\rm amb}}$.

Third, when the B-T overlap $\Sigma_{{\rm B} \cap {\rm T}} =
\Sigma_{\rm bullet} \cap \Sigma_{\rm target}$ is not empty (i.e.,
$\Sigma_{{\rm B} \cap {\rm T}} \ne \varnothing$), the metric
action $S_{\rm met}[g_{\mu \nu}]$ can be desirable. Depending on
the ``mode of existence" (mathematical/physical), the overlapping
metric action $S_{\rm met}^{\rm (ovlp)}[g_{\mu \nu}]$ has two
different implications:
\begin{itemize}
\item (a) An overlapping action $S_{\rm met}^{\rm (ovlp)}[g_{\mu
\nu}]$ has only the ``mathematical existence (ME)" unlike the
original action $S_{\rm emb}^{\rm (3 br)}[f^A]$: this overlapping
action is called the ``ME action" $S_{\rm met}^{\rm (ME)}[g_{\mu
\nu}]$. Due to its mathematical existence, this ME action $S_{\rm
met}^{\rm (ME)}[g_{\mu \nu}]$ cannot affect the occurrence of any
element $f_{\rm sol}^A$ of the target $\Sigma_{\rm target} = \{
f_{\rm sol}^A \}$ through the B-T overlap $\Sigma_{{\rm B} \cap
{\rm T}}^{\rm (ovlp) \cdot ME}$. In other words, irrespective of
the B-T overlap $\Sigma_{{\rm B} \cap {\rm T}}^{\rm (ovlp) \cdot
ME}$, every element $f_{\rm sol}^A$ of the target $\Sigma_{\rm
target}$ still can occur in the ambient spacetime
$\mathbb{M}^{D_{\rm amb}}$ as a motion of the space 3-brane.
Therefore, the ME action $S_{\rm met}^{\rm (ME)}[g_{\mu \nu}]$
making the overlap $\Sigma_{{\rm B} \cap {\rm T}}^{\rm (ovlp)
\cdot ME}$ is a ``mere tool" for knowing the target $\Sigma_{\rm
target}$ of the true theory $S_{\rm emb}^{\rm (3 br)}[f^A]$.
\item (b) An overlapping action $S_{\rm met}^{\rm (ovlp)}[g_{\mu
\nu}]$ has the ``physical existence (PE)" like the original action
$S_{\rm emb}^{\rm (3 br)}[f^A]$: this overlapping action is called
the ``PE action" $S_{\rm met}^{\rm (PE)}[g_{\mu \nu}]$. Due to its
physical existence, this PE action $S_{\rm met}^{\rm (PE)}[g_{\mu
\nu}]$ allows (forbids) the occurrence of an element $f_{\rm
sol}^A$ of the target $\Sigma_{\rm target}$, when this 4DL
solution $f_{\rm sol}^A$ does (not) belong to the B-T bullet
$\Sigma_{{\rm B} \cap {\rm T}}^{\rm (ovlp) \cdot PE}$. In other
words, only the element $f_{\rm ove}^A$ of the B-T overlap
$\Sigma_{{\rm B} \cap {\rm T}}^{\rm (ovlp) \cdot PE} = \{ f_{\rm
ove}^A \}$ ($\subset \Sigma_{\rm target}$) can occur in the
ambient spacetime $\mathbb{M}^{D_{\rm amb}}$ as a motion of the
space 3-brane. Since this decrease in ``set of possible motions"
from $\Sigma_{\rm target}$ to $\Sigma_{{\rm B} \cap {\rm T}}^{\rm
(ovlp) \cdot PE}$ ($\subset \Sigma_{\rm target}$) is similarly
found for constitutive equations (e.g., Ohm's law), the PE action
$S_{\rm met}^{\rm (PE)}[g_{\mu \nu}]$ making the overlap
$\Sigma_{{\rm B} \cap {\rm T}}^{\rm (ovlp) \cdot PE}$ produces a
``constitutive equation" specific to the space 3-brane in
$\mathbb{M}^{D_{\rm amb}}$.
\end{itemize}

Our AAT method using the metric action $S_{\rm met}^{\rm
(ovlp)}[g_{\mu \nu}]$ seems similarly found in General Relativity:
in General Relativity, the locally inertial coordinates (LIC)
$\xi^{\hat{\alpha}}$ are studied by solving the PDE $\partial_\mu
\xi^{\hat{\alpha}} \partial_\nu \xi^{\hat{\beta}}
\eta_{\hat{\alpha} \hat{\beta}} = {\bf g}_{\mu \nu}^{\rm sol}$
with $(\delta S_{\rm EH} / \delta {\bf g}_{\mu \nu})[{\bf g}_{\mu
\nu}^{\rm sol}] = 0$ (this situation corresponds to
Eqs.~(\ref{sol-met-set}) and (\ref{bullet-set})). Moreover, these
LIC $\xi^{\hat{\alpha}}$ are similar to the embedding $f^A$ of the
action $S_{\rm emb}^{\rm (3 br)}[f^A]$, because (i)
$\xi^{\hat{\alpha}}$ appear in the ``GR metric" ${\bf g}_{\mu \nu}
= \partial_\mu \xi^{\hat{\alpha}} \partial_\nu \xi^{\hat{\beta}}
\eta_{\hat{\alpha} \hat{\beta}}$ like $f^A$ in $\gamma_{\mu \nu} =
\partial_\mu f^A \partial_\nu f^B \eta_{AB}^{\rm bulk}$, and (ii)
$\xi^{\hat{\alpha}}$ form an immersion of $x^\mu$ like $f^A$ due
to ${\rm rank}(\partial_\mu \xi^{\hat{\alpha}}) = {\rm
rank}(\partial_\mu f^A) = 4$, where $\partial_\mu
\xi^{\hat{\alpha}}$ is the vierbein. Then, due to these {\it
similarities} between $\xi^{\hat{\alpha}}$ and $f^A$, the
analogical reasoning can support that the embedding $f^A$ has the
metric action $S_{\rm met}^{\rm (ovlp)}[g_{\mu \nu}]$ like the LIC
$\xi^{\hat{\alpha}}$ having $S_{\rm EH} [{\bf g}_{\mu \nu}]$. For
the use of these actions $S_{\rm met}^{\rm (ovlp)}[g_{\mu \nu}]$
and $S_{\rm EH} [{\bf g}_{\mu \nu}]$, see between
Eqs.~(\ref{sol-subst}) and (\ref{equiv-var}).

Mathematically, the AAT method using the overlapping action
$S_{\rm met}^{\rm (ovlp)}[g_{\mu \nu}]$ consists of two main
steps: (i) finding a solution $g_{\mu \nu}^{\rm sol}$ of $(\delta
S_{\rm met}^{\rm (ovlp)} / \delta g_{\mu \nu}) [g_{\mu \nu}] = 0$
as in Eq.~(\ref{sol-met-set}), and next (ii) finding a solution
$f_{\rm bul}^A$ of the rifle PDE $\partial_\mu f^A \partial_\nu
f^B \eta_{AB}^{\rm bulk} = g_{\mu \nu}^{\rm sol}$ as in
Eq.~(\ref{bullet-set}).

Since this ``two-step AAT method" is a method for solving the two
coupled equations $(\delta S_{\rm met}^{\rm (ovlp)} / \delta
g_{\mu \nu}) [g_{\mu \nu}] = 0$ and $\partial_\mu f^A \partial_\nu
f^B \eta_{AB}^{\rm bulk} = g_{\mu \nu}$, we can try a different
method, i.e., the insertion of the latter equation $\partial_\mu
f^A \partial_\nu f^B \eta_{AB}^{\rm bulk} = g_{\mu \nu}$ into the
former 
\beq \label{sol-subst} \left. \frac{\delta S_{\rm met}^{\rm
(ovlp)}}{\delta g_{\mu \nu}} \, \right|_{\rm repl} ~ \stackrel{\rm
def}{=} ~ \frac{\delta S_{\rm met}^{\rm (ovlp)}}{\delta g_{\mu
\nu}} \, [\partial_\mu f^A \partial_\nu f^B \eta_{AB}^{\rm bulk}]
~ = ~ 0 ~, \eeq 
where the symbol $|_{\rm repl}$ denotes the replacement $g_{\mu
\nu} \Rrightarrow \partial_\mu f^A \partial_\nu f^B \eta_{AB}^{\rm
bulk}$. This new equation $\delta S_{\rm met}^{\rm (ovlp)} /
\delta g_{\mu \nu} |_{\rm repl} = 0$ is called the
``replaced-equation" (cf.~Eqs.~(\ref{OQ-eq-repl}) and
(\ref{U-eqs-repl})).

Because solving Eq.~(\ref{sol-subst}) is the same as solving the
rifle PDE $\partial_\mu f^A \partial_\nu f^B \eta_{AB}^{\rm bulk}
= g_{\mu \nu}^{\rm sol}$ of the two-step AAT method, the solution
set of Eq.~(\ref{sol-subst}) 
\beq \label{sol-ovlp} \Sigma_{\rm ovlp}^{\rm (sol)} \,
\stackrel{\rm def}{=} \, \{ \, f^A \,| ~ \delta S_{\rm met}^{\rm
(ovlp)} / \delta g_{\mu \nu} |_{\rm repl} = 0 \, \} \eeq 
is equal to the bullet set $\Sigma_{\rm bullet}^{\rm (ovlp)}$ of
the overlapping action $S_{\rm met}^{\rm (ovlp)}[g_{\mu \nu}]$,
namely, 
\beq \label{sol-ovlp=bullet} \Sigma_{\rm ovlp}^{\rm (sol)} ~ = ~
\Sigma_{\rm bullet}^{\rm (ovlp)} ~. \eeq 

In Eq.~(\ref{sol-subst}), the replacement $|_{\rm repl}$ was
applied after the functional derivative $\delta / \delta g_{\mu
\nu}$. Here, we apply the replacement $|_{\rm repl}$ before the
derivative $\delta / \delta g_{\mu \nu}$. This produces a new {\it
functional} of the embedding $f^A$ 
\beq \label{new-action} \widetilde{S}_{\rm met}^{\rm (ovlp)}[f^A]
~ \stackrel{\rm def}{=} ~ S_{\rm met}^{\rm (ovlp)}[g_{\mu \nu}]
|_{\rm repl} ~ = ~ S_{\rm met}^{\rm (ovlp)}[\partial_\mu f^A
\partial_\nu f^B \eta_{AB}^{\rm bulk}] ~, \eeq 
implying the Lagrangian $\widetilde{{\cal L}}_{\rm met}^{\rm
(ovlp)}$ of this new functional $\widetilde{S}_{\rm met}^{\rm
(ovlp)}[f^A]$ satisfies (cf.~Eq.~(\ref{final-eff-S})) 
\beq \label{new-Lag} \widetilde{{\cal L}}_{\rm met}^{\rm
(ovlp)}(\partial_\mu f^A, \ldots \,) ~ \stackrel{\rm def}{=} ~
{\cal L}_{\rm met}^{\rm (ovlp)} (g_{\mu \nu}, \ldots \,) |_{\rm
repl} ~. \eeq 
Of course, it is possible that the original action $S_{\rm
emb}^{\rm (3 br)}[f^A]$ in Eq.~(\ref{final-eff-S}) takes the form
of the new functional $\widetilde{S}_{\rm met}^{\rm (ovlp)}[f^A] =
S_{\rm met}^{\rm (ovlp)}[\partial_\mu f^A \partial_\nu f^B
\eta_{AB}^{\rm bulk}]$ (see Eq.~(\ref{S-emb=eq})).

After the replacement in Eq.~(\ref{new-action}) 
\beq \label{g=ind} g_{\mu \nu} \, = \, \partial_\mu f^A
\partial_\nu f^B \eta_{AB}^{\rm bulk} ~, \eeq 
the functional derivative $\delta / \delta g_{\mu \nu}$ in
Eq.~(\ref{sol-subst}) is replaced with $\delta / \delta
(\partial_\mu f^A \partial_\nu f^B \eta_{AB}^{\rm bulk})$. Thus,
$\frac{\delta S_{\rm met}^{\rm (ovlp)}}{\delta g_{\mu \nu}} |_{\rm
repl} = 0$ in Eq.~(\ref{sol-subst}) is expressed as 
\beq \label{var-new-action} \frac{\delta \widetilde{S}_{\rm
met}^{\rm (ovlp)}}{\delta (\partial_\mu f^A \partial_\nu f^B
\eta_{AB}^{\rm bulk})} \, [f^A] \, = \, 0 ~, \eeq 
which is different from the {\it usual} variational equation 
\beq \label{usual-var} \frac{\delta \widetilde{S}_{\rm met}^{\rm
(ovlp)}}{\delta f^A} \, [f^A] \, = \, 0 ~. \eeq 

Since $\frac{\delta \widetilde{S}_{\rm met}^{\rm (ovlp)}}{\delta
(\partial_\mu f^A \partial_\nu f^B \eta_{AB}^{\rm bulk})} = 0$ in
Eq.~(\ref{var-new-action}) is the same as $\frac{\delta S_{\rm
met}^{\rm (ovlp)}}{\delta g_{\mu \nu}} |_{\rm repl} = 0$ in
Eq.~(\ref{sol-subst}), the solution set $\widetilde{\Sigma}_{\rm
ovlp}^{\rm (sol)}$ of the former equation 
\beq \label{sol-ovlp2} \widetilde{\Sigma}_{\rm ovlp}^{\rm (sol)}
\, \stackrel{\rm def}{=} \, \{ \, f^A \,| ~ \delta
\widetilde{S}_{\rm met}^{\rm (ovlp)} / \delta (\partial_\mu f^A
\partial_\nu f^B \eta_{AB}^{\rm bulk}) = 0 \, \} \eeq 
satisfies 
\beq \label{sol-ovlp2=1} ~~~~~ \widetilde{\Sigma}_{\rm ovlp}^{\rm
(sol)} \, = \, \Sigma_{\rm ovlp}^{\rm (sol)} \, = \, \Sigma_{\rm
bullet}^{\rm (ovlp)} ~~~ ({\rm see ~ Eq.}~(\ref{sol-ovlp=bullet}))
~. \eeq 

Due to the equality $\widetilde{\Sigma}_{\rm ovlp}^{\rm (sol)} =
\Sigma_{\rm bullet}^{\rm (ovlp)}$ in Eq.~(\ref{sol-ovlp2=1}), the
bullet $\Sigma_{\rm bullet}^{\rm (ovlp)}$ is also produced by the
odd-looking equation $\delta \widetilde{S}_{\rm met}^{\rm (ovlp)}
/ \delta (\partial_\mu f^A \partial_\nu f^B \eta_{AB}^{\rm bulk})
= 0$ in Eqs.~(\ref{var-new-action}) and (\ref{sol-ovlp2}).
Therefore, this equation $\delta \widetilde{S}_{\rm met}^{\rm
(ovlp)} / \delta (\partial_\mu f^A \partial_\nu f^B \eta_{AB}^{\rm
bulk}) = 0$ can replace the equation of motion $\delta S_{\rm
emb}^{\rm (3 br)} / \delta f^A = 0$ {\it within the overlapping
B-T overlap $\Sigma_{{\rm B} \cap {\rm T}}^{\rm (ovlp)} =
\Sigma_{\rm bullet}^{\rm (ovlp)} \cap \Sigma_{\rm target}$} ($\ne
\varnothing$){\,---\,}the ``equivalence" between these two
equations within $\Sigma_{{\rm B} \cap {\rm T}}^{\rm (ovlp)}$.
This conclusion becomes more evident, when we compare the bullet
$\Sigma_{\rm bullet}^{\rm (ovlp)} = \widetilde{\Sigma}_{\rm
ovlp}^{\rm (sol)} = \{ f^A | \, \delta \widetilde{S}_{\rm
met}^{\rm (ovlp)} / \delta (\partial_\mu f^A \partial_\nu f^B
\eta_{AB}^{\rm bulk}) = 0 \, \}$ with the target $\Sigma_{\rm
target} = \{ f^A | \, \delta S_{\rm emb}^{\rm (3 br)} / \delta f^A
= 0 \, \}$ in Eq.~(\ref{target-set}).

In the above conclusion, we should be careful in interpreting the
odd-looking equation $\delta \widetilde{S}_{\rm met}^{\rm (ovlp)}
/ \delta (\partial_\mu f^A \partial_\nu f^B \eta_{AB}^{\rm bulk})
= 0$: this odd-looking equation must not be interpreted as the
{\it equation of motion} for the space 3-brane, because (i) the
space 3-brane already has its own equation of motion $\delta
S_{\rm emb}^{\rm (3 br)} / \delta f^A = 0$, and (ii) the solution
set $\widetilde{\Sigma}_{\rm ovlp}^{\rm (sol)}$ ($= \Sigma_{\rm
bullet}^{\rm (ovlp)}$) in Eq.~(\ref{sol-ovlp2}) may not be equal
to the target $\Sigma_{\rm target}$ (see Eq.~(\ref{eq<emb})).
Then, the odd-looking equation $\delta \widetilde{S}_{\rm
met}^{\rm (ovlp)} / \delta (\partial_\mu f^A \partial_\nu f^B
\eta_{AB}^{\rm bulk}) = 0$ may be interpreted, at best, as the
constitutive equation specific to the space 3-brane (see above).
Despite this, if we try to know the target $\Sigma_{\rm target}$
by using the new functional $\widetilde{S}_{\rm met}^{\rm
(ovlp)}[f^A]$, the odd-looking equation $\delta \widetilde{S}_{\rm
met}^{\rm (ovlp)} / \delta (\partial_\mu f^A \partial_\nu f^B
\eta_{AB}^{\rm bulk}) = 0$ should be used rather than the usual
variational equation $\delta \widetilde{S}_{\rm met}^{\rm (ovlp)}
/ \delta f^A = 0$.

However, in General Relativity, it is well known that $(\delta
S_{\rm EH} / \delta {\bf g}_{\mu \nu}) |_{\, {\bf g}_{\mu \nu} =
\, \partial_\mu \xi^{\hat{\alpha}} \partial_\nu \xi^{\hat{\beta}}
\eta_{\hat{\alpha} \hat{\beta}}} = 0$ {\it if and only if} $\delta
\widetilde{S}_{\rm EH} / \delta (\partial_\mu \xi^{\hat{\alpha}})
= 0$, where $\partial_\mu \xi^{\hat{\alpha}}$ is the vierbein
satisfying ${\bf g}_{\mu \nu} = \partial_\mu \xi^{\hat{\alpha}}
\partial_\nu \xi^{\hat{\beta}} \eta_{\hat{\alpha} \hat{\beta}}$.
(For the mathematical proof, see Ref.~\cite{GR-Wein}.) From the
similarities of $f^A$ to $\xi^{\hat{\alpha}}$ (see above), we
easily confirm the equivalence that 
\beq \label{equiv-var} \frac{\delta \widetilde{S}_{\rm met}^{\rm
(ovlp)}}{\delta (\partial_\mu f^A \partial_\nu f^B \eta_{AB}^{\rm
bulk})} \, [f^A] \, = \, 0 ~~~~ {\it if ~ and ~ only ~ if} ~~~~
\frac{\delta \widetilde{S}_{\rm met}^{\rm (ovlp)}}{\delta
(\partial_\mu f^A)} \, [f^A] \, = \, 0 ~. \eeq 

For the special case of 
\beq \label{S-emb=eq} ~~~~~~~~~~~~~~~~ S_{\rm emb}^{\rm (3
br)}[f^A] \, = \, \widetilde{S}_{\rm met}^{\rm (ovlp)}[f^A] \, =
\, S_{\rm met}^{\rm (ovlp)}[\partial_\mu f^A \partial_\nu f^B
\eta_{AB}^{\rm bulk}] ~~~~~ ({\rm see ~ Eq.} ~ (\ref{new-action}))
~, \eeq 
the usual variational equation $\delta \widetilde{S}_{\rm
met}^{\rm (ovlp)} / \delta f^A = 0$ in Eq.~(\ref{usual-var})
becomes the equation of motion for the space 3-brane. According to
\bea \label{var-rel} \delta \widetilde{S}_{\rm met}^{\rm (ovlp)} /
\delta f^A ~ &=& ~ \left[ \delta \widetilde{S}_{\rm met}^{\rm
(ovlp)} / \delta (\partial_\mu f^C \partial_\nu f^D \eta_{CD}^{\rm
bulk}) \right] \times \left[ \delta (\partial_\mu f^M \partial_\nu
f^N \eta_{MN}^{\rm bulk}) / \delta f^A \right] \\ ~ &\equiv& ~ -
\, 2 \, \partial_\mu \left\{ \left[ \delta \widetilde{S}_{\rm
met}^{\rm (ovlp)} / \delta (\partial_\mu f^C \partial_\nu f^D
\eta_{CD}^{\rm bulk}) \right] \partial_\nu f^M \eta_{AM}^{\rm
bulk} \right\} \, , \eea 
the odd-looking equation $\delta \widetilde{S}_{\rm met}^{\rm
(ovlp)} / \delta (\partial_\mu f^A \partial_\nu f^B \eta_{AB}^{\rm
bulk}) = 0$ is not a necessary but {\it sufficient} condition for
the equation of motion $\delta \widetilde{S}_{\rm met}^{\rm
(ovlp)} / \delta f^A = 0$.

This means 
\beq \label{eq<emb} ~~~~~~~ \widetilde{\Sigma}_{\rm ovlp}^{\rm
(sol) \cdot sp} ~ \subsetneqq ~ \Sigma_{\rm target}^{\rm sp} ~~~ (
{\rm thus} ~~ \Sigma_{{\rm B} \cap {\rm T}}^{\rm (ovlp) \cdot sp}
\, \subsetneqq \, \Sigma_{\rm target}^{\rm sp} ) ~, \eeq 
where the superscript ``sp" denotes the special case $S_{\rm
emb}^{\rm (3 br)}[f^A] = \widetilde{S}_{\rm met}^{\rm
(ovlp)}[f^A]$, and $\Sigma_{\rm target}^{\rm sp} \stackrel{\rm
def}{=} \{ f_{\rm sol}^A | \, (\delta \widetilde{S}_{\rm met}^{\rm
(ovlp)} / \delta f^A)[f_{\rm sol}^A] = 0 \, \}$ is the target set
for this special case. The inequality $\widetilde{\Sigma}_{\rm
ovlp}^{\rm (sol) \cdot sp} \subsetneqq \Sigma_{\rm target}^{\rm
sp}$ in Eq.~(\ref{eq<emb}) supports the above statement that
$\delta \widetilde{S}_{\rm met}^{\rm (ovlp)} / \delta
(\partial_\mu f^A \partial_\nu f^B \eta_{AB}^{\rm bulk}) = 0$ must
not be interpreted as the equation of motion for the space
3-brane.

Due to $\widetilde{\Sigma}_{\rm ovlp}^{\rm (sol) \cdot sp}
\subsetneqq \Sigma_{\rm target}^{\rm sp}$, the special case
$S_{\rm emb}^{\rm (3 br)}[f^A] = \widetilde{S}_{\rm met}^{\rm
(ovlp)}[f^A]$ in Eq.~(\ref{S-emb=eq}) does not have the ``defect"
of 
\beq \label{defect} \widetilde{\Sigma}_{\rm ovlp}^{\rm (sol)} - \,
\Sigma_{\rm target} ~ \ne ~ \varnothing ~, \eeq 
which means that the bullet set $\widetilde{\Sigma}_{\rm
ovlp}^{\rm (sol)}$ in Eq.~(\ref{sol-ovlp2}) contains elements
outside the target set $\Sigma_{\rm target} = \{ f_{\rm sol}^A
\}$. This suggests defining the ``contained metric action" $S_{\rm
met}^{\rm (cont)}[g_{\mu \nu}]$ as an overlapping action $S_{\rm
met}^{\rm (ovlp)}[g_{\mu \nu}]$ satisfying
$\widetilde{\Sigma}_{\rm ovlp}^{\rm (sol)} \subsetneqq \Sigma_{\rm
target}$ (i.e., without the defect of $\widetilde{\Sigma}_{\rm
ovlp}^{\rm (sol)} - \Sigma_{\rm target} \ne \varnothing$). The
overlapping action $S_{\rm met}^{\rm (ovlp)}[g_{\mu \nu}]$ of the
special case in Eq.~(\ref{S-emb=eq}) is an example of the
contained action.

As implied in Eq.~(\ref{defect}), for evaluating an ME action
$S_{\rm met}^{\rm (ME)}[g_{\mu \nu}]$, we may use the
``bullet-target (B-T) difference" 
\beq \label{BT-diff} \Delta_{\rm BT} \, \stackrel{\rm def}{=} \,
\Sigma_{\rm bullet} - \, \Sigma_{\rm target} \eeq 
together with the B-T overlap $\Sigma_{{\rm B} \cap {\rm T}} =
\Sigma_{\rm bullet} \cap \Sigma_{\rm target}$ in
Eq.~(\ref{MI-BT}). For example, a ``large" B-T overlap
$\Sigma_{{\rm B} \cap {\rm T}}$ and a ``small" B-T difference
$\Delta_{\rm BT}$ can result in a ``good" ME action $S_{\rm
met}^{\rm (ME)}[g_{\mu \nu}]$.

\section{\label{sec: form-met} The Symmetries and the Forms of
the Overlapping Metric Action in the AAT Method}

Now, in terms of symmetries, we study the {\it forms} of the
overlapping metric action $S_{\rm met}^{\rm (ovlp)}[g_{\mu \nu}]$
(see Table \ref{sym-AAT}). The B-T overlap $\Sigma_{{\rm B} \cap
{\rm T}}^{\rm (ovlp)} = \Sigma_{\rm bullet}^{\rm (ovlp)} \cap
\Sigma_{\rm target}$ represents the maximum knowledge which we can
obtain about the target $\Sigma_{\rm target} = \{ f_{\rm sol}^A
\}$ of the original action $S_{\rm emb}^{\rm (3 br)}[f^A]$ by
using the chosen metric action $S_{\rm met}^{\rm (ovlp)}[g_{\mu
\nu}]$.

\begin{table}
\centering \caption{Symmetries in the Aim-At-Target (AAT) Method}
\vspace*{0.2cm}
\begin{tabular}{ccccc} \hline \vspace*{-0.35cm} \\ {\sc Action} & &
$ISO(1, D_{\rm amb}-1)$ & & ${\rm Diff}(4)$ \vspace*{0.05cm} \\
\hline \vspace*{-0.33cm} \\ $S_{\rm emb}^{\rm (3 br)}[f^A]$ & &
{\sf O} & & {\sf O} \vspace*{0.05cm} \\ \vspace*{-0.33cm} \\
$S_{\rm met}^{\rm (ovlp)} [g_{\mu \nu}]$ & & {\sf O} & & {\sf O}
(\,{\sf X}$^{\dagger \,}$) \vspace*{0.13cm} \\ \hline
\end{tabular} \label{sym-AAT} \\ \vspace*{0.2cm}
~~~~~~~~~~~~~~~~ {\small (\,{\sf O}\,: preserving\,,\, {\sf X}\,:
breaking\,)} \\ \vspace*{0.06cm} \begin{tabular}{l}
~~~~~~~~~~~~~~~~~~ {\small $^\dagger$ The ${\rm Diff}(4)$
invariance may be broken by an ME action
$S_{\rm met}^{\rm (ME)}[g_{\mu \nu}]$,} \\
~~~~~~~~~~~~~~~~~~~~ {\small as said in the text.} \\
\end{tabular}
\end{table}

First, we consider the symmetries of the target $\Sigma_{\rm
target} = \{ f_{\rm sol}^A \}$ of the original action $S_{\rm
emb}^{\rm (3 br)}[f^A]$, as follows: since the original action
$S_{\rm emb}^{\rm (3 br)}[f^A]$ is invariant under $ISO(1, D_{\rm
amb}-1)$ and ${\rm Diff}(4)$, the definition of the invariance of
this action $S_{\rm emb}^{\rm (3 br)}[f^A]$ implies that the
solution set $\Sigma_{\rm target} = \{ f_{\rm sol}^A \}$ of the
action $S_{\rm emb}^{\rm (3 br)}[f^A]$ is also invariant under
$ISO(1, D_{\rm amb}-1)$ and ${\rm Diff}(4)$ (see Sec.~\ref{sec:
3br-eff-action}). By definition, the induced-metric set
$\Sigma_{\rm ind} = \{ \gamma_{\mu \nu} \}$ in Eq.~(\ref{set-ind})
is invariant under $ISO(1, D_{\rm amb}-1)$ and ${\rm Diff}(4)$.
Note $\gamma_{\mu \nu}^{\, \prime} = \gamma_{\mu \nu}$ under
$ISO(1, D_{\rm amb}-1)$, and $\gamma_{\rho \sigma}^{\, \prime} =
\frac{\partial x^\mu}{\partial x^{\prime \rho}} \frac{\partial
x^\nu}{\partial x^{\prime \sigma}} \, \gamma_{\mu \nu}$ under
${\rm Diff}(4)$.

Next, we consider the symmetries of the bullet $\Sigma_{\rm
bullet}^{\rm (ovlp)} = \{ f_{\rm bul}^A \}$ of the overlapping
action $S_{\rm met}^{\rm (ovlp)}[g_{\mu \nu}]$, as follows: since
$g_{\mu \nu} = \partial_\mu f^A \partial_\nu f^B \eta_{AB}^{\rm
bulk}$ in Eq.~(\ref{g=ind}) is already invariant under $ISO(1,
D_{\rm amb}-1)$, the action $S_{\rm met}^{\rm (ovlp)}[g_{\mu
\nu}]$ has the $ISO(1, D_{\rm amb}-1)$ invariance. This implies
its solution-metric set $\Sigma_{\rm met}^{\rm (sol) \cdot ovlp} =
\{ g_{\mu \nu}^{\rm sol} \}$ also has the $ISO(1, D_{\rm amb}-1)$
invariance.

Then, since both $g_{\mu \nu}^{\rm sol}$ and $\partial_\mu f^A
\partial_\nu f^B \eta_{AB}^{\rm bulk}$ are invariant under $ISO(1,
D_{\rm amb}-1)$, $\Sigma_{\rm bullet}^{\rm (ovlp)} \owns f^{\prime
A} = \Lambda_B^A f^B + c^{\, A}$ is equivalent to $\Sigma_{\rm
bullet}^{\rm (ovlp)} \owns f^A$, which means the bullet
$\Sigma_{\rm bullet}^{\rm (ovlp)} = \{ f_{\rm bul}^A \}$ has the
$ISO(1, D_{\rm amb}-1)$ invariance like the target $\Sigma_{\rm
target} = \{ f_{\rm sol}^A \}$. Therefore, since the intersection
of two $\textsc{g}$-invariant sets is $\textsc{g}$-invariant
($\textsc{g}$: a group), the B-T overlap $\Sigma_{{\rm B} \cap
{\rm T}}^{\rm (ovlp)} = \Sigma_{\rm bullet}^{\rm (ovlp)} \cap
\Sigma_{\rm target}$ is invariant under $ISO(1, D_{\rm amb}-1)$.

Before studying the ${\rm Diff}(4)$ symmetry properties, we need
to re-consider the (i) mathematical and (ii) physical existences
of the overlapping action $S_{\rm met}^{\rm (ovlp)}[g_{\mu \nu}]$
(see the summary of the AAT method in Sec.~\ref{sec: AAT-method}):

First, we study an ME action $S_{\rm met}^{\rm (ME)}[g_{\mu
\nu}]$, which is a mere tool for knowing the target $\Sigma_{\rm
target}$. Since this mere tool $S_{\rm met}^{\rm (ME)}[g_{\mu
\nu}]$ cannot forbid any element $f_{\rm sol}^A$ of the target set
$\Sigma_{\rm target} = \{ f_{\rm sol}^A \}$, we can use a ${\rm
Diff}(4)$-breaking or a ${\rm Diff}(4)$-preserving action $S_{\rm
met}^{\rm (ME)}[g_{\mu \nu}]$ as long as this ME action $S_{\rm
met}^{\rm (ME)}[g_{\mu \nu}]$ provides a considerable information
about the target $\Sigma_{\rm target}$.

For example, we can use a ${\rm Diff}(4)$-breaking ME action
$S_{\rm met}^{\rm (ME)}[g_{\mu \nu}]$ as a mere tool for
$\Sigma_{\rm target}$. Since this action $S_{\rm met}^{\rm
(ME)}[g_{\mu \nu}]$ is not invariant under ${\rm Diff}(4)$ unlike
the original action $S_{\rm emb}^{\rm (3 br)}[f^A]$, the solution
set $\Sigma_{\rm met}^{\rm (sol) \cdot ME}$ of the metric action
$S_{\rm met}^{\rm (ME)}[g_{\mu \nu}]$ has an element $g_{\mu
\nu}^{\rm sol \natural}$ satisfying 
\beq \label{ME-sym-br} ~~~~~~~ g_{\mu \nu}^{\rm sol \natural} \,
\in \, \Sigma_{\rm met}^{\rm (sol) \cdot ME} ~~~ {\rm but} ~~~
g_{\rho \sigma}^{\rm sol \natural \, \prime} \, \not \in \,
\Sigma_{\rm met}^{\rm (sol) \cdot ME} ~~~~ {\rm for ~ an ~
element} ~ \Phi_{\rm 4D}^\natural ~ {\rm of} ~ {\rm Diff}(4) ~,
\eeq 
where $g_{\rho \sigma}^{\rm sol \natural \, \prime} =
\frac{\partial x^\mu}{\partial x^{\prime \rho}} \frac{\partial
x^\nu}{\partial x^{\prime \sigma}} \, g_{\mu \nu}^{\rm sol
\natural}$ with $x^\prime = \Phi_{\rm 4D}^\natural(x)$. This means
the solution-metric set $\Sigma_{\rm met}^{\rm (sol) \cdot ME} =
\{ g_{\mu \nu}^{\rm sol \cdot ME} \}$ breaks the ${\rm Diff}(4)$
invariance.

Despite this, if the ${\rm Diff}(4)$-breaking set $\Sigma_{\rm
met}^{\rm (sol) \cdot ME} = \{ g_{\mu \nu}^{\rm sol \cdot ME} \}$
contains a ``${\rm Diff}(4)$ gauge slice" $\Sigma_{\rm ind}^{\rm
(GS)}$ of the ${\rm Diff}(4)$-preserving set $\Sigma_{\rm ind} =
\{ \gamma_{\mu \nu} \}$, we can still know the target $\Sigma_{\rm
target} = \{ f_{\rm sol}^A \}$ by, for example, (i) finding a
solution $f_{\rm sol}^A$ ($\in \Sigma_{\rm target}$) of the rifle
PDE $\partial_\mu f^A \partial_\nu f^B \eta_{AB}^{\rm bulk} =
g_{\mu \nu}^{\rm sol \cdot ME} \in \Sigma_{\rm ind}^{\rm (GS)}$,
and (ii) applying ${\rm Diff}(4)$ to this solution $f_{\rm
sol}^A$, which forms its ``${\rm Diff}(4)$ gauge orbit" $\langle
f_{\rm sol}^A \rangle_{\rm diff}$. This aspect is similarly found
in a gauge theory, where the gauge invariance is broken by adding
a gauge-fixing term.

Thus, the breaking of the ${\rm Diff}(4)$ invariance by the ME
action $S_{\rm met}^{\rm (ME)}[g_{\mu \nu}]$ may not be a serious
problem for knowing the target $\Sigma_{\rm target}$ (see the
symbol {\sf X} in Table \ref{sym-AAT}). Of course, we can use a
${\rm Diff}(4)$-preserving ME action $S_{\rm met}^{\rm
(ME)}[g_{\mu \nu}]$ as another mere tool for the target
$\Sigma_{\rm target}$.

Next, we study a PE action $S_{\rm met}^{\rm (PE)}[g_{\mu \nu}]$,
which has the physical existence unlike the ME action $S_{\rm
met}^{\rm (ME)}[g_{\mu \nu}]$. Then, since the PE action $S_{\rm
met}^{\rm (PE)}[g_{\mu \nu}]$ produces the constitutive equation,
only the element $f_{\rm ove}^A$ of the B-T overlap $\Sigma_{{\rm
B} \cap {\rm T}}^{\rm (PE)} = \{ f_{\rm ove}^A \}$ ($\subset
\Sigma_{\rm target}$) can be a motion of the space 3-brane in
$\mathbb{M}^{D_{\rm amb}}$, as said in Sec.~\ref{sec: AAT-method}.

The PE action $S_{\rm met}^{\rm (PE)}[g_{\mu \nu}]$ can determine
the ${\rm Diff}(4)$ symmetry property of the B-T overlap
$\Sigma_{{\rm B} \cap {\rm T}}^{\rm (PE)} = \Sigma_{\rm
bullet}^{\rm (PE)} \cap \Sigma_{\rm target}$ through its bullet
$\Sigma_{\rm bullet}^{\rm (PE)}$: for example, we consider a ${\rm
Diff}(4)$-breaking PE action $S_{\rm met}^{\rm (PE)}[g_{\mu
\nu}]$, whose solution-metric set $\Sigma_{\rm met}^{\rm (sol)
\cdot PE}$ has an element $g_{\mu \nu}^{\rm sol \sharp}$
satisfying 
\beq \label{PE-sym-br} ~~~~~~~ g_{\mu \nu}^{\rm sol \sharp} \, \in
\, \Sigma_{\rm met}^{\rm (sol) \cdot PE} ~~~ {\rm but} ~~~ g_{\rho
\sigma}^{\rm sol \sharp \, \prime} \, \not \in \, \Sigma_{\rm
met}^{\rm (sol) \cdot PE} ~~~~ {\rm for ~ an ~ element} ~
\Phi_{\rm 4D}^\sharp ~ {\rm of} ~ {\rm Diff}(4) ~, \eeq 
where $g_{\rho \sigma}^{\rm sol \sharp \, \prime} = \frac{\partial
x^\mu}{\partial x^{\prime \rho}} \frac{\partial x^\nu}{\partial
x^{\prime \sigma}} \, g_{\mu \nu}^{\rm sol \sharp}$ with $x^\prime
= \Phi_{\rm 4D}^\sharp(x)$. This means the breaking of the ${\rm
Diff}(4)$ invariance by the solution-metric set $\Sigma_{\rm
met}^{\rm (sol) \cdot PE} = \{ g_{\mu \nu}^{\rm sol \cdot PE} \}$.

Suppose that a 4DL solution $f_{\rm sol \sharp}^A$ ($\in
\Sigma_{\rm target}$) of the original action $S_{\rm emb}^{\rm (3
br)}[f^A]$ satisfies 
\beq \label{f*-in-B} ~~~~~~~ \partial_\mu f_{\rm sol \sharp}^A \,
\partial_\nu f_{\rm sol \sharp}^B \, \eta_{AB}^{\rm bulk} \, = \,
g_{\mu \nu}^{\rm sol \sharp} ~~ ({\rm i.e.,} ~ f_{\rm sol
\sharp}^A \in \Sigma_{\rm bullet}^{\rm (PE)}) ~, \eeq 
which means the induced metric $\gamma_{\mu \nu}^\sharp =
\partial_\mu f_{\rm sol \sharp}^A \partial_\nu f_{\rm sol \sharp}^B
\eta_{AB}^{\rm bulk}$ (cf.~Eq.~(\ref{ind-metric})) has the
equality 
\beq \label{gam*=g*} \gamma_{\mu \nu}^\sharp \, = \, g_{\mu
\nu}^{\rm sol \sharp} ~. \eeq 
Then, the transformed 4DL solution $f_{\rm sol \sharp}^{\prime
A}(x^\prime) = f_{\rm sol \sharp}^A(x)$ with $x^\prime = \Phi_{\rm
4D}^\sharp(x)$ satisfies 
\beq \label{f*-B2} \partial_\rho^{\, \prime} f_{\rm sol
\sharp}^{\prime A} \, \partial_\sigma^{\, \prime} f_{\rm sol
\sharp}^{\prime B} \, \eta_{AB}^{\rm bulk} \, = \, g_{\rho
\sigma}^{\rm sol \sharp \, \prime} ~, \eeq 
where $f_{\rm sol \sharp}^{\prime A}$ ($\in \Sigma_{\rm target}$)
is an element of the ${\rm Diff}(4)$ gauge orbit $\langle f_{\rm
sol \sharp}^A \rangle_{\rm diff}$.

Due to $g_{\rho \sigma}^{\rm sol \sharp \, \prime} \not \in
\Sigma_{\rm met}^{\rm (sol) \cdot PE}$ in Eq.~(\ref{PE-sym-br}),
the transformed 4DL solution $f_{\rm sol \sharp}^{\prime A}$ ($\in
\Sigma_{\rm target}$) does not belong to the bullet $\Sigma_{\rm
bullet}^{\rm (PE)}$ (i.e., $f_{\rm sol \sharp}^{\prime A} \not \in
\Sigma_{\rm bullet}^{\rm (PE)}$) unlike the original 4DL solution
$f_{\rm sol \sharp}^A$ in Eq.~(\ref{f*-in-B}). Thus, like the
bullet $\Sigma_{\rm bullet}^{\rm (PE)}$, the B-T overlap
$\Sigma_{{\rm B} \cap {\rm T}}^{\rm (PE)}$ describing the space
3-brane breaks the ${\rm Diff}(4)$ invariance, because 
\beq \label{br-ovlp} \Sigma_{{\rm B} \cap {\rm T}}^{\rm (PE)} \,
\owns \, f_{\rm sol \sharp}^A ~~~ {\rm but} ~~~ \Sigma_{{\rm B}
\cap {\rm T}}^{\rm (PE)} \, \not \owns \, f_{\rm sol
\sharp}^{\prime A} ~. \eeq 
To sum up, the ${\rm Diff}(4)$-breaking PE action $S_{\rm
met}^{\rm (PE)}[g_{\mu \nu}]$ may imply the ${\rm
Diff}(4)$-breaking B-T overlap $\Sigma_{{\rm B} \cap {\rm T}}^{\rm
(PE)}$.

However, the ${\rm Diff}(4)$-breaking B-T overlap $\Sigma_{{\rm B}
\cap {\rm T}}^{\rm (PE)}$ can cause a physical problem of being
contrary to the observed General Relativity: since only the
element $f_{\rm ove}^A$ of the B-T overlap $\Sigma_{{\rm B} \cap
{\rm T}}^{\rm (PE)} = \{ f_{\rm ove}^A \}$ can occur in the
ambient spacetime $\mathbb{M}^{D_{\rm amb}}$ as a motion of the
space 3-brane (see Sec.~\ref{sec: AAT-method}), the latter result
$\Sigma_{{\rm B} \cap {\rm T}}^{\rm (PE)} \not \owns f_{\rm sol
\sharp}^{\prime A}$ in Eq.~(\ref{br-ovlp}) forbids $\gamma_{\rho
\sigma}^{\sharp \, \prime} = \partial_\rho^{\, \prime} f_{\rm sol
\sharp}^{\prime A} \partial_\sigma^{\, \prime} f_{\rm sol
\sharp}^{\prime B} \eta_{AB}^{\rm bulk}$ to occur in
$\mathbb{M}^{D_{\rm amb}}$ {\it unlike} the former $\Sigma_{{\rm
B} \cap {\rm T}}^{\rm (PE)} \owns f_{\rm sol \sharp}^A$, which
allows $\gamma_{\mu \nu}^\sharp = \partial_\mu f_{\rm sol
\sharp}^A \partial_\nu f_{\rm sol \sharp}^B \eta_{AB}^{\rm bulk}$
to occur in $\mathbb{M}^{D_{\rm amb}}$. Thus, due to the
approximation ${\bf g}_{\mu \nu} \approx \gamma_{\mu \nu}$ in
Eq.~(\ref{GR-met=ind}), ${\bf g}_{\rho \sigma}^{\sharp \, \prime}$
($\approx \gamma_{\rho \sigma}^{\sharp \, \prime}$) cannot occur
in $\mathbb{M}^{D_{\rm amb}}$ unlike ${\bf g}_{\mu \nu}^\sharp$
($\approx \gamma_{\mu \nu}^\sharp$). This means that the primed GR
metric ${\bf g}_{\rho \sigma}^{\sharp \, \prime}$ cannot be a
solution of General Relativity unlike the unprimed one ${\bf
g}_{\mu \nu}^\sharp$. As a result, General Relativity should be a
${\rm Diff}(4)$-breaking theory, which is falsified by
observations.

Therefore, it is natural to use only a ${\rm Diff}(4)$-preserving
PE action $S_{\rm met}^{\rm (PE)}[g_{\mu \nu}]$ which produces the
${\rm Diff}(4)$-preserving B-T overlap $\Sigma_{{\rm B} \cap {\rm
T}}^{\rm (PE)}$. In addition, since it is not compulsory that the
ME action $S_{\rm met}^{\rm (ME)}[g_{\mu \nu}]$ breaks the ${\rm
Diff}(4)$ invariance, we can choose to use a ${\rm
Diff}(4)$-preserving ME action $S_{\rm met}^{\rm (ME)}[g_{\mu
\nu}]$. To sum up, we use only a ${\rm Diff}(4)$-invariant case of
the overlapping action $S_{\rm met}^{\rm (ovlp)}[g_{\mu \nu}]$,
irrespective of whether it is an ME or PE action.

For a ${\rm Diff}(4)$-invariant overlapping action $S_{\rm
met}^{\rm (ovlp)}[g_{\mu \nu}]$, due to the definition
$\Sigma_{{\rm B} \cap {\rm T}}^{\rm (ovlp)} \stackrel{\rm def}{=}
\Sigma_{\rm bullet}^{\rm (ovlp)} \cap \Sigma_{\rm target}$, every
element $f_{\rm ove}^A$ of the B-T overlap $\Sigma_{{\rm B} \cap
{\rm T}}^{\rm (ovlp)} = \{ f_{\rm ove}^A \}$ satisfies 
\beq \label{ove=bul=sol} ~~~~~~~~~ f_{\rm ove}^A \, = \, f_{\rm
bul}^A \, = \, f_{\rm sol}^A ~~~~~~ {\rm with} ~~ f_{\rm bul}^A \,
\in \, \Sigma_{\rm bullet}^{\rm (ovlp)} ~~ {\rm and} ~~ f_{\rm
sol}^A \, \in \, \Sigma_{\rm target} ~, \eeq 
which results in 
\beq \label{df2=g-sol=ind} \partial_\mu f_{\rm ove}^A \,
\partial_\nu f_{\rm ove}^B \, \eta_{AB}^{\rm bulk} \, = \, g_{\mu
\nu}^{\rm sol} \, = \, \gamma_{\mu \nu} ~. \eeq 

From Eq.~(\ref{df2=g-sol=ind}), we obtain the equality for the
solution metric 
\beq \label{met-sol=ind} ~~~~~~~~~~~~~~~~ g_{\mu \nu}^{\rm sol} \,
= \, \gamma_{\mu \nu} ~~~ {\rm for ~ every ~ element} ~ f_{\rm
ove}^A ~ {\rm of} ~ \Sigma_{{\rm B} \cap {\rm T}}^{\rm (ovlp)} ~~~
({\rm cf.~Eq.}~(\ref{GR-met=ind})) ~. \eeq 
This equality $g_{\mu \nu}^{\rm sol} = \gamma_{\mu \nu}$ within
the B-T overlap $\Sigma_{{\rm B} \cap {\rm T}}^{\rm (ovlp)}$ means
that, below the ``metric cutoff\," $\Lambda_{\rm met}$, the metric
$g_{\mu \nu}$ can describe the ``emergent field" $\gamma_{\mu
\nu}$ ($= \partial_\mu f_{\rm sol}^A \partial_\nu f_{\rm sol}^B
\eta_{AB}^{\rm bulk}$), which is derived from the locations ($\in
\mathbb{M}^{D_{\rm amb}}$) of space quanta occupying
$\mathbb{M}^{D_{\rm amb}}$.

Moreover, due to the equality $g_{\mu \nu}^{\rm sol} = \gamma_{\mu
\nu}$, the spacetime ${\cal S}_{\rm met}^{\rm \, 4D}$ having this
metric $g_{\mu \nu}^{\rm sol}$ is exactly the same as the world
volume ${\cal WV}_{\rm sq}$ of the space 3-brane, i.e., 
\beq \label{ST-met=WV} ~~~~~~~~~~~~~~~~~~ {\cal S}_{\rm met}^{\rm
\, 4D} \, = \, {\cal WV}_{\rm sq} ~~~ {\rm for ~ every ~ element}
~ f_{\rm ove}^A ~ {\rm of} ~ \Sigma_{{\rm B} \cap {\rm T}}^{\rm
(ovlp)} ~~~ ({\rm cf.~Eq.}~(\ref{S-GR=WV})) ~. \eeq 
This equality ${\cal S}_{\rm met}^{\rm \, 4D} = {\cal WV}_{\rm
sq}$ within the B-T overlap $\Sigma_{{\rm B} \cap {\rm T}}^{\rm
(ovlp)}$ means that, below the cutoff $\Lambda_{\rm met}$, the
spacetime ${\cal S}_{\rm met}^{\rm \, 4D}$ with the metric $g_{\mu
\nu}^{\rm sol}$ can describe the ``emergent spacetime" ${\cal
WV}_{\rm sq}$, which is formed by the world lines ${\cal WL}_{\rm
sq}$ ($\subset \mathbb{M}^{D_{\rm amb}}$) of many space quanta in
$\mathbb{M}^{D_{\rm amb}}$.

In sum, by Eqs.~(\ref{met-sol=ind}) and (\ref{ST-met=WV}), we have
the equality for the two spacetime manifolds 
\beq \label{equal-ST} ~~~~~~~~~ ( {\cal S}_{\rm met}^{\rm \, 4D},
\, g_{\mu \nu}^{\rm sol} ) \, = \, ( {\cal WV}_{\rm sq}, \,
\gamma_{\mu \nu} ) ~~~~~~ {\rm within} ~~ \Sigma_{{\rm B} \cap
{\rm T}}^{\rm (ovlp)} ~~ ({\rm below} ~ \Lambda_{\rm met}) ~. \eeq
Exact values for the spacetime measurements are provided by the
exact or true spacetime $({\cal WV}_{\rm sq}, \gamma_{\mu \nu})$,
which is the 4D emergent spacetime occupying the ambient spacetime
$\mathbb{M}^{D_{\rm amb}}$.

Now, for the ${\rm Diff}(4)$-preserving overlapping action $S_{\rm
met}^{\rm (ovlp)}[g_{\mu \nu}]$, we consider the form of its
Lagrangian ${\cal L}_{\rm met}^{\rm (ovlp)}(\Lambda_{\rm met};
g_{\mu \nu})$ more closely: as in usual effective theories, this
Lagrangian ${\cal L}_{\rm met}^{\rm (ovlp)}$ having its own UV
cutoff $\Lambda_{\rm met}$ (cf.~Eq.~(\ref{met-action})) can be
expressed as 
\beq \label{met-series} {\cal L}_{\rm met}^{\rm
(ovlp)}(\Lambda_{\rm met}; g_{\mu \nu}) ~ = ~ \sum \, c_k \,
\frac{{\cal O}_k }{\, \Lambda_{\rm met}^{d_k - 4}} ~, \eeq 
where the coefficient $c_k$ has no mass dimension, and the local
operator ${\cal O}_k$ of mass dimension $d_k$ consists of the
metric $g_{\mu \nu}$ and its derivatives. To make ${\cal L}_{\rm
met}^{\rm (ovlp)}(\Lambda_{\rm met}; g_{\mu \nu})$ invariant under
${\rm Diff}(4)$, we assume every operator ${\cal O}_k$ is
invariant under ${\rm Diff}(4)$. Since the ${\rm Diff}(4)$
invariance of the overlapping action $S_{\rm met}^{\rm
(ovlp)}[g_{\mu \nu}]$ is shared by General Relativity, we easily
expect this metric action $S_{\rm met}^{\rm (ovlp)}[g_{\mu \nu}]$
to contain the Einstein-Hilbert action (see
Eq.~(\ref{gen-met-Lag})).

Generally speaking, since the metric Lagrangian ${\cal L}_{\rm
met}^{\rm (ovlp)}$ having the derivatives of $g_{\mu \nu}$ can
contain at least one dimensionful parameter (say, $\xi_{\rm met}$)
to maintain its mass dimension $[\, {\cal L}_{\rm met}^{\rm
(ovlp)} \,] = 4$, the Lagrangian ${\cal L}_{\rm met}^{\rm (ovlp)}$
becomes the function of the parameter $\xi_{\rm met}$, which has a
Laurent series for $\xi_{\rm met}$. Thus, this Laurent series with
$\xi_{\rm met} = \Lambda_{\rm met}$ can lead to the series like
Eq.~(\ref{met-series}), even when the effective-theory nature of
${\cal L}_{\rm met}^{\rm (ovlp)}(\Lambda_{\rm met}; g_{\mu \nu})$
is not considered.

If we (i) observe at an energy $E_{\rm obs}$ ($\lesssim
\Lambda_{\rm met}$), and (ii) neglect all the operators with $d_k
\ge d_{\, \rm negl}$, then the error $\varepsilon_{\rm negl}$ has
a size of $O(E_{\rm obs}/\Lambda_{\rm met})^{d_{\rm negl} - 4}$,
implying 
\beq \label{ind-d-negl} d_{\, \rm negl} ~ \approx ~ 4 \, + \,
\frac{\log \varepsilon_{\rm negl}}{\, \log ( E_{\rm
obs}/\Lambda_{\rm met} )} ~. \eeq 
This leads to the approximate predictive power that a computation
with the error $\varepsilon_{\rm negl}$ requires only a finite
number of operators ${\cal O}_k$ up to the maximally allowed mass
dimension $d_{\rm max}$ ($< d_{\, \rm negl}$).

When the operator ${\cal O}_k$ in Eq.~(\ref{met-series}) contains
$N_\partial$ derivatives $\partial_\alpha$ and $N_g$ metrics
$g_{\mu \nu}$, the ${\rm Diff}(4)$ invariance requires the
operator ${\cal O}_k$ to possess $\frac{1}{2} N_\partial + N_g$
inverse metrics $g^{\mu \nu}$ for contraction. The mass dimension
of ${\cal O}_k$ satisfies 
\beq \label{ind-d-k} ~~~~~~~~ d_k \, = \, [\, {\cal O}_k] \, = \,
[\, (\hat{g}^{-1})^{ \frac{1}{2} N_\partial + N_g} \times
\partial^{ N_\partial} \times \hat{g}^{N_g} \,] \, = \, [ \,
(\hat{g} \, dX dX)^{- \frac{1}{2} N_\partial } \, ] \, = \,
N_\partial ~, \eeq 
where the four symbols have correspondences $\hat{g}^{-1}
\leftrightarrow g^{\mu \nu}$, $\partial \leftrightarrow
\partial_\mu$, $\hat{g} \leftrightarrow g_{\mu \nu}$ and
$dX \leftrightarrow dx^\mu$.

Since the number $\frac{1}{2} N_\partial + N_g$ of inverse metrics
$g^{\mu \nu}$ should be an integer ($\ge 0$), 
\beq \label{even-N} N_\partial \, = \, 2 \times ({\rm integer}) ~,
\eeq 
implying $d_k$ is an {\it even} integer due to $d_k = N_\partial$
in Eq.~(\ref{ind-d-k}). Thus, the ``overlapping Lagrangian" ${\cal
L}_{\rm met}^{\rm (ovlp)}(\Lambda_{\rm met}; g_{\mu \nu})$ in
Eq.~(\ref{met-series}) has the derivative expansion 
\beq \label{ind-deriv-exp} {\cal L}_{\rm met}^{\rm
(ovlp)}(\Lambda_{\rm met}; g_{\mu \nu}) ~ = \, \sum_{d_k \, : \,
{\rm even}} \, c_{\, d_k} \, \Lambda_{\rm met}^4 \, O \left(
\frac{\partial}{\, \Lambda_{\rm met}} \right)^{d_k} \, , \eeq 
where $d_k$ are non-negative even integers.

The overlapping Lagrangian ${\cal L}_{\rm met}^{\rm
(ovlp)}(\Lambda_{\rm met}; g_{\mu \nu})$ in
Eq.~(\ref{ind-deriv-exp}) can have the form of 
\beq \label{gen-met-Lag} ~~~~~~ {\cal L}_{\rm met}^{\rm (ovlp)} \,
= \, c_{\, 0 \,} \Lambda_{\rm met}^4 + c_{\, 2 \,} \Lambda_{\rm
met}^2 R + c_{\, 4}^{(1)} R^{\, 2} + c_{\, 4}^{(2)} R_{\mu \nu}
R^{\, \mu \nu} + c_{\, 4}^{(3)} g^{\mu \nu} \nabla_\mu R \,
\nabla_\nu R + \cdots , \eeq 
where all the coefficients (e.g., $c_{\, 0}$, $c_{\, 2}$) are
dimensionless, and both of the covariant derivative $\nabla_\mu$
and the curvature quantities (e.g., $R$) are built from the metric
$g_{\mu \nu}$ (see Ref.~\cite{GR-Carr}).

\section{\label{sec: EFT-univ} The Effective Theory for the
Universe: the Inclusion of Matter}

According to observations, our universe contains various particles
(e.g., leptons) which are different in kind from space quanta. To
distinguish those particles from the space quanta, we coin a new
term {\bf occupant quantum} (OQ) denoting any particle which (i)
differs from space quanta, and (ii) occupies the space 3-brane
without departing from it (i.e., the confinement of the occupant
quantum to the space 3-brane).

To sum up, our universe can be regarded as a {\it composite
system} which consists of space quanta and occupant quanta, moving
within the ambient spacetime $\mathbb{M}^{D_{\rm amb}}$.

Since space quantum is more fundamental than graviton, there can
be a scenario that every particle of the Standard Model (SM) is a
bound state of occupant quanta. However, there can be another
scenario that each SM particle is identified with a single
occupant quantum. Besides these, there can be various other
scenarios.

Despite this, from now on, we will consider only the {\it
low-energy} spectrum (e.g., the SM particles) of occupant quanta
which can be observed at low enough energies: since each of these
observable occupant quanta is confined to the world volume ${\cal
WV}_{\rm sq}$ of the space 3-brane, it is described by a function
$\Psi_{\rm OQ}$ whose domain is the world volume ${\cal WV}_{\rm
sq}$. For a brane-chart $x^\mu$ of ${\cal WV}_{\rm sq}$, the
``brane-field" $\Psi_{\rm OQ}$ on ${\cal WV}_{\rm sq}$ is
represented as the function $\Psi_{\rm OQ}(x^\mu)$ of the four
coordinates $x^\mu$.

The {\it value} $\Psi_{\rm OQ}(x^\mu(p))$ at a point $p \in {\cal
WV}_{\rm sq}$ is either (i) a ``brane-tensor of a type" (e.g., a
scalar) of ${\cal WV}_{\rm sq}$, or (ii) a ``brane-spinor" (e.g.,
a Weyl spinor) of the ``brane Lorentz group" $SO(1,3)$ at the
point $p$. The vierbein $e_\mu^a$ satisfying $e_\mu^a e_\nu^b
\eta_{ab} = \gamma_{\mu \nu}$ can be used in the action for
brane-spinors. Suppose that the bosons and fermions of the
Standard Model are described by their corresponding brane-fields
$\Psi_{\rm OQ}^{\rm (SM)}$. Like the induced metric $\gamma_{\mu
\nu}(x)$, all the SM brane-fields $\Psi_{\rm OQ}^{\rm (SM)}(x)$
are invariant (i.e., ``bulk-scalars") under every
B$\Rightarrow$${\rm B}^\prime$ transformation $Y^A \rightarrow
Y^{\prime A} \in ISO(1, D_{\rm amb} - 1)$ between the bulk-charts
$Y^A$ and $Y^{\prime A}$ of $\mathbb{M}^{D_{\rm amb}}$.

The action $S_{\rm OQ}^{\rm (3 br)}$ for the observable occupant
quanta $\Psi_{\rm OQ}(x)$ can be expressed as 
\beq \label{br-OQ-act} S_{\rm OQ}^{\rm (3 br)}[\Psi_{\rm OQ}, f^A]
~ \stackrel{\rm def}{=} \, \int_{{\cal WV}_{\rm sq}} d^{\, 4} x
\sqrt{|\, {\rm det}(\gamma_{\mu \nu})|} ~ {\cal L}_{\rm OQ}^{\rm
(3 br)} (\Psi_{\rm OQ}, \partial_\mu f^A, \ldots \,) ~, \eeq 
where $\gamma_{\mu \nu} = \partial_\mu f^A \partial_\nu f^B
\eta_{AB}^{\rm bulk}$. This action $S_{\rm OQ}^{\rm (3
br)}[\Psi_{\rm OQ}, f^A]$ is assumed to be invariant under $ISO(1,
D_{\rm amb} - 1)$ and ${\rm Diff}(4)$ like the 3-brane action
$S_{\rm emb}^{\rm (3 br)}[f^A]$. Of course, the action $S_{\rm
OQ}^{\rm (3 br)}$ in Eq.~(\ref{br-OQ-act}) may depend on a
``bulk-field" $\Psi_{\rm bulk}(Y^A)$ of the bulk spacetime
$\mathbb{M}^{D_{\rm amb}}$, whose field point $Y^A$ should satisfy
$Y^A = f^A(x^\mu)$. For example, when $\Psi_{\rm bulk}(Y^A)$ is a
bulk-tensor (e.g., a $D_{\rm amb}$-dimensional vector), it can
appear in the action $S_{\rm OQ}^{\rm (3 br)}$ through its
pullback $(f^*\Psi_{\rm bulk})(x^\mu)$ at sufficiently low
energies.

Finally, the {\bf ``original" universe action} $S_{\rm univ}^{\rm
(3 br)}[f^A, \Psi_{\rm OQ}]$ at low energies is written as 
\beq \label{br-univ-act} S_{\rm univ}^{\rm (3 br)}[f^A, \Psi_{\rm
OQ}] ~ = ~ S_{\rm emb}^{\rm (3 br)}[f^A] ~ + ~ S_{\rm OQ}^{\rm (3
br)}[\Psi_{\rm OQ}, f^A] ~, \eeq 
where the integral for the 3-brane action $S_{\rm emb}^{\rm (3
br)}[f^A]$ shares the same set ${\cal WV}_{\rm sq}$ with that for
$S_{\rm OQ}^{\rm (3 br)}[\Psi_{\rm OQ}, f^A]$ in
Eq.~(\ref{br-OQ-act}).

For the original action $S_{\rm univ}^{\rm (3 br)}[f^A, \Psi_{\rm
OQ}]$ in Eq.~(\ref{br-univ-act}), its {\bf universe target}
$\Sigma_{\rm target}^{\rm (univ)} = \{ f_{\rm sol}^A \}$ is
defined as 
\beq \label{U-target} ~~~~ \Sigma_{\rm target}^{\rm (univ)} \,
\stackrel{\rm def}{=} \, \{\, f_{\rm sol}^A : {\rm 4DL ~
embedding} ~ | ~\, (\delta S_{\rm univ}^{\rm (3 br)} / \delta
f^A)[f_{\rm sol}^A, \Psi_{\rm OQ}^{\rm sol}] = 0 ~ \} ~~~ \subset
~ {\cal F}_{\rm space} ~, \eeq 
where ($f_{\rm sol}^A, \Psi_{\rm OQ}^{\rm sol}$) is a solution of
the coupled Euler-Lagrange (E-L) equations 
\beq \label{br-eqs} (\delta S_{\rm univ}^{\rm (3 br)} / \delta
f^A)[f^A, \Psi_{\rm OQ}] \, = \, 0 ~~~~ {\rm and} ~~~ (\delta
S_{\rm univ}^{\rm (3 br)} / \delta \Psi_{\rm OQ})[f^A, \Psi_{\rm
OQ}] \, = \, 0 ~. \eeq 
Although the element $f_{\rm sol}^A$ of the set $\Sigma_{\rm
target}^{\rm (univ)} = \{ f_{\rm sol}^A \}$ satisfies the
different equation (i.e., $\delta S_{\rm univ}^{\rm (3 br)} /
\delta f^A = 0$) from $\delta S_{\rm emb}^{\rm (3 br)} / \delta
f^A = 0$ for the target $\Sigma_{\rm target}$ in
Eq.~(\ref{target-set}), the solution $f_{\rm sol}^A$ in
Eq.~(\ref{U-target}) is still called a ``4D-Lorentzian (4DL)
solution."

Since the universe target $\Sigma_{\rm target}^{\rm (univ)}$ in
Eq.~(\ref{U-target}) is defined similarly to the target
$\Sigma_{\rm target}$, we can similarly apply the AAT method in
order to know the universe target $\Sigma_{\rm target}^{\rm
(univ)} = \{ f_{\rm sol}^A \}$, as follows: as in Sec.~\ref{sec:
AAT-method}, the knowledge about the universe target $\Sigma_{\rm
target}^{\rm (univ)}$ is related to the {\bf universe
induced-metric set} 
\beq \label{U-set-ind} \Sigma_{\rm ind}^{\rm (univ)} \,
\stackrel{\rm def}{=} \, \{\, \gamma_{\mu \nu} \, | ~ \gamma_{\mu
\nu} = \partial_\mu f_{\rm sol}^A \, \partial_\nu f_{\rm sol}^B \,
\eta_{AB}^{\rm bulk} ~~ {\rm for ~ every} ~ f_{\rm sol}^A \in
\Sigma_{\rm target}^{\rm (univ)} ~ \} ~. \eeq 

To study this universe induced-metric set $\Sigma_{\rm ind}^{\rm
(univ)} = \{ \gamma_{\mu \nu} \}$ as in Sec.~\ref{sec:
AAT-method}, we impose three requirements on the {\bf
``overlapping" universe action} $S_{\rm univ}^{\rm (ovlp)} =
\int_{{\cal S}_{\rm univ}^{\rm 4D}} d^{\, 4} x  \, \widehat{{\cal
L}}_{\rm univ}^{\rm \, (ovlp)}$:
\begin{itemize}
\item For the study of $\Sigma_{\rm ind}^{\rm (univ)} = \{
\gamma_{\mu \nu} \}$, the overlapping action $S_{\rm univ}^{\rm
(ovlp)}$ is a functional of the 4D Lorentzian metric $g_{\mu \nu}$
on the 4D manifold ${\cal S}_{\rm univ}^{\rm \, 4D}$.
\item The spacetime ${\cal S}_{\rm univ}^{\rm \, 4D}$ for the
action $S_{\rm univ}^{\rm (ovlp)} = \int_{{\cal S}_{\rm univ}^{\rm
4D}} d^{\, 4} x \sqrt{|\, {\rm det}(g_{\mu \nu})|} \, {\cal
L}_{\rm univ}^{\rm (ovlp)}$ satisfies 
\beq \label{equal-U-ST} ~~~~~~~~~~~~ {\cal S}_{\rm univ}^{\rm \,
4D} \, = \, {\cal WV}_{\rm sq} ~~~~~~ ({\rm cf.~Eqs.} ~
(\ref{ST-met=WV}) ~ {\rm and} ~ (\ref{equal-ST-U})) ~. \eeq 
\item The solution $g_{\mu \nu}^{\rm sol \cdot U}$ (called the
``U-metric") of the equation $\delta S_{\rm univ}^{\rm (ovlp)} =
0$ (see Eq.~(\ref{U-cartridge})) satisfies 
\beq \label{equal-U-met} ~~~~~~~~~ g_{\mu \nu}^{\rm sol \cdot U}
\, = \, \gamma_{\mu \nu} ~~~~~~ ({\rm cf.~Eqs.} ~
(\ref{met-sol=ind}) ~ {\rm and} ~ (\ref{equal-ST-U})) ~. \eeq 
Thus, since this induced metric $\gamma_{\mu \nu}$ depends on the
observable occupant quanta through the 4DL solution $f_{\rm
sol}^A$ due to Eqs.~(\ref{U-target}) and (\ref{U-set-ind}), it is
natural to assume that the overlapping universe action $S_{\rm
univ}^{\rm (ovlp)}$ depends on these occupant quanta.
\end{itemize}

Therefore, we consider the overlapping universe action of the form
\beq \label{univ-act} S_{\rm univ}^{\rm (ovlp)}[g_{\mu \nu},
\psi_{\rm oq}] ~ \stackrel{\rm def}{=} ~ S_{\rm met}^{\rm
(ovlp)}[g_{\mu \nu}] ~ + ~ S_{\rm OQ}^{\rm (ovlp)}[\psi_{\rm oq},
g_{\mu \nu}] ~, \eeq 
where the ``occupant-quantum (OQ) action" 
\beq \label{OQ-act} S_{\rm OQ}^{\rm (ovlp)}[\psi_{\rm oq}, g_{\mu
\nu}] ~ \stackrel{\rm def}{=} \, \int_{{\cal S}_{\rm univ}^{\rm
4D}} d^{\, 4} x \sqrt{|\, {\rm det}(g_{\mu \nu})|} ~ {\cal L}_{\rm
OQ}^{\rm (ovlp)}(\psi_{\rm oq}, g_{\mu \nu}, \ldots \,) ~, \eeq 
and the metric action 
\beq \label{univ-met-act} S_{\rm met}^{\rm (ovlp)}[g_{\mu \nu}] ~
\stackrel{\rm def}{=} \, \int_{{\cal S}_{\rm univ}^{\rm 4D}} d^{\,
4} x \sqrt{|\, {\rm det}(g_{\mu \nu})|} ~ {\cal L}_{\rm met}^{\rm
(ovlp)}(\Lambda_{\rm met}; g_{\mu \nu}) ~. \eeq 
Because this metric action $S_{\rm met}^{\rm (ovlp)}[g_{\mu \nu}]$
defined for ${\cal S}_{\rm univ}^{\rm 4D}$ will be chosen to be
invariant under $ISO(1, D_{\rm amb} - 1)$ and ${\rm Diff}(4)$ (see
below), its Lagrangian ${\cal L}_{\rm met}^{\rm
(ovlp)}(\Lambda_{\rm met}; g_{\mu \nu})$ in
Eq.~(\ref{univ-met-act}) has the same form as the Lagrangian in
Eqs.~(\ref{ind-deriv-exp}) and (\ref{gen-met-Lag}){---\,}we use
the same notations.

Like $g_{\mu \nu}$ describing $\gamma_{\mu \nu}$ through $g_{\mu
\nu}^{\rm sol \cdot U} = \gamma_{\mu \nu}$, each
``occupant-quantum (OQ) field" $\psi_{\rm oq}$ in the overlapping
action $S_{\rm univ}^{\rm (ovlp)}[g_{\mu \nu}, \psi_{\rm oq}]$
describes its counterpart $\Psi_{\rm OQ}$ through 
\beq \label{equal-oq} ~~~~~~~~~~~~~~~~ \psi_{\rm oq}^{\rm sol} \,
= \, \Psi_{\rm OQ}^{\rm sol} ~~~~~~ ({\rm see ~ Eqs.} ~
(\ref{U-eqs}) ~ {\rm and} ~ (\ref{U-cartridge}) ) ~, \eeq 
where $\psi_{\rm oq}^{\rm sol}$ is a part of the solution ($g_{\mu
\nu}^{\rm sol \cdot U}, \psi_{\rm oq}^{\rm sol}$) of the coupled
E-L equations 
\beq \label{U-eqs} (\delta S_{\rm univ}^{\rm (ovlp)} / \delta
g_{\mu \nu})[g_{\mu \nu}, \psi_{\rm oq}] \, = \, 0 ~~~~ {\rm and}
~~~ (\delta S_{\rm univ}^{\rm (ovlp)} / \delta \psi_{\rm
oq})[g_{\mu \nu}, \psi_{\rm oq}] \, = \, 0 ~. \eeq 

When $\delta S_{\rm univ}^{\rm (ovlp)} / \delta \psi_{\rm oq} = 0$
in Eq.~(\ref{U-eqs}) is compared with $\delta S_{\rm univ}^{\rm (3
br)} / \delta \Psi_{\rm OQ} = 0$ in Eq.~(\ref{br-eqs}), we can
find a simple method for achieving the above equality $\psi_{\rm
oq}^{\rm sol} = \Psi_{\rm OQ}^{\rm sol}$ under the assumption
$g_{\mu \nu}^{\rm sol \cdot U} = \gamma_{\mu \nu}$ in
Eq.~(\ref{equal-U-met}), as follows: the original OQ action
$S_{\rm OQ}^{\rm (3 br)}[\Psi_{\rm OQ}, f^A]$ in
Eq.~(\ref{br-OQ-act}) can satisfy, at least at low enough
energies, 
\beq \label{OQ-act-repl} S_{\rm OQ}^{\rm (3 br)}[\Psi_{\rm OQ},
f^A] ~ = \, \widetilde{S}_{\rm OQ}^{\rm (ovlp)}[\Psi_{\rm OQ},
f^A] ~ \stackrel{\rm def}{=} \, S_{\rm OQ}^{\rm (ovlp)}[\Psi_{\rm
OQ}, \partial_\mu f^A \partial_\nu f^B \eta_{AB}^{\rm bulk}] ~,
\eeq 
which is obtained by the replacements (i) $\psi_{\rm oq}
\Rrightarrow \Psi_{\rm OQ}$ and (ii) $g_{\mu \nu} \Rrightarrow
\partial_\mu f^A \partial_\nu f^B \eta_{AB}^{\rm bulk}$ in the
overlapping OQ action $S_{\rm OQ}^{\rm (ovlp)}[\psi_{\rm oq},
g_{\mu \nu}]$ in Eq.~(\ref{OQ-act}).

Due to Eq.~(\ref{OQ-act-repl}), the solution $\Psi_{\rm OQ}^{\rm
sol}$ of $(\delta S_{\rm OQ}^{\rm (3 br)} / \delta \Psi_{\rm
OQ})[\Psi_{\rm OQ}, f_{\rm sol}^A] = 0$ from Eq.~(\ref{br-eqs}) is
also a solution of the replaced-equation from Eq.~(\ref{U-eqs}) 
\beq \label{OQ-eq-repl} \delta S_{\rm OQ}^{\rm (ovlp)} / \delta
\psi_{\rm oq} |_{\rm repl} \, \stackrel{\rm def}{=} \, (\delta
S_{\rm OQ}^{\rm (ovlp)} / \delta \psi_{\rm oq})[\psi_{\rm oq},
\partial_\mu f_{\rm sol}^A \partial_\nu f_{\rm sol}^B
\eta_{AB}^{\rm bulk}] \, = \, 0 ~, \eeq 
which contains $f_{\rm sol}^A$ unlike other replaced-equations in
Eqs.~(\ref{sol-subst}) and (\ref{U-eqs-repl}) due to the
assumption $g_{\mu \nu}^{\rm sol \cdot U} = \gamma_{\mu \nu}$. In
this manner, the equality $\psi_{\rm oq}^{\rm sol} = \Psi_{\rm
OQ}^{\rm sol}$ in Eq.~(\ref{equal-oq}) is achieved.

For this equality $\psi_{\rm oq}^{\rm sol} = \Psi_{\rm OQ}^{\rm
sol}$, the OQ field $\psi_{\rm oq}$ shares the same $ISO(1, D_{\rm
amb} - 1)$ and ${\rm Diff}(4)$ symmetry properties with its
corresponding brane-field $\Psi_{\rm OQ}$. For example, the OQ
field $\psi_{\rm oq}$ is a ${\rm Diff}(4)$-tensor or
$SO(1,3)$-spinor of the spacetime ${\cal S}_{\rm univ}^{\rm 4D}$
like its counterpart $\Psi_{\rm OQ}$. Of course, the equality
$\psi_{\rm oq}^{\rm sol} = \Psi_{\rm OQ}^{\rm sol}$ may have a
limited validity like $g_{\mu \nu}^{\rm sol} = \gamma_{\mu \nu}$
in Eq.~(\ref{met-sol=ind}), which is valid only for the B-T
overlap $\Sigma_{{\rm B} \cap {\rm T}}^{\rm (ovlp)}$ ($\subset
\Sigma_{\rm target}$).

Due to Eq.~(\ref{OQ-act-repl}), the original universe action
$S_{\rm univ}^{\rm (3 br)}[f^A, \Psi_{\rm OQ}]$ in
Eq.~(\ref{br-univ-act}) can satisfy 
\beq \label{new-br-univ-act} ~~~~~~ S_{\rm univ}^{\rm (3 br)}[f^A,
\Psi_{\rm OQ}] ~ = \, S_{\rm emb}^{\rm (3 br)}[f^A] \, + \,
\widetilde{S}_{\rm OQ}^{\rm (ovlp)}[\Psi_{\rm OQ}, f^A] ~~~~ {\rm
at ~ low ~ enough ~ energies} \, . \eeq 
Hamilton's principle $\delta S_{\rm univ}^{\rm (3 br)} / \delta
f^A = 0$ gives the equation of motion for the space 3-brane 
\beq \label{univ-f-eq} ~ \partial_\mu (\, {\cal T}_{\rm 3 br}
\sqrt{|\, {\rm det}(\gamma_{\rho \sigma})|} \, \gamma^{\mu \nu}
\partial_\nu f^B \eta_{AB}^{\rm bulk} \,) \, + \, \cdots ~ = ~
\partial_\mu ( \sqrt{|\, {\rm det}(\gamma_{\rho \sigma})|} ~
T_{\rm OQ}^{\mu \nu} \partial_\nu f^B \eta_{AB}^{\rm bulk} \,) ~,
\eeq 
where 
\beq \label{br-OQ-EM} T_{{\rm OQ} \, \mu \nu} ~ \stackrel{\rm
def}{=} \, - \, \frac{2}{\sqrt{|\, {\rm det}(\gamma_{\alpha
\beta})|} \,} \, \frac{\delta \widetilde{S}_{\rm OQ}^{\rm
(ovlp)}}{\delta \gamma^{\mu \nu}} ~. \eeq 
Since the OQ action $\widetilde{S}_{\rm OQ}^{\rm (ovlp)}[\Psi_{\rm
OQ}, f^A]$ is added to the 3-brane action $S_{\rm emb}^{\rm (3
br)}[f^A]$, the equation of motion in Eq.~(\ref{univ-f-eq}) is
changed from Eq.~(\ref{f-eq}).

As in Sec.~\ref{sec: AAT-method}, for the overlapping universe
action $S_{\rm univ}^{\rm (ovlp)}[g_{\mu \nu}, \psi_{\rm oq}]$ in
Eq.~(\ref{univ-act}), its {\bf universe cartridge} $\Sigma_{\rm
univ}^{\rm (sol)} = \{ g_{\mu \nu}^{\rm sol \cdot U} \}$ is
defined as 
\beq \label{U-cartridge} \Sigma_{\rm univ}^{\rm (sol)} \,
\stackrel{\rm def}{=} \, \{\, g_{\mu \nu}^{\rm sol \cdot U} \, | ~
(\delta S_{\rm univ}^{\rm (ovlp)} / \delta g_{\mu \nu})[g_{\mu
\nu}^{\rm sol \cdot U}, \psi_{\rm oq}^{\rm sol}] = 0 ~ \} ~, \eeq
where ($g_{\mu \nu}^{\rm sol \cdot U}, \psi_{\rm oq}^{\rm sol}$)
is the solution of the coupled E-L equations in Eq.~(\ref{U-eqs}).

Then, the {\bf universe bullet} $\Sigma_{\rm bullet}^{\rm (univ)}
= \{ f_{\rm bul}^A \}$ of the overlapping action $S_{\rm
univ}^{\rm (ovlp)}[g_{\mu \nu}, \psi_{\rm oq}]$ is defined as 
\beq \label{U-bullet} \Sigma_{\rm bullet}^{\rm (univ)} \,
\stackrel{\rm def}{=} \, \{\, f^A \,| ~\, \partial_\mu f^A
\partial_\nu f^B \eta_{AB}^{\rm bulk} = g_{\mu \nu}^{\rm sol
\cdot U} ~~ {\rm for ~ every} ~ g_{\mu \nu}^{\rm sol \cdot U} \in
\Sigma_{\rm univ}^{\rm (sol)} ~ \} ~~~ \subset ~ {\cal F}_{\rm
space} ~. \eeq 
Like the universe target $\Sigma_{\rm target}^{\rm (univ)} = \{
f_{\rm sol}^A \}$ in Eq.~(\ref{U-target}), the universe bullet
$\Sigma_{\rm bullet}^{\rm (univ)} = \{ f_{\rm bul}^A \}$ depends
on the occupant quanta through the U-metric $g_{\mu \nu}^{\rm sol
\cdot U}$ in Eq.~(\ref{U-bullet}), because this solution metric
$g_{\mu \nu}^{\rm sol \cdot U}$ depends on the occupant quanta
$\psi_{\rm oq}$ through, e.g., the $\psi_{\rm oq}$-dependent
equation $(\delta S_{\rm univ}^{\rm (ovlp)} / \delta g_{\mu
\nu})[g_{\mu \nu}, \psi_{\rm oq}] = 0$ in Eq.~(\ref{U-eqs}).

Because the action $S_{\rm univ}^{\rm (ovlp)}[g_{\mu \nu},
\psi_{\rm oq}]$ in Eq.~(\ref{univ-act}) is an {\it overlapping}
one, the ``universe B-T overlap" $\Sigma_{\rm B \cap T}^{\rm
(univ)} \stackrel{\rm def}{=} \Sigma_{\rm bullet}^{\rm (univ)}
\cap \Sigma_{\rm target}^{\rm (univ)}$ is not the empty set, i.e.,
\beq \label{U-ove} \Sigma_{\rm B \cap T}^{\rm (univ)} ~ \ne ~
\varnothing ~, \eeq 
where $\Sigma_{{\rm B} \cap {\rm T}}^{\rm (univ)} = \{ f_{\rm
ove}^A \}$ is assumed to contain a low-energy motion (e.g.,
$|\partial| \ll \Lambda_{\rm met}$) which the space 3-brane can
perform in the ambient spacetime $\mathbb{M}^{D_{\rm amb}}$. As in
Sec.~\ref{sec: AAT-method}, the universe B-T overlap $\Sigma_{{\rm
B} \cap {\rm T}}^{\rm (univ)} = \{ f_{\rm ove}^A \}$ is the
maximum knowledge which we can obtain about the universe target
$\Sigma_{\rm target}^{\rm (univ)} = \{ f_{\rm sol}^A \}$ by using
the overlapping universe action $S_{\rm univ}^{\rm (ovlp)}[g_{\mu
\nu}, \psi_{\rm oq}]$.

Until now, we have presented the ``two-step AAT method" for the
overlapping universe action $S_{\rm univ}^{\rm (ovlp)}[g_{\mu
\nu}, \psi_{\rm oq}]$ (cf.~Sec.~\ref{sec: AAT-method}):
\begin{itemize}
\item Step 1: finding a solution $g_{\mu \nu}^{\rm sol \cdot U}$
of the coupled equations $(\delta S_{\rm univ}^{\rm (ovlp)} /
\delta g_{\mu \nu})[g_{\mu \nu}, \psi_{\rm oq}] = 0$ and $(\delta
S_{\rm univ}^{\rm (ovlp)} / \delta \psi_{\rm oq})[g_{\mu \nu},
\psi_{\rm oq}] = 0$ in Eq.~(\ref{U-eqs}), and next
\item Step 2: finding a solution $f_{\rm sol}^A$ of the new rifle
PDE $\partial_\mu f^A \partial_\nu f^B \eta_{AB}^{\rm bulk} =
g_{\mu \nu}^{\rm sol \cdot U}$.
\end{itemize}

Instead of this two-step AAT method, as in Sec.~\ref{sec:
AAT-method}, we try another method of eliminating the metric
$g_{\mu \nu}$ from those E-L equations $\delta S_{\rm univ}^{\rm
(ovlp)} / \delta g_{\mu \nu} = 0$ and $\delta S_{\rm univ}^{\rm
(ovlp)} / \delta \psi_{\rm oq} = 0$ by inserting the PDE
$\partial_\mu f^A \partial_\nu f^B \eta_{AB}^{\rm bulk} = g_{\mu
\nu}$ into them. Namely, we solve the coupled replaced-equations
\beq \label{U-eqs-repl} ~~~~~~~~~~ \delta S_{\rm univ}^{\rm
(ovlp)} / \delta g_{\mu \nu} |_{\rm repl} \, = \, 0 ~~~~ {\rm and}
~~~ \delta S_{\rm univ}^{\rm (ovlp)} / \delta \psi_{\rm oq} |_{\rm
repl} \, = \, 0 ~~~~~ ({\rm cf.~Eq.}~(\ref{sol-subst})) ~, \eeq 
where $\delta S_{\rm univ}^{\rm (ovlp)} / \delta Z |_{\rm repl}
\stackrel{\rm def}{=} (\delta S_{\rm univ}^{\rm (ovlp)} / \delta
Z)[\partial_\mu f^A \partial_\nu f^B \eta_{AB}^{\rm bulk}, \,
\psi_{\rm oq}]$ for $Z = g_{\mu \nu}, \, \psi_{\rm oq}$. The
former replaced-equation $\delta S_{\rm univ}^{\rm (ovlp)} /
\delta g_{\mu \nu} |_{\rm repl} = 0$ is expressed as $\delta
\widetilde{S}_{\rm univ}^{\rm (ovlp)} / \delta(\partial_\mu f^A
\partial_\nu f^B \eta_{AB}^{\rm bulk}) = 0$, where
$\widetilde{S}_{\rm univ}^{\rm (ovlp)}[f^A, \psi_{\rm oq}]
\stackrel{\rm def}{=} S_{\rm univ}^{\rm (ovlp)}[\partial_\mu f^A
\partial_\nu f^B \eta_{AB}^{\rm bulk}, \psi_{\rm oq}]$
(cf.~Eqs.~(\ref{new-action}) and (\ref{var-new-action})).

Since solving the coupled replaced-equations in
Eq.~(\ref{U-eqs-repl}) is the same as solving the new rifle PDE
$\partial_\mu f^A \partial_\nu f^B \eta_{AB}^{\rm bulk} = g_{\mu
\nu}^{\rm sol \cdot U}$ of the two-step AAT method, the solution
set for Eq.~(\ref{U-eqs-repl}) 
\beq \label{U-sol-ovlp} \Sigma_{\rm univ}^{\rm (sol) } \,
\stackrel{\rm def}{=} \, \{\, f^A \, | ~\, \delta S_{\rm
univ}^{\rm (ovlp)} / \delta g_{\mu \nu} |_{\rm repl} = 0 ~~ {\rm
and} ~\, \delta S_{\rm univ}^{\rm (ovlp)} / \delta \psi_{\rm oq}
|_{\rm repl} = 0 ~ \} \eeq 
is equal to the universe bullet $\Sigma_{\rm bullet}^{\rm (univ)}
= \{ f_{\rm bul}^A \}$ in Eq.~(\ref{U-bullet}), namely, 
\beq \label{U-sol=bullet} ~~~~~~~~~~~ \Sigma_{\rm univ}^{\rm (sol)
} ~ = ~ \Sigma_{\rm bullet}^{\rm (univ)} ~~~~~ ({\rm
cf.~Eq.}~(\ref{sol-ovlp=bullet})) ~. \eeq 
This equality $\Sigma_{\rm univ}^{\rm (sol) } = \Sigma_{\rm
bullet}^{\rm (univ)}$ means that the universe bullet $\Sigma_{\rm
bullet}^{\rm (univ)} = \{ f_{\rm bul}^A \}$ is also produced by
the coupled replaced-equations $\delta S_{\rm univ}^{\rm (ovlp)} /
\delta g_{\mu \nu} |_{\rm repl} = \delta S_{\rm univ}^{\rm (ovlp)}
/ \delta \psi_{\rm oq} |_{\rm repl} = 0$.

Therefore, these replaced-equations $\delta S_{\rm univ}^{\rm
(ovlp)} / \delta g_{\mu \nu} |_{\rm repl} = \delta S_{\rm
univ}^{\rm (ovlp)} / \delta \psi_{\rm oq} |_{\rm repl} = 0$ can be
used instead of the E-L equations $\delta S_{\rm univ}^{\rm (3
br)} / \delta f^A = \delta S_{\rm univ}^{\rm (3 br)} / \delta
\Psi_{\rm OQ} = 0$ in Eq.~(\ref{br-eqs}) {\it within the universe
B-T overlap $\Sigma_{\rm B \cap T}^{\rm (univ)}$} (see below
Eq.~(\ref{sol-ovlp2=1})). In other words, within this B-T overlap
$\Sigma_{\rm B \cap T}^{\rm (univ)}$, the replaced-equations from
$S_{\rm univ}^{\rm (ovlp)}[g_{\mu \nu}, \psi_{\rm oq}]$ are
``equivalent" to the E-L equations from $S_{\rm univ}^{\rm (3
br)}[f^A, \Psi_{\rm OQ}]$.

When a ``single" overlapping action $S_{\rm univ}^{\rm
(ovlp)}[g_{\mu \nu}, \psi_{\rm oq}]$ is discovered as a result of
investigation, we can assume that the replaced-equations from this
discovered action $S_{\rm univ}^{\rm (ovlp)}[g_{\mu \nu},
\psi_{\rm oq}]$ are applied, at least, to many and various motions
which the space 3-brane can perform in the ambient spacetime
$\mathbb{M}^{D_{\rm amb}}$ at low enough energies. Namely, the AAT
method using the single discovered action $S_{\rm univ}^{\rm
(ovlp)}[g_{\mu \nu}, \psi_{\rm oq}]$ is valid for those many and
various low-energy motions of the space 3-brane. (For a further
study, see our next paper \cite{Bae}.) Of course, the discovered
action $S_{\rm univ}^{\rm (ovlp)}[g_{\mu \nu}, \psi_{\rm oq}]$ can
change, depending on observation energies.

Suppose a low-energy motion of the space 3-brane is described by a
4DL solution $f_{\rm sol}^A(x^\mu)$. Then, each momentum $p^{\,
\mu}$ in the Fourier transform of $f_{\rm sol}^A(x^\mu)$ satisfies
$|\, p^{\, \mu} | \ll \Lambda_{\rm cont}$ for all $\mu$. In this
Fourier-transform context, the low-energy motion $f_{\rm
sol}^A(x^\mu)$ is expressed as 
\beq \label{LE-motion} |\partial| ~ \ll ~ \Lambda_{\rm cont} ~.
\eeq 
For this low-energy motion $f_{\rm sol}^A$ of $|\partial| \ll
\Lambda_{\rm cont}$, the U-metric $g_{\mu \nu}^{\rm sol \cdot U}$
($= \partial_\mu f_{\rm sol}^A \partial_\nu f_{\rm sol}^B
\eta_{AB}^{\rm bulk}$ by Eq.~(\ref{equal-U-met})) is also
expressed as $|\partial| \ll \Lambda_{\rm cont}$.

As in Sec.~\ref{sec: form-met}, we choose the ``invariant case"
that the overlapping action $S_{\rm univ}^{\rm (ovlp)}[g_{\mu
\nu}, \psi_{\rm oq}]$ in Eq.~(\ref{univ-act}) is invariant under
$ISO(1, D_{\rm amb} - 1)$ and ${\rm Diff}(4)$ like the original
one $S_{\rm univ}^{\rm (3 br)}[f^A, \Psi_{\rm OQ}]$, irrespective
of whether $S_{\rm univ}^{\rm (ovlp)}[g_{\mu \nu}, \psi_{\rm oq}]$
is an ME or PE action. Thus, the metric Lagrangian ${\cal L}_{\rm
met}^{\rm (ovlp)}(\Lambda_{\rm met}; g_{\mu \nu})$ in $S_{\rm
univ}^{\rm (ovlp)}[g_{\mu \nu}, \psi_{\rm oq}]$ shares the same
form with that in Eqs.~(\ref{ind-deriv-exp}) and
(\ref{gen-met-Lag}). Then, the ${\rm Diff}(4)$-invariant universe
action $S_{\rm univ}^{\rm (ovlp)}[g_{\mu \nu}, \psi_{\rm oq}]$ can
contain (i) the Einstein-Hilbert action and (ii) the action for
matter (i.e., occupant quanta), both of which are the essential
parts of General Relativity.

Within the region $|\partial| \ll \Lambda_{\rm cont}$ in
Eq.~(\ref{LE-motion}), we have the equality for the two spacetime
manifolds (see Eqs.~(\ref{equal-U-ST}) and (\ref{equal-U-met})) 
\beq \label{equal-ST-U} ({\cal S}_{\rm univ}^{\rm \, 4D}, \,
g_{\mu \nu}^{\rm sol \cdot U}) ~ = ~ ({\cal WV}_{\rm sq}, \,
\gamma_{\mu \nu}) ~. \eeq 
Note the 4D emergent spacetime $({\cal WV}_{\rm sq}, \gamma_{\mu
\nu})$ is determined by the 4DL solution $f_{\rm sol}^A$ for the
E-L equations $\delta S_{\rm univ}^{\rm (3 br)} / \delta f^A =
\delta S_{\rm univ}^{\rm (3 br)} / \delta \Psi_{\rm OQ} = 0$ in
Eq.~(\ref{U-target}). This emergent manifold $({\cal WV}_{\rm sq},
\gamma_{\mu \nu})$ occupying $\mathbb{M}^{D_{\rm amb}}$ is the
exact or true spacetime which provides exact values for our
spacetime measurements (see below Eq.~(\ref{equal-ST})).

As said below Eq.~(\ref{univ-met-act}), since the metric
Lagrangian ${\cal L}_{\rm met}^{\rm (ovlp)}(\Lambda_{\rm met};
g_{\mu \nu})$ in Eq.~(\ref{univ-met-act}) shares the same form
with that in Eqs.~(\ref{ind-deriv-exp}) and (\ref{gen-met-Lag}),
we use the same notations. Due to the power-law behaviors
$(\partial / \Lambda_{\rm met})^{d_k}$ in
Eqs.~(\ref{ind-deriv-exp}) and (\ref{gen-met-Lag}), the most
dominant term in ${\cal L}_{\rm met}^{\rm (ovlp)}(\Lambda_{\rm
met}; g_{\mu \nu})$ for $|\partial| \ll \Lambda_{\rm met}$ is the
$d_k = 0$ Lagrangian ${\cal L}_{\rm met}^{(0)} \stackrel{\rm
def}{=} c_{\, 0 \,} \Lambda_{\rm met}^4$, which contributes to a
cosmological constant. Moreover, the next dominant term is the
$d_k = 2$ Lagrangian ${\cal L}_{\rm met}^{(2)} \stackrel{\rm
def}{=} c_{\, 2 \,} \Lambda_{\rm met}^2 R$, which contains only
the two-derivative terms of the metric $g_{\mu \nu}$.

As assumed before, the AAT method using the overlapping universe
action $S_{\rm univ}^{\rm (ovlp)}[g_{\mu \nu}, \psi_{\rm oq}]$ is
applied to various low-energy motions (i.e., $|\partial| \ll
\Lambda_{\rm cont}$) of the space 3-brane. Within the region
$|\partial| \ll \Lambda_{\rm cont}$, we can find a low-energy
region $|\partial| \ll \Lambda_{\rm met}$ ($\lesssim
O(\Lambda_{\rm cont}$)) in which the overlapping action $S_{\rm
univ}^{\rm (ovlp)}[g_{\mu \nu}, \psi_{\rm oq}]$ has the
approximation 
\beq \label{2d-approx-U} S_{\rm univ}^{\rm (ovlp)}[g_{\mu \nu},
\psi_{\rm oq}] ~ \approx ~ S_{\rm univ}^{(\le \, 2)}[g_{\mu \nu},
\psi_{\rm oq}^{\rm low}] ~, \eeq 
where 
\bea \label{2d-act-U} ~~ S_{\rm univ}^{(\le \, 2)}[g_{\mu \nu},
\psi_{\rm oq}^{\rm low}] ~ &\stackrel{\rm def}{=}& ~ S_{\rm
met}^{(\le \, 2)}[g_{\mu \nu}] \, + \, S_{\rm OQ}^{\rm
(low)}[\psi_{\rm oq}^{\rm low}, g_{\mu \nu}] ~, \\
\label{2-der-act} S_{\rm met}^{(\le \, 2)} [g_{\mu \nu}] ~
&\stackrel{\rm def}{=}& \, \int_{{\cal S}_{\rm univ}^{\rm 4D \,
(\le \, 2)}} d^{\, 4} x \sqrt{|\, {\rm det}(g_{\mu \nu})|} ~
\left( c_{\, 0 \,} \Lambda_{\rm met}^4 + c_{\, 2 \,} \Lambda_{\rm
met}^2 R \, \right) \, . \eea 

In Eq.~(\ref{2d-act-U}), the new OQ action $S_{\rm OQ}^{\rm
(low)}[\psi_{\rm oq}^{\rm low}, g_{\mu \nu}]$ is the low-energy
approximation of its full theory $S_{\rm OQ}^{\rm
(ovlp)}[\psi_{\rm oq}, g_{\mu \nu}]$ in Eq.~(\ref{OQ-act}).
Namely, $S_{\rm OQ}^{\rm (low)}[\psi_{\rm oq}^{\rm low}, g_{\mu
\nu}]$ contains only the low-dimension interactions of $S_{\rm
OQ}^{\rm (ovlp)}[\psi_{\rm oq}, g_{\mu \nu}]$ which are not
negligible in the low-energy region $|\partial| \ll \Lambda_{\rm
met}$. Of course, some heavy OQ fields (say, $\psi_{\rm oq}^{\rm
heavy}$) appearing in $S_{\rm OQ}^{\rm (ovlp)}[\psi_{\rm oq},
g_{\mu \nu}]$ may be decoupled from its low-energy approximation
$S_{\rm OQ}^{\rm (low)}[\psi_{\rm oq}^{\rm low}, g_{\mu \nu}]$. In
Eq.~(\ref{2-der-act}), the integral for $S_{\rm met}^{(\le \,
2)}[g_{\mu \nu}]$ undergoes the replacement ${\cal S}_{\rm
univ}^{\rm \, 4D} \Rrightarrow {\cal S}_{\rm univ}^{\rm \, 4D \,
(\le \, 2)}$, which is also undergone by the integral for $S_{\rm
OQ}^{\rm (low)}[\psi_{\rm oq}^{\rm low}, g_{\mu \nu}]$.

Due to the approximate equality $S_{\rm univ}^{\rm (ovlp)} \approx
S_{\rm univ}^{(\le \, 2)}$ in Eq.~(\ref{2d-approx-U}), the
solution ($g_{\mu \nu}^{\rm sol \cdot U}, \, \psi_{\rm oq}^{\rm
sol}$) of the ``exact" equations $\delta S_{\rm univ}^{\rm (ovlp)}
/ \delta g_{\mu \nu} = 0$ and $\delta S_{\rm univ}^{\rm (ovlp)} /
\delta \psi_{\rm oq} = 0$ in Eq.~(\ref{U-eqs}) has the approximate
equalities 
\beq \label{approx-g-OQ-U} g_{\mu \nu}^{\rm sol \cdot U} ~ \approx
~ g_{\mu \nu}^{\rm sol \, (\le \, 2)} ~~~~ {\rm and} ~~~~
\psi_{\rm oq}^{\rm sol} ~ \approx ~ \psi_{\rm oq}^{\rm sol \, (\le
\, 2)} ~, \eeq 
where ($g_{\mu \nu}^{\rm sol \, (\le \, 2)}, \, \psi_{\rm oq}^{\rm
sol \, (\le \, 2)}$) is the solution of the ``approximate"
equations $\delta S_{\rm univ}^{(\le \, 2)} / \delta g_{\mu \nu} =
0$ and $\delta S_{\rm univ}^{(\le \, 2)} / \delta \psi_{\rm oq} =
0$. In addition, $g_{\mu \nu}^{\rm sol \cdot U} \approx g_{\mu
\nu}^{\rm sol \, (\le \, 2)}$ in Eq.~(\ref{approx-g-OQ-U}) implies
the approximate equality for the spacetime 
\beq \label{approx-ST-U} {\cal S}_{\rm univ}^{\rm \, 4D} ~ \approx
~ {\cal S}_{\rm univ}^{\rm \, 4D \, (\le \, 2)} ~. \eeq 

In the low-energy region $|\partial| \ll \Lambda_{\rm met}$, the
approximate equation $\delta S_{\rm univ}^{(\le \, 2)} / \delta
g_{\mu \nu} = 0$ is the same as Einstein's equation with the three
parameters $c_{\, 0}$, $c_{\, 2}$ and $\Lambda_{\rm met}$ 
\beq \label{Eins-eq} G_{\mu \nu} \, - \, \frac{c_{\, 0 \,}
\Lambda_{\rm met}^2}{2 c_{\, 2 \,}} ~ g_{\mu \nu} ~\, = ~\,
\frac{1}{2 c_{\, 2 \,} \Lambda_{\rm met}^2} ~ T_{\mu \nu}^{\rm
(low)} ~, \eeq 
where $1/(2 c_2 \Lambda_{\rm met}^2)$ corresponds to $8 \pi G_{\rm
N}$ of the ordinary Einstein's equation, and 
\beq \label{OQ-EM} T_{\mu \nu}^{\rm (low)} ~ \stackrel{\rm def}{=}
\, - \, \frac{2}{\sqrt{|\, {\rm det}(g_{\alpha \beta})|} \,} \,
\frac{\delta S_{\rm OQ}^{\rm (low)}}{\delta g^{\mu \nu}} ~. \eeq
For $\gamma_{\mu \nu} = g_{\mu \nu}$, the tensor $T_{{\rm OQ} \,
\mu \nu}$ in Eq.~(\ref{br-OQ-EM}) satisfies $T_{{\rm OQ} \, \mu
\nu} \approx T_{\mu \nu}^{\rm (low)}$ at the low energies
$|\partial| \ll \Lambda_{\rm met}$.

Suppose that the ``${\rm scalar} \times g_{\mu \nu}$" term in
Eq.~(\ref{Eins-eq}) is negligible as in the observed $\Lambda$CDM
model \cite{Cosm-Obs}. Then, a spherical massive object can
produce the Schwarzschild metric $g_{\mu \nu}^{\rm (S)}$, which
leads to the gravitational potential $\phi_{\rm grav} = -
\,(g_{00}^{\rm (S)} + 1)/2$ in the Newtonian limit
\cite{{GR-Schu},{GR-Wein},{GR-Carr}}. Since this potential
$\phi_{\rm grav}$ ($\propto 1 / c_{\, 2 \,} \Lambda_{\rm met}^2$)
depends on $c_{\, 2 \,} \Lambda_{\rm met}^2$ strongly, the value
of $c_{\, 2 \,} \Lambda_{\rm met}^2$ can be easily determined by
the comparison with observed data.

In Eq.~(\ref{2-der-act}), when the coefficient $c_{\, 2 \,}
\Lambda_{\rm met}^2$ in the integrand satisfies 
\beq \label{cutoff} ~~~~~~~~~~~~~~~ c_{\, 2 \,} \Lambda_{\rm
met}^2 \, = \, 1 / 16 \pi G_{\rm N} ~~~~ ({\rm i.e.,} ~\,
\Lambda_{\rm met} = M_{\rm P}/ \sqrt{16 \pi c_{\, 2}} ~) ~, \eeq
the approximate metric action $S_{\rm met}^{(\le \, 2)}[g_{\mu
\nu}]$ may not be distinguished from the Einstein-Hilbert action
$S_{\rm EH}^{\rm (DE)}[{\bf g}_{\mu \nu}]$ with a dark energy (DE)
density $\rho_{\rm DE}$ 
\beq \label{EH+DE} S_{\rm EH}^{\rm (DE)}[\, {\bf g}_{\mu \nu}] ~
\stackrel{\rm def}{=} \, \int_{{\cal S}_{\rm GR}} d^{\, 4} x
\sqrt{|\, {\rm det}({\bf g}_{\mu \nu})|} ~ ( \textit{\textbf{R}} /
16 \pi G_{\rm N} - \, \rho_{\rm DE} ) ~, \eeq 
where the Ricci scalar $\textit{\textbf{R}}$ of General Relativity
(GR) is built from the GR metric ${\bf g}_{\mu \nu}$.

To be more concrete, we consider in what situation General
Relativity is valid: since using General Relativity of the
$(\partial/M_{\rm P})^{d_k \le 2}$ terms means neglecting all the
higher-order $(\partial/M_{\rm P})^{d_k \ge 4}$ terms of the
Lagrangian ${\cal L}_{\rm met}^{\rm (ovlp)}(\Lambda_{\rm met};
g_{\mu \nu})$ in Eq.~(\ref{univ-met-act}), it is important to
estimate the size of $\partial$. As a measure of $|\partial|$,
Kretschmann scalar $K \stackrel{\rm def}{=}
\textit{\textbf{R}}^{\mu \nu \rho \sigma} \textit{\textbf{R}}_{\mu
\nu \rho \sigma}$ is used due to $K = O(\partial^{\, 4})$.

In this context, we deal with an extremely strong gravity related
to a Schwarzschild black hole of mass $M_{\rm bh}$, whose
Kretschmann scalar is $K |_{{\rm at} \, r} = 48 M_{\rm bh}^2 /
M_{\rm P}^{4 \,} r^6$ \cite{GR-Carr}. Outside the Schwarzschild
radius $R_{\rm S} = 2 M_{\rm bh} / M_{\rm P}^2$ (i.e., $r > R_{\rm
S}$), the scalar satisfies $K |_{{\rm at} \, r} < K |_{{\rm at} \,
R_{\rm S}}$, which leads to $|\partial|/M_{\rm P} \lesssim M_{\rm
P} / M_{\rm bh}$ due to $K |_{{\rm at} \, r} = O(\partial^{\, 4})$
and $K |_{{\rm at} \, R_{\rm S}} = O(M_{\rm P} / M_{\rm bh})^4
M_{\rm P}^4$. Then, for $M_{\rm bh} \gtrsim M_\odot$ ($\approx
10^{38} M_{\rm P}$), the result $|\partial|/M_{\rm P} \ll 1$
implies that General Relativity is valid outside the event horizon
at $r = R_{\rm S}$.

Meanwhile, inside this event horizon, there is a radius $R_{\rm
\infty}$ satisfying $K |_{{\rm at} \, R_{\rm \infty}} = O(M_{\rm
P}^4)$, which produces $R_{\rm \infty} = O(M_{\rm bh}/M_{\rm
P})^{1/3} M_{\rm P}^{-1}$ ($M_{\rm P}^{-1} \ll R_{\rm \infty} \ll
R_{\rm S}$). For $r \lesssim R_{\rm \infty}$, $|\partial|/M_{\rm
P} \gtrsim 1$ (i.e., $d_{\rm max} \rightarrow \infty$) implies
that General Relativity is not valid far inside the event horizon.
Similarly, for $r \ll R_{\rm cont}$ with $K |_{{\rm at} \, R_{\rm
cont}} = O(\Lambda_{\rm cont})^4$, $|\partial|/\Lambda_{\rm cont}
\gg 1$ implies that the continuum approximation of the quasi-3D
object breaks down{---\,}the above black hole may not have the
singularity at its center $r=0$.

To sum up, in the low-energy region $|\partial| \ll \Lambda_{\rm
met}$, General Relativity can be a good approximation of the
overlapping universe action $S_{\rm univ}^{\rm (ovlp)}[g_{\mu
\nu}, \psi_{\rm oq}]$, which is an essential part of the AAT
method for studying the original universe action $S_{\rm
univ}^{\rm (3 br)}[f^A, \Psi_{\rm OQ}]$ (i.e., the principle
governing the motions of the space 3-brane). Note that our
spacetime ${\cal WV}_{\rm sq}$ ($\approx {\cal S}_{\rm GR}$) can
have its own gravity (i.e., General Relativity) although the
ambient spacetime ${\cal S}^{D_{\rm amb}}$ does not have any
``bulk gravity" (i.e., ${\cal S}^{D_{\rm amb}} =
\mathbb{M}^{D_{\rm amb}}$).

In the case of 
\beq \label{met-2=EH} S_{\rm met}^{(\le \, 2)}[{\bf g}_{\mu \nu}]
\, = \, S_{\rm EH}^{\rm (DE)}[{\bf g}_{\mu \nu}] ~, \eeq 
the solution metric ${\bf g}_{\mu \nu}^{\rm sol}$ and the
spacetime ${\cal S}_{\rm GR}$ of General Relativity have the
equalities 
\bea \label{GR-met=g-2} ~~~~~~~~~ {\bf g}_{\mu \nu}^{\rm sol} \,
&=& \, g_{\mu \nu}^{\rm sol \, (\le \, 2)} ~~ (\approx \, g_{\mu
\nu}^{\rm sol \cdot U} = \gamma_{\mu \nu} ~~ {\rm due ~ to ~
Eqs.}~(\ref{equal-ST-U}) ~ {\rm and} ~ (\ref{approx-g-OQ-U})) ~, \\
\label{GR-ST=ST-2} {\cal S}_{\rm GR} \, &=& \, {\cal S}_{\rm
univ}^{\rm \, 4D \, (\le \, 2)} ~~ (\approx \, {\cal S}_{\rm
univ}^{\rm \, 4D} = {\cal WV}_{\rm sq} ~~ {\rm due ~ to ~
Eqs.}~(\ref{equal-ST-U}) ~ {\rm and} ~ (\ref{approx-ST-U})) ~.
\eea 

According to Eqs.~(\ref{GR-met=g-2}) and (\ref{GR-ST=ST-2}), the
spacetime (${\cal S}_{\rm GR}, {\bf g}_{\mu \nu}^{\rm sol}$) of
General Relativity can be at least a good approximation of the
exact or true spacetime (${\cal WV}_{\rm sq}, \gamma_{\mu \nu}$),
which is formed by many space quanta occupying the ambient
spacetime $\mathbb{M}^{D_{\rm amb}}$. This supports the
space-quantum hypothesis in Eq.~(\ref{sq-hyp}). If the exact
equality $S_{\rm univ}^{\rm (ovlp)} = S_{\rm univ}^{(\le \, 2)}$
really happens instead of the approximate one in
Eq.~(\ref{2d-approx-U}), the ``approximate equality" signs
$\approx$ in Eqs.~(\ref{approx-g-OQ-U}), (\ref{approx-ST-U}),
(\ref{GR-met=g-2}) and (\ref{GR-ST=ST-2}) are replaced with the
equality signs $=$.

Finally, until now, we have considered only the special situation
that the ambient spacetime ${\cal S}^{D_{\rm amb}}$ is the flat
manifold $\mathbb{M}^{D_{\rm amb}} = (\mathbb{R}^{D_{\rm amb}},
\eta_{AB}^{\rm bulk})$, in which the inertial bulk observer
$O_{\rm bulk}$ uses the inertial bulk-coordinates $Y^A$ (see
Sec.~\ref{sec: cont-approx}).

However, the ambient spacetime ${\cal S}^{D_{\rm amb}}$ can be a
{\it curved} manifold having a general bulk metric $g_{AB}^{\rm
bulk}$, implying the replacements 
\bea \label{rep-S} \mathbb{M}^{D_{\rm amb}} ~~~ & \Rrightarrow &
~~~ {\cal S}^{D_{\rm amb}} ~, \\ \label{rep-g} \eta_{AB}^{\rm
bulk} ~~~ & \Rrightarrow & ~~~ g_{AB}^{\rm bulk} ~, \\
\label{rep-diff} ISO(1, D_{\rm amb} - 1) ~~~ & \Rrightarrow & ~~~
{\rm Diff}(D_{\rm amb}) ~. ~~~~~ \eea 
For these replacements, our previous studies can be extended
similarly.

The topology of the ambient spacetime ${\cal S}^{D_{\rm amb}}$ may
be, for example, $\mathbb{R}^{D_{\rm amb}}$ or $\mathbb{R}^4
\times \mathbb{T}^{D_{\rm amb} - 4}$, where $\mathbb{T}^{D_{\rm
amb} - 4}$ is a ($D_{\rm amb} - 4\,$)-dimensional spacelike torus.
For $D_{\rm amb}=4$, when the topology of ${\cal S}^{D_{\rm amb}}$
is $\mathbb{R}^4$, the topology of the space 3-brane can be
$\mathbb{R}^3$, implying the world volume ${\cal WV}_{\rm sq}$ of
this space 3-brane may be spatially flat. Then, due to ${\cal
S}_{\rm GR} \approx {\cal WV}_{\rm sq}$ in Eq.~(\ref{GR-ST=ST-2}),
the corresponding spacetime ${\cal S}_{\rm GR}$ of General
Relativity may be spatially flat, which can agree with the
observed $\Lambda$CDM model \cite{Cosm-Obs}. For $D_{\rm amb} \ge
5$, the same conclusions can be reached even for the topology
$\mathbb{R}^4 \times \mathbb{T}^{D_{\rm amb} - 4}$ of ${\cal
S}^{D_{\rm amb}}$, when the size of this torus $\mathbb{T}^{D_{\rm
amb} - 4}$ is much smaller than the distance $d_{\rm \, sq}$
between space quanta{\,---\,}at low energies $\ll \Lambda_{\rm
cont}$, the bulk spacetime ${\cal S}^{D_{\rm amb}}$ can be
observed as if its topology were $\mathbb{R}^4$.


~\\

\section*{Acknowledgements}

We would like to thank Prof.~K.~Y.~Lee, Prof.~I.~Park,
Dr.~J.-H.~Park, Dr.~H.-S.~Lee, and Prof.~E.~J.~Chun for reading
this paper and expressing their opinions.

~\\~\\





%
%
\end{document}